\documentclass[a4paper,12pt]{report}
\pdfoutput=1
\usepackage[comma]{natbib}

\usepackage{graphicx}
\usepackage{amsbsy}
\usepackage{amscd}
\usepackage{amsgen}
\usepackage{amsmath}
\usepackage{amsopn}
\usepackage{amstext}
\usepackage{amsxtra}
\usepackage{amssymb}
\usepackage{graphics}
\usepackage{makeidx}
\makeindex

\usepackage{makeglos}
\makeglossary

 \newtheorem{definition}{Definition}[chapter]
 \newtheorem{remark}{Remark}[chapter]
 \newtheorem{theorem}{Theorem}[chapter]
\newtheorem{lemma}{Lemma}[chapter]

\newtheorem{proposition}{Proposition}[chapter]
\newtheorem{example}{Example}[chapter]

\newcommand{\proof } {\noindent{\bf{Proof.}} }
\newcommand{\tr} {\noindent{\rm{tr}} }
\renewcommand{\baselinestretch}{1.00}

\begin{document}


\begin{titlepage}
\begin{center}
\vspace*{1in}
{\LARGE Topics in Estimation of Quantum Channels}
\par
\vspace{0.3in}
{\large Caleb J. O'Loan}
\par
\vspace{0.3in}
\begin{figure}[htp]
\centering\includegraphics[totalheight=0.3\textheight]{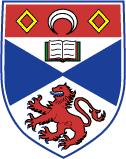}
\end{figure}
\par
\vspace{0.5in}
A Thesis submitted \\
to the University of St Andrews \\
in application for the degree of \\
 Doctor of Philosophy
\par
\vspace{0.2in}
8th December 2009

\end{center}

\end{titlepage}

\pagenumbering{roman}

\chapter*{Declaration}
I, Caleb J O'Loan, hereby certify that this thesis, which is approximately $23,000$ words in length, has been written by me, that it is the record of work carried out by me and that it has not been submitted in any previous application for a higher degree. 
\\
\\
I was admitted as a research student in September, 2005 and as a candidate for the degree of Ph.D. in September, 2006; the higher study for which this is a record was carried out in the University of St Andrews between 2005 and 2009. 
\\
\\
date ...............      $\qquad$ signature of candidate ..........................   
\\
\\
I hereby certify that the candidate has fulfilled the conditions of the Resolution and Regulations appropriate for the degree of иии in the University of St Andrews and that the candidate is qualified to submit this thesis in application for that degree. 
\\
\\
date ...............       $\qquad$ signature of supervisor ..........................  
\\
\\
In submitting this thesis to the University of St Andrews we understand that we are giving permission for it to be made available for use in accordance with the regulations of the University Library for the time being in force, subject to any copyright vested in the work not being affected thereby.  We also understand that the title and the abstract will be published, and that a copy of the work may be made and supplied to any bona fide library or research worker, that my thesis will be electronically accessible for personal or research use, and that the library has the right to migrate my thesis into new electronic forms as required to ensure continued access to the thesis. We have obtained any third-party copyright permissions that may be required in order to allow such access. 
\\
\\
The following is an agreed request by candidate and supervisor regarding the electronic publication of this thesis:
\\
\\
Access to Printed copy and electronic publication of thesis through the University of St Andrews.
\\
\\
date ...............        
\\
\\
signature of candidate .......................... 
\\
\\
\\
signature of supervisor ..........................  



\chapter*{Acknowledgements} 
This thesis would not have been completed without the patient and helpful supervision of Prof.\ Peter E.\ Jupp. Many thanks go to him. I am also grateful to Dr Terry Rudolph for good advice, for the continued support of my family, and to Richard Gormley for checking the final manuscript.

\tableofcontents

\begin{abstract}
A quantum channel is a mapping which sends density matrices to density matrices. The estimation of quantum channels is of great importance to the field of quantum information.
In this thesis two topics related to estimation of quantum channels are investigated.
The first of these is the upper bound of \cite{saromilb06} on the Fisher information obtainable by measuring the output of a channel.
Two questions raised by Sarovar and Milburn about their bound are answered. A Riemannian metric on the space of quantum states is introduced, related to the construction of the Sarovar and Milburn bound.   
Its properties are characterized. 
 
The second topic investigated is the estimation of unitary channels. 
The situation is considered in which an experimenter has several
non-identical unitary channels that have the same parameter.
It is shown that it is possible to improve estimation using the channels
together, analogous to the case of identical unitary channels.
Also, a new method of phase estimation is given based on a method sketched 
by \cite{kit95}.
Unlike other phase estimation procedures which perform similarly,
this procedure requires only very basic experimental resources. 

\end{abstract}

\pagenumbering{arabic}

\chapter{Mathematical Background}    \label{ch:mathback}

\section{Overview}
This thesis is concerned with estimation of quantum channels. Almost every protocol in quantum information uses quantum channels. They are used in important protocols such as teleportation,  Deutsch's algorithm, the Grover search algorithm and the Shor factorization algorithm \cite[Chapters 5 and 7]{bellac06}. 
In theory it is assumed that a channel is known precisely, yet in practice this will not generally be the case. Thus the estimation of quantum channels is of fundamental importance to the field of quantum information. 

Chapter \ref{ch:mathback} contains the mathematical and quantum-theoretic background needed to understand the thesis.
Chapters \ref{ch:SM1} and \ref{ch:smqi} are concerned with the upper bound of \cite{saromilb06} on the  Fisher information obtainable by measuring the output of a channel.
Chapters \ref{ch:niprl} and 5 consider estimation of unitary channels.
 
In Chapter  \ref{ch:mathback} definitions are given of fundamental objects such as quantum systems,  quantum states, quantum measurements and combined systems.
Quantum channels are defined and quantum channel estimation introduced.
A brief historical background is given of the key developments in channel estimation.
Chapter  \ref{ch:mathback}  contains also a few small, new results.
 
Chapter  \ref{ch:SM1} considers work by \cite{saromilb06}, who introduced an upper bound on the Fisher information obtained by measuring the output states of quantum channels. They showed that for certain channels, called quasi-classical channels, their bound is attainable.
They asked (i) whether their bound is attainable more generally; (ii) whether or not it is possible to find an explicit expression for measurements attaining this bound. 
Both of these questions  are answered in Chapter \ref{ch:SM1}.

In the process of answering the previous questions, Chapter \ref{ch:SM1} shows that Sarovar and Milburn's work leads to a new Riemannian metric on the space of quantum states.
Chapter \ref{ch:smqi}  considers the questions: What are the properties of this new metric? Is it well defined? 
 
 Chapters $4$ and $5$ are concerned with the cost of estimation of unitary channels. It is known that when there are $n$ identical copies of a unitary channel, there exists \citep{kahn07} an estimation procedure such that the cost function (a function of the expected fidelity, see (\ref{eq.fhuu3}))   is $O(1/n^2)$, instead of the usual $O(1/n)$.
Chapter \ref{ch:niprl} considers the question: If there are $n$ unitary channels which are not identical, but have the same parameter, is an analogous speed-up possible? 
 
  \cite{kit95} sketched an iterative method for phase estimation such that the cost function is $O((\log n/n)^2)$.
  This method requires only a single copy of a unitary channel and basic measurements.
  In Chapter \ref{ch:ipe} it is shown that several attempts to give a detailed method for iterative phase estimation have been unsuccessful.
  There have been other successful iterative methods, but these require an extra rotation gate capable of performing arbitrary rotations with almost perfect accuracy.
  Thus Chapter \ref{ch:ipe} seeks to answer the question: Does a complete iterative phase estimation method exist which requires only a single copy of the unitary and basic measurements?

\section{System} \label{sysa}
A {\it quantum system} is a physical system that obeys the laws of quantum mechanics.
The {\it state} of a quantum system (or {\it quantum state}, or {\it quantum state of a system}) is a quantification of the system, which, if known, allows an experimenter to make accurate predictions about the results of any future measurements on that system \citep{gill01t}.
Since measurement results are probabilistic, knowledge of a quantum state means that, given any measurement, it is possible to work out the long-term relative frequency of the observed outcomes.  

A quantum system is represented by a complex Hilbert space $\mathcal{H}$ of dimension $d$,\index{Hilbert Space}
with a Hermitian inner product.
The dimension $d$ is given by the maximum number of distinguishable states in the system. For the spin of an electron, or the polarisation of a photon,
$\mathcal{H} = \mathbb{C}^2$. This is because there are only two distinguishable states: spin up and spin down. Any other quantum state can be represented as a complex linear combination of these two states.

In this thesis only finite dimensional complex vector spaces are considered.
Any column vector in a complex vector space is denoted by $| \psi \rangle$, the symbol $\psi$ is a label, while $| \cdot \rangle$ denotes that the object is a complex column vector. This representation of complex vectors is called {\it Dirac notation}.
Given a vector
\begin{equation}
| \psi \rangle = \left( \begin{array}{c}
\psi_1  \\
\psi_2 \\
. \\
. \\
\psi_n 
\end{array} \right),
\label{eq.vecpsi}
\end{equation}
 its dual $\langle \psi |$ is defined as (with $\psi_j^*$ denoting the complex conjugate of $\psi_j$) 
 \begin{equation}
\langle \psi | = \left( \psi_1^*, \psi_2^*, \dots, \psi_n^*  \right).
\label{eq.vecpsid}
\end{equation}

 An {\it inner product} \index{inner product} is a bilinear map that maps a pair of complex vectors to a complex number.
Given the vectors $| \psi \rangle$ and $| \phi \rangle$,  the inner product between these vectors is denoted by $\langle \psi | \phi \rangle$. There are many different inner products. In this thesis, only  the following inner product will be used

\begin{eqnarray}
 \langle \psi | \phi \rangle &=& \sum_{i=1}^n \psi_i^* \phi_i  \label{innerp1}\\
& =&  \left( \psi_1^*, \psi_2^*, \dots, \psi_n^*  \right) \left( \begin{array}{c}
\phi_1  \\
\phi_2 \\
. \\
. \\
\phi_n 
\end{array} \right).
\end{eqnarray}

The {\it norm} of a vector $ | \psi \rangle$ can be defined as $\| \psi \| = \sqrt{\langle \psi | \psi \rangle }$.
Vectors with norm equal to one are defined as {\it unit vectors}.
Vectors $| \psi \rangle$ and $| \phi \rangle$ are {\it orthogonal} if their inner product is zero.

Given the vectors $| \psi \rangle$ and $| \phi \rangle$, the  {\it outer product}  $| \phi \rangle \langle  \psi |$ is given by
\begin{eqnarray}
| \phi \rangle \langle  \psi | & =&  \left( \begin{array}{c}
\phi_1  \\
\phi_2 \\
. \\
. \\
\phi_n 
\end{array} \right)
 \left( \psi_1^*, \psi_2^*, \dots, \psi_n^*  \right)\\
 \label{eq,op1}
 & =&  \left( \begin{array}{cccc}
\phi_1  \psi_1^*& \phi_1 \psi_2^*& \dots & \phi_1 \psi_n^*\\
\phi_2  \psi_1^*& \phi_2 \psi_2^*& \dots & \phi_2 \psi_n^*\\
\cdots & \cdots &\cdots & \cdots \\
\cdots & \cdots &\cdots & \cdots \\
\phi_n  \psi_1^*& \phi_n \psi_2^*& \dots & \phi_n \psi_n^*
\end{array} \right) .
 \label{eq,op2}
\end{eqnarray}

In Dirac notation $\langle \psi | \phi \rangle$ represents a complex number, and $| \phi \rangle \langle  \psi |$ a matrix.

A set of vectors $| u_1 \rangle, \dots, | u_n \rangle$ is {\it orthonormal} \index{orthonormal basis} if the vectors are normalized and orthogonal, i.e.\ $ \langle u_i | u_j \rangle = \delta_{ij}$. Given a set of orthonormal vectors $| u_1 \rangle, \dots, | u_n \rangle$ in a vector space $V$, such that $n = \mathrm{dim} \, V$, this set of vectors forms an orthonormal basis of $V$.
Any vector $| v \rangle$ in $V$ can be written as a scalar multiple of these vectors, i.e.
\begin{equation}
| v \rangle  = \sum_{i=1}^n v_i | u_i \rangle, \quad v_i = \langle u_i | v \rangle \in \mathbb{C}.
\end{equation}

\begin{lemma}
Given an orthonormal basis $| u_1 \rangle, \dots, | u_n \rangle$ for a vector space $V$, 
\begin{equation}
\sum_{i=1}^n | u_i \rangle \langle u_i | = \mathbb{I}_n.
\end{equation}
 \end{lemma}
  This is called the {\it completeness relation} \cite[p.\ 67]{chuang00}.
 \\
 \\
 \proof
 If   $| u_1 \rangle, \dots, | u_n \rangle$ is an orthonormal basis for $V$, then any $|v \rangle \in  V$ can be written as 
 $| v \rangle  = \sum_i v_i | u_i \rangle$, where $v_i = \langle u_i | v \rangle$.
 Now,
 \begin{equation}
\left(  \sum_{i=1}^n | u_i \rangle \langle u_i |  \right) | v \rangle = \sum_{i=1}^n v_i | u_i \rangle = | v \rangle.
 \end{equation}
Since this holds for all $| v \rangle$, the result follows.

The {\it Hermitian transpose} of a matrix $A$, denoted by $A^\dagger$, is the matrix found by taking the transpose of $A$ and replacing each entry with its complex conjugate $\left([A^\dagger]_{ij} = [A]_{ji}^* \right)$.
A {\it Hermitian matrix}  (also called a {\it self-adjoint matrix}) \index{Hermitian matrix} \index{self adjoint matrix} is a matrix which is equal to its Hermitian transpose, i.e.\ $B$ is Hermitian if $B^\dagger = B$.
The Pauli matrices, given in (\ref{eg:Paulim}), are examples of Hermitian matrices.

Any matrix $B$ that is Hermitian can be diagonalized, that is written in the form $U D U^\dagger$,
where $D$ is a diagonal matrix and $U$ is a unitary matrix, or equivalently in terms of its eigenvalues $\{ a_i \}$ 
and eigenvectors $\{ | w_i \rangle \}$ as
\begin{equation*}
B = \sum_i a_i | w_i \rangle \langle w_i |.
\end{equation*}

\section{States} \label{mb:states}
{\it Pure states} \index{pure state} will now be introduced. These form a subset of the set of all quantum states.
A pure state of dimension $d$ can be represented by a $d$-dimensional complex unit vector $| \psi \rangle$. 
For real $\theta$, the vectors $| \psi \rangle$ and $e^{i\theta}| \psi \rangle$ represent the same state.

 More generally, a $d$ dimensional quantum state is represented by a $d \times d$ matrix $\rho$, also called a {\it density matrix}. \index{density matrix}This is a linear operator
which acts on a complex Hilbert space $\mathcal{H}$, is non-negative ($v^T \rho v \geq 0$ for all $v \in \mathbb{R}^d$) and has trace $1$.
A consequence of being non-negative is that $\rho$ is self-adjoint.
The set of states in a complex Hilbert space  $\mathcal{H}$ will be denoted by $S(\mathcal{H})$.
 
 A pure state can be referred to either by its state vector $| \psi \rangle$, or by its density matrix $\rho = | \psi \rangle \langle \psi |$.
For example, 
\begin{equation}
| \psi \rangle =\frac{1}{\sqrt{2} }\left( \begin{array}{c}
1 \\
1 
\end{array} \right),
\quad 
\rho = | \psi \rangle \langle \psi | =  \left( \begin{array}{cc}
1/2 & 1/2 \\
1/2 & 1/2 
\end{array} \right).
\label{eq.purest}
\end{equation}

States which are not pure (have rank greater than one) are called {\it mixed states}. \index{mixed states} A simple test for whether a state $\rho$ is pure or mixed is to take the trace of 
$\rho^2$. For pure states $\tr \{ \rho^2 \} = 1$; for mixed states $\tr \{ \rho^2 \} < 1$.

Examples of $2$-dimensional mixed states are
\begin{equation}
\rho_1= \left( \begin{array}{cc}
3/4 & 0 \\
0 & 1/4 
\end{array} \right),
\quad 
\rho_2 = \left( \begin{array}{cc}
1/2 & -1/6 \\
-1/6 & 1/2 
\end{array} \right).
\label{rho2andf}
\end{equation}

A mixed state can be expressed as a mixture of pure states in many different ways. For instance, the state $\rho_1$, given in (\ref{rho2andf}), can be written as 
\begin{eqnarray*}
\rho_1 &=& 3/4 \left( \begin{array}{cc}
1 & 0 \\
0 & 0 
\end{array} \right)
+ 1/4   \left( \begin{array}{cc}
0 & 0 \\
0 & 1 
\end{array} \right)
\\
& =& 1/2 \left( \begin{array}{cc}
3/4 & \sqrt{3}/4 \\
\sqrt{3}/4  & 1/4 
\end{array} \right)
+ 1/2 \left( \begin{array}{cc}
3/4 & -\sqrt{3}/4  \\
-\sqrt{3}/4  & 1/4 
\end{array} \right) 
\\
&=&  1/4 \left( \begin{array}{cc}
1/2 &1/2 \\
1/2 & 1/2
\end{array} \right)
+
1/4 \left( \begin{array}{cc}
1/2 & -1/2  \\
-1/2  & 1/2 
\end{array} \right) +
 1/2 \left( \begin{array}{cc}
1 & 0 \\
0 & 0
\end{array} \right).
\end{eqnarray*}

The set of $2$-dimensional states, 
which is of great importance to the theory of quantum information, will now be investigated.
A $2$-dimensional quantum state is called a {\it qubit}. \index{qubit}This is because 
qubits are the quantum analogue of `bits' (binary digits). In quantum information qubits are used
to store and transmit information.
Pure qubits  are often expressed in the basis
\begin{equation}
| 0 \rangle = \left( \begin{array}{c}
1\\
0
\end{array} \right), \qquad 
| 1 \rangle = \left( \begin{array}{c}
0\\
1
\end{array} \right).
\end{equation}
The state $| \psi \rangle$, given in (\ref{eq.purest}), can be expressed as $(| 0 \rangle + | 1 \rangle)/\sqrt{2}$.  

Any $2$-dimensional quantum state can be written, with specific values of $x$, $y$ and $z$, as
\begin{equation}
\rho = 1/2 \left( \begin{array}{cc}
1+z & x - i y \\
x+iy & 1-z 
\end{array} \right),
\label{eq.par2st}
\end{equation}
where $x^2 + y^2 + z^2 \leq 1$. The set of pure states corresponds to those states for which $x^2 + y^2 + z^2 =1$.
Any $2$-dimensional state can be thought of as being a point with Cartesian co-ordinates $(x,y,z)$ contained within a unit ball, known as the {\it Bloch ball} or  {\it Poincar\'e ball}. The points on the surface of the ball correspond to pure states; the points within the ball to mixed states.

Alternatively, the state (\ref{eq.par2st}) can be written in terms of the identity and Pauli matrices \index{Pauli matrices}
\begin{eqnarray}
\mathbb{I} &=& \left( \begin{array}{cc}
1 & 0 \\
0 & 1 
\end{array} \right)\\
\sigma_x =  \left( \begin{array}{cc}
0 & 1 \\
1  & 0 
\end{array} \right)
\quad 
\sigma_y &=&  \left( \begin{array}{cc}
0 & -i  \\
i  & 0 
\end{array} \right) 
\quad
\sigma_z =  \left( \begin{array}{cc}
1 &0\\
0 & -1
\end{array} \right),
\label{eg:Paulim}
\end{eqnarray}
as
\begin{equation}
\rho= \frac{1}{2}( \mathbb{I} + x \sigma_x + y \sigma_y + z\sigma_z), \quad x, y, z \in \mathbb{R}. 
\end{equation}

\section{Measurements} \label{measurements}
In this section measurements \index{measurements} are introduced.
When a state is measured, a single result, $m$, is observed out of a set $\Omega$ of possible outcomes. Associated with each outcome $m$ is a matrix $M_m$.
The set of matrices $\{ M_m \}$ constitute a {\it measurement}.
The following conditions are imposed on $M_m$
\begin{equation}
M_m^\dagger = M_m,  \quad M_m \geq 0, \quad \sum_{m \in \Omega} M_m = \mathbb{I}.
\label{mocond}
\end{equation}
A measurement of this form is called a {\it Positive Operator Valued Measure}, or {\it POVM } for short. \index{POVM}

Given a state $\rho$ and a measurement $M = \{ M_m \}$, the result $m$ is observed with probability given by the Born rule
\begin{equation}
p(m)  = \tr \{ \rho M_m \}.
\label{bornrule}
\end{equation}
The $p(m)$ defined in (\ref{bornrule}) satisfies  
\begin{enumerate}
\item[(i)] $p(m) \geq 0$, \quad as $\rho \geq 0$ and $M_m \geq 0$, 
\item[(ii)] $\sum_{m \in \Omega} p(m) = 1$, \quad as
\end{enumerate}
\begin{eqnarray*}
\sum_{m \in \Omega}p(m) &=& \sum_{m \in \Omega}  \tr \{ \rho M_m \} \\
 &=&  \tr \{ \rho  \sum_{m \in \Omega}  M_m \} \\
 &=&  \tr \{ \rho \mathbb{I} \} \\
 &=& 1. \qquad 
 \end{eqnarray*}

More generally, \citep[p.\  23]{busch95} one can consider a non-empty set $\Omega$ and a sigma algebra $\mathcal{F}$, which is a collection of subsets of elements of $\Omega$ that obeys certain rules.
Together $\Omega$ and $\mathcal{F}$ give, what is known as, a measured space $(\Omega, \mathcal{F} )$.
A POVM over a measured space $(\Omega, \mathcal{F} )$ is a set $\{ M(A_i) \}_{A_i \in \mathcal{F}}$ of operators on $\mathcal{H}$ such that
\begin{eqnarray}
M(A_i) &\geq& 0, \quad \mathrm{for \, all} \, A_i \in \mathcal{F}, \\
M(\cup_i A_i ) &=& \sum_i M(A_i), \quad \mathrm{for \, disjoint} \, A_i,\ \\
M( \Omega) &=& \mathbb{I}.
\end{eqnarray}
Applying the measurement $M$ to a state $\rho$ yields outcome $i$ with probability
\begin{equation}
p(i)= \tr \{ \rho M(A_i) \}.
\end{equation}

The most commonly used measurements are {\it Projection Valued Measures} (abbreviated to PVMs). \index{measurements!projective}
A {\it PVM}  is a POVM with elements, usually  written as $P_m$, which satisfy  $P_m P_{m'} = \delta_{m m'} P_m$. (These measurements are also called {\it projective measurements}.) A PVM $\{ P_m \}$ is associated with an {\it observable}, $M$, a Hermitian operator on $\mathcal{H}$ \citep[p.\  87]{chuang00}. The observable has spectral decomposition
\begin{equation*}
M = \sum_{m \in \Omega} m P_m.
\end{equation*}
A simple example of a $2$-dimensional observable is $\sigma_x$ 
\begin{equation*}
\sigma_x = (+1) \left( \begin{array}{cc}
1/2& 1/2\\
1/2 & 1/2 
\end{array} \right) +
(-1)  \left( \begin{array}{cc}
1/2& -1/2\\
-1/2 & 1/2 
\end{array} \right) .
\end{equation*}
Measuring this observable corresponds to using the PVM 
\begin{equation}
M^x= ( M_0, \mathbb{I} - M_0 ), \quad M_0 = \left( \begin{array}{cc}
1/2& 1/2\\
1/2 & 1/2 
\end{array} \right). 
\label{eq.measinx}
\end{equation}
The term `measuring in $x$'  refers to  using the PVM $M^x$ (for which $M_0 = (\mathbb{I} + \sigma_x)/2$). Similarly, the term  `measuring in $y$' refers to  using the PVM $M^y= ( M_0, \mathbb{I} - M_0 )$, with $M_0 = (\mathbb{I} + \sigma_y)/2$, and  `measuring in $z$' to  using the PVM $M^z= ( M_0, \mathbb{I} - M_0 )$, with $M_0 = (\mathbb{I} + \sigma_z)/2$.

A  POVM gives only information about data from a measurement. To describe how a state is changed by a measurement it is necessary to use {\it instruments}.  More information on instruments is given in \cite{bngilljupp03}.

\section{Combined systems}
The {\it tensor product} \index{tensor product} is a mathematical operation which can be used to combine vector spaces to form a larger vector space. Given two vector spaces $V$ and $W$, it is possible to combine them to form the vector space $V \otimes W$, with  $\mathrm{dim}(V \otimes W) =  \mathrm{dim} \, V \times \mathrm{dim} \, W$. 
Given vectors $| v \rangle   \in V$ and $| w \rangle   \in W$, the vector $| v \rangle  \otimes | w \rangle  \in V \otimes W$.
The vector $| v \rangle  \otimes | w \rangle$ is computed from $| v \rangle$ and $| w \rangle$ in the following way  
 \begin{equation*}
| v \rangle  = 
 \left( \begin{array}{c}
v_1  \\
v_2 \\
. \\
. \\
v_m 
\end{array} \right) ,
| w \rangle = \left( \begin{array}{c}
w_1  \\
w_2 \\
. \\
. \\
w_n 
\end{array} \right),  \quad 
| v \rangle \otimes | w \rangle = \left( \begin{array}{c}
v_1w_1  \\
v_1 w_2 \\
. \\
v_1w_n \\
v_2 w_1 \\
. \\
v_m w_n 
\end{array} \right). 
\end{equation*}
Similarly, given two matrices 
\begin{equation*}
A = \left( \begin{array}{cc}
A_{11} & A_{12} \\
A_{21} & A_{22} 
\end{array} \right),
\quad 
B = \left( \begin{array}{cc}
B_{11} & B_{12} \\
B_{21} & B_{22} 
\end{array} \right),
\end{equation*}
their tensor product is equal to
\begin{eqnarray*}
A \otimes B &=& \left( \begin{array}{cc}
A_{11} B & A_{12}  B \\
A_{21} B & A_{22 }  B 
\end{array} \right) 
\\
&=& 
\left( \begin{array}{cccc}
A_{11} B_{11} & A_{11}  B_{12} & A_{12} B_{11} & A_{12}  B_{12} \\
A_{11} B_{21} & A_{11}  B_{22} & A_{12} B_{21} & A_{12}  B_{22} \\
A_{21} B_{11} & A_{21}  B_{12} & A_{22} B_{11} & A_{22}  B_{12} \\
A_{21} B_{21} & A_{21}  B_{22} & A_{22} B_{21} & A_{22}  B_{22} 
\end{array} \right).
\end{eqnarray*}
More generally, for an $M \times N$ matrix $A$ and a $P \times Q$ matrix $B$,
$A \otimes B$ is an $M P \times N Q$ matrix with entries
\begin{eqnarray*}
A \otimes B &=& \left( \begin{array}{ccc}
A_{11} B & \cdots & A_{1N}  B \\
\cdots  & \cdots  & \cdots  \\
A_{M1} B & \cdots & A_{MN }  B 
\end{array} \right) 
\\
&=& 
\left( \begin{array}{cccc}
A_{11} B_{11} & A_{11}  B_{12} & \cdots & A_{1N}  B_{1Q} \\
A_{11} B_{21} & A_{11}  B_{22} &  \cdots & A_{1N}  B_{2Q} \\
 \cdots&  \cdots &  \cdots &  \cdots \\
A_{M1} B_{P1} & A_{M1}  B_{P2} &  \cdots & A_{MN}  B_{PQ} 
\end{array} \right).
\end{eqnarray*}

The following rules hold for tensor products
\begin{eqnarray*}
(A \otimes B) | v \rangle \otimes | w \rangle &=& A  | v \rangle \otimes B | w \rangle \\
A \otimes (B + C) &=& A \otimes B + A\otimes C \\
(A \otimes B)(C \otimes D) &=& (AC \otimes BD) \\
\tr (A \otimes B)  &=& (\tr A) (\tr B ) \\
 (A \otimes B)^\dagger &=&  A^\dagger \otimes B^\dagger.
 \end{eqnarray*}
Given two different quantum systems represented by  Hilbert spaces $\mathcal{H}_A$ and $\mathcal{H}_B$, the combined system is represented by the tensor product of these Hilbert spaces, i.e.\ $\mathcal{H}_A \otimes \mathcal{H}_B$, which will be labelled $\mathcal{H}_{A,B}$.  

If $\rho^A \in S(\mathcal{H}_A)$ and $\rho^B \in S(\mathcal{H}_B)$, then
the {\it composite state} in $S(\mathcal{H}_{A,B})$ is $\rho^{A, B} = \rho^A \otimes \rho^B$.
Given two pure states represented by $| \psi^A \rangle$ and $| \psi^B \rangle$, the composite system is in state represented by $| \psi^{A, B} \rangle = | \psi^A \rangle \otimes | \psi^B \rangle$.  For the sake of brevity the sign $\otimes$ will usually be omitted, and the composite state written as $| \psi^A \psi^B \rangle$.
Of special interest are states, especially pure ones, which exist in the composite system but cannot be written in the form $| \psi^A  \psi^B \rangle$. 
Examples of such states are the {\it Bell states}, 
\begin{eqnarray}
| \phi^+ \rangle &=& \frac{1}{\sqrt{2}} ( | 00\rangle + |11\rangle)
\label{eq.bellstate}
\\
| \phi^- \rangle &=& \frac{1}{\sqrt{2}} ( | 00\rangle - |11\rangle)
\label{eq.bellstate2}
\\
| \psi^+ \rangle &=& \frac{1}{\sqrt{2}} ( | 01\rangle + |10\rangle)
\label{eq.bellstate3}
\\
| \psi^- \rangle &=& \frac{1}{\sqrt{2}} ( | 01\rangle - |10\rangle).
\label{eq.bellstate4}
\end{eqnarray}
\index{entanglement}
\begin{definition}
A pure state $| \psi^{A,B} \rangle \in \mathcal{H}_{A,B}$, which cannot be written as  
$| \psi^A \rangle \otimes | \psi^B \rangle$ is said to be {\it entangled}. 
More generally, we can consider entangled mixed states. A state $\rho \in S(\mathcal{H}_{A,B})$ which cannot be written as a mixture of separable pure states $\ | \psi_i^A \rangle  \otimes  | \psi_i^B \rangle  \in \mathcal{H}_{A,B}$, i.e.\ as
\begin{equation*}
\rho = \sum_i p_i  | \psi_i^A \rangle \langle \psi_i^A | \otimes  | \psi_i^B \rangle \langle \psi_i^B |,
\end{equation*}
is said to be an {\it  entangled state}.
 \end{definition}
 
\begin{definition}
States that are not entangled are said to be {\it separable}. \index{separable states}
\end{definition}
 
 \subsection{Partial trace} \label{sec:pt}
 The {\em partial trace} \index{partial trace} is an important operation when considering combined systems.
Given a density matrix $\rho^{AB} \in S(\mathcal{H}_{A, B})$,
the state of system $\mathcal{H}_A$ is found by taking the partial trace over $\mathcal{H}_B$,
 \begin{equation}
 \rho_A  = \tr_B \{ \rho^{AB} \},
 \end{equation}
where the partial trace $\tr_B$ is defined as
 \begin{eqnarray*}
 \tr_B \{ | \phi_1 \rangle \langle  \phi_2 | \otimes | \psi_1 \rangle \langle  \psi_2 | \}  &=& | \phi_1 \rangle \langle  \phi_2 |   \tr \{  | \psi_1 \rangle \langle  \psi_2 | \} \\
 &=& | \phi_1 \rangle \langle  \phi_2 |   \langle  \psi_2 | \psi_1 \rangle.
 \end{eqnarray*}
A state found by taking the partial trace of a larger state on a combined system is known as a {\it reduced state}.
 
 The density matrix for the Bell state $| \phi^+ \rangle$, given in (\ref{eq.bellstate}), is
 \begin{equation*}
 \rho  = | \phi^+ \rangle \langle \phi^+| = \frac{1}{2}\left( | 00\rangle \langle 0 0 | +  | 00\rangle \langle1 1 | +  | 11\rangle \langle 0 0 | +  | 11\rangle \langle 11 | \right).
 \end{equation*}
 Taking the partial trace over  $\mathcal{H}_B$ gives 
  \begin{eqnarray*}
 \rho_A = \tr_B (\rho)  &=& \frac{1}{2}\bigg( | 0\rangle \langle  0 |  \tr ( | 0 \rangle \langle 0 |)+  | 1\rangle \langle 0 | \tr ( | 1 \rangle \langle 0|) \nonumber \\
&+& \quad \,\,\,| 0\rangle \langle  1  |\tr ( | 0 \rangle \langle 1 |) +  | 1\rangle \langle 1 |\tr ( | 1 \rangle \langle 1 |) \bigg )\\
 &=& \frac{1}{2}( | 0\rangle \langle  0 |  +  | 1\rangle \langle 1 | ).
 \end{eqnarray*}
 Note that, although the composite state is pure, the reduced state is mixed.
 This is one of the interesting properties of entanglement.
 
 If $\rho^{AB} = \rho \otimes \sigma$ then
   \begin{equation*}
 \rho_A  = \tr_B \{ \rho \otimes \sigma\} = \rho \, \tr\{ \sigma \} = \rho,
 \end{equation*}
 as would be expected. Similarly, $\rho_B = \tr_A \{  \rho^{AB} \} = \sigma$.
 
 \section{Measurements on several copies of a state}
Given a product state of the form $\rho^{\otimes n} = \rho^{(1)} \otimes  \cdots \otimes \rho^{(n)} $,
where $\rho^{(j)}$ denotes the $j$th copy of $\rho$, 
 there are several types of measurements that can be performed.
 In this section the most common types of measurements will be defined: {\it collective measurements},
 {\it separable measurements}, {\it LOCC}, {\it adaptive measurements} and {\it separate measurements}.
 This section is similar to \citep[Section 1.2.6]{ballester05}.

\subsection{Collective measurements}
This is the most general type of measurement. If $\mathrm{dim} (\rho) = d$ then $\rho^{\otimes n}$ acts on $\mathbb{C}^{d^n}$.
Collective measurements are POVMs whose elements are $d^n \times d^n$ complex matrices satisfying (\ref{mocond}), \citep{massarr00,gill08}.

\subsection{Separable measurements}
These form a smaller class of measurements. Separable measurements are POVMs whose elements can be expressed as
\begin{equation}
M_{m} = \sum_{i=1}^k  M_{m_i}^{(1)} \otimes \cdots \otimes M_{m_i}^{(n)}, \quad M_{m_i}^{(j)} \geq 0.
\label{eq.sepm}
\end{equation}
The elements $M_m$ must also satisfy (\ref{mocond}).
These measurements do not have a clear physical meaning.

\subsection{LOCC} \label{ssec:adaptm}
Another class of measurements are {\it Local Operations and Classical Communication} (LOCC) \citep[p.\ 573]{chuang00}. These form a smaller class of measurements
than separable measurements as there exist separable measurements that are not LOCC \citep{bennet99}. Unlike separable measurements LOCC has a clear physical meaning. Consider the situation in which there are $n$ experimenters, each with a single copy of $\rho$. Each experimenter can only measure his own copy of $\rho$ but is allowed to communicate with the other experimenters. 
LOCC measurements are easier to perform than some collective measurements, though the latter may often lead to a far more accurate estimate.

 \subsection{Adaptive measurements} \label{ssec:adaptm}
 In the field of quantum statistical inference one often comes across the term {\it adaptive} measurements.\index{measurements!adaptive} 
 Measurements of this type were introduced by \cite{nag88,nag89} because the `optimal' measurements 
on a single quantum state often depend on the unknown state itself. (\cite{nag89} is included in \cite[pp.125-132]{hayashi05}. See also \cite{barndorffnielsengill00} for a discussion of the adaptive measurement strategy.)
This  dilemma of the optimal estimation strategy depending on the unknown parameter
was described by \cite{coch73} as
`You tell me the value of the parameter $\theta$ and I promise to design the best experiment 
for estimating $\theta$.'

In an {\it adaptable} measurement procedure, $n'$ measurements (with $n'$ small) are performed on the first $n'$ copies of $\rho$ to get a rough estimate $\hat \rho$ of the state. Next the POVM which is optimal for $\hat \rho^{\otimes n-n'}$ is used on $\rho^{\otimes n-n'}$.
An adaptive measurement may be collective but not separable, or separable but not LOCC, or simply LOCC, depending on what the optimal measurment is for $\hat \rho^{\otimes n-n'}$.

An example is now given of an adaptive measurement. (The following procedure is LOCC.)
Consider the state
 \begin{equation*}
\rho_\phi =  \left( \begin{array}{cc}
\cos^2(\theta/2) & \sin(\theta/2)\cos(\theta/2) e^{-i\phi}  \\
\sin(\theta/2)\cos(\theta/2)e^{i\phi}   & \sin^2(\theta/2)   
\end{array} \right),
\label{eq.adaptex}
\end{equation*}
where $\theta$ is known. The case  $\theta = \pi/2$ is special,  
as there exists an optimal POVM that does not depend on $\phi$. When $\theta \neq \pi/2$, the optimal POVM 
depends on $\phi$, and one such POVM is
 \begin{equation*}
M = (M_0, \mathbb{I} - M_0), \qquad M_0 =  \frac{1}{2}\left( \begin{array}{cc}
1 & -i e^{-i \phi} \\
i e^{i \phi} & 1 
\end{array} \right).
\label{eq.adaptexp}
\end{equation*}
If $\rho_\phi$ is measured in $x$ (see after (\ref{eq.measinx})),
outcome $0$ is observed with probability $p(0;\phi) = (1 + \sin \theta \cos \phi)/2$, and $1$ with probability $p(1;\phi) = (1 - \sin \theta \cos \phi)/2$.
Put $N=n'/2$ and let $N_{x=0}$ be the number of times that outcome $0$ is observed when $\rho_\phi$ is measured $N$ times in $x$. This gives an estimate $N_{x=0}/N$ of $p(0;\phi)$, and since $\theta$ is known, an estimate of  $\cos\phi$.

If $\rho_\phi$ is measured in $y$, outcome $0$ is observed with probability $p(0;\phi) = (1 + \sin \theta \sin \phi)/2$, and outcome $1$ with probability $p(1;\phi) = (1 - \sin \theta \sin \phi)/2$.
Put $N=n'/2$ and let $N_{y=0}$ be the number of times that outcome $0$ is observed when $\rho_\phi$ is measured $N$ times in $y$.
This gives an estimate $N_{y=0}/N$ of $p(0;\phi)$, and since $\theta$ is known, an estimate of  $\sin\phi$.

Using estimates of $\cos\phi$ and $\sin\phi$, an estimate $\hat \phi$ of $\phi$ is obtained. The `optimal' POVM,
 \begin{equation*}
M = (M_0, \mathbb{I} - M_0), \qquad M_0 =  \frac{1}{2}\left( \begin{array}{cc}
1 & -i e^{-i \hat \phi} \\
i e^{i \hat \phi} & 1 
\end{array} \right)
\label{eq.adaptexp2}
\end{equation*}
is used on the remaining $n - n'$ copies of $\rho_\phi$.
Using this measurement, outcome $0$ is observed with probability 
$p(0;\phi) = (1 + \sin \theta \sin(\phi - \hat \phi))/2$, and outcome $1$ with probability $p(1;\phi) = (1 - \sin \theta \sin(\phi - \hat \phi))/2$. Provided that $ \hat \phi - \phi \in [ - \pi/2, \pi/2]$, the estimate $p(0;\hat\phi)$ of $p(0;\phi)$ can be used to get a more accurate estimate $\hat \phi' $ of $\phi$, namely
\begin{equation*}
\hat \phi' = \hat \phi + \mathrm{arcsin}\left( \frac{2 p(0;\hat\phi) - 1}{\sin \theta} \right).
\end{equation*}
It has been shown by \cite{fuji06} that, for an adaptive quantum estimation scheme, the sequence of maximum likelihood estimators is strongly consistent and asymptotically efficient.

 \subsection{Separate measurements}
 These form the smallest class of measurements. A separate measurement is LOCC with no communication. That is, there are $n$ experimenters, each with a copy of $\rho$, and no communication is allowed between them.
  \index{measurements!separate} 
  
\section{Quantum Channels} \label{sec:qch}
A {\it quantum channel} \index{quantum channel} is a trace-preserving completely-positive map (TP-CP map) sending density matrices to density matrices. It can be thought of as the quantum analogue of a stochastic mapping.
A mapping  $\mathcal{F}$  is {\it positive} if for all $A \geq 0$, $\mathcal{F}(A) \geq 0$.
A mapping $\mathcal{F}$ is {\it completely positive} if for all positive integers $k$ and $B \geq 0$, $(\mathbb{I}_k \otimes \mathcal{F})(B)$ is positive \cite[p.\ 367]{chuang00}.

The mathematical formalism for a quantum channel is originally due to \cite{choi75}.
He showed that a linear map $\Phi$ from $\mathcal{M}_n$ to $\mathcal{M}_m$ ($\mathcal{M}_m$ is the set of $m \times m$ complex matrices) is completely positive if and only if it can be written in the form $\Phi (A) = \sum_k E_k A E_k^\dagger$ where $E_k$ are $m \times n$ matrices. 
For the map $\Phi$ to be a quantum channel, it is further required that the mapping is trace-preserving.
The map $\Phi$ is trace-preserving if and only if $\sum_k E_k^\dagger E_k = \mathbb{I}_n$.
Such a set of matrices $E = \{ E_k \}$ are known as a set of {\it Kraus operators}.

Thus, any quantum channel can be represented using Kraus operators $E_k$ as \index{Kraus operators} \citep{kraus83,chuang00,beng06} 
\begin{equation}
\rho_0 \mapsto \sum_k E_k \rho_0 E_k^\dagger,
\label{eq.krausdec}
\end{equation}
where
\begin{equation}
\sum_k E_k^\dagger E_k = \mathbb{I}_n.
\end{equation}
The form (\ref{eq.krausdec}) for a general quantum channel can be derived as follows.
 Consider the composite state formed by the input state $\rho_0 \in S(\mathcal{H})$  and the environment $\rho_{env}\in S(\mathcal{H}_{env})$.
 Put
 \begin{equation*}
 \rho = \rho_0 \otimes \rho_{env}.
 \end{equation*}
Suppose that $\rho$ undergoes unitary evolution, i.e.\index{unitary channel}
 \begin{eqnarray*}
 \rho &\mapsto& U \rho U^\dagger, \\
 &=& U(\rho_0 \otimes \rho_{env})  U^\dagger,
 \end{eqnarray*}
 where $U^\dagger U = U U^\dagger = \mathbb{I}$. It is assumed that $\rho_{env}$ is a pure state, with  $\rho_{env} = | 0 \rangle \langle 0 |$, where $| 0 \rangle, |1 \rangle , \dots, |d-1 \rangle$, form a basis of $\mathcal{H}_{env}$.
 (This is the only place in this thesis where $| 0 \rangle$ does not refer to $(1,0)^T$.)
  It is found that $\rho_0$ has undergone the following transformation
  \begin{eqnarray*}
 \rho_0  &\mapsto& \tr_{env} \{ U(\rho_0 \otimes | 0 \rangle \langle 0 | )  U^\dagger \}  \\
 &=&\sum_k  \langle k | U| 0 \rangle \rho_0 \langle 0 |  U^\dagger | k \rangle \\
  &=&\sum_k  E_k  \rho_0 E_k^\dagger, \quad E_k = \langle k | U| 0 \rangle.
 \end{eqnarray*}
 
Some examples will now be given of parametric families of quantum channels. 
A unitary channel is a mapping which transforms a state $\rho \in S(\mathbb{C}^d)$ to the state $U \rho U^\dagger  \in S(\mathbb{C}^d)$, where $U$ is a $d \times d$ complex unitary matrix.  
Chapter \ref{ch:ipe} considers the problem of estimating the parameter $\theta$ in a unitary channel acting on $\mathcal{H} = \mathbb{C}^2$, with unitary matrix
\begin{equation}
U_\theta = \left( \begin{array}{cc}
1 & 0 \\
0  & e^{i 2 \pi \theta} 
\end{array} \right).
\label{eq.kitu}
\end{equation}
If this channel acts on the state $\rho$, given in  (\ref{eq.purest}), it produces the output state 
\begin{equation*}
U_\theta \rho U_\theta^\dagger  = 1/2 \left( \begin{array}{cc}
1 & e^{-i 2 \pi \theta} \\
e^{i 2 \pi \theta}  & 1 
\end{array} \right).
\end{equation*}
Examples of non-unitary channels (channels with at least two non-zero Kraus operators $E_k$) are now considered. 

The family of depolarizing channels $\mathcal{E}: S(\mathbb{C}^d) \rightarrow S(\mathbb{C}^d)$ \index{depolarizing channel} is the set of mappings \citep[p.\ 378]{chuang00}
\begin{equation}
 \rho_0 \mapsto (1 - \epsilon) \rho_0 + \frac{\epsilon}{d} \mathbb{I}_d, \quad  0 < \epsilon < 1.
\label{eq.depngd}
\end{equation}
A depolarizing channel describes the process in which with probability $1 -\epsilon$ the state is left unchanged, and with probability $\epsilon$ is replaced by the completely mixed state $\mathbb{I}_d/d$.

The family of $2$-dimensional depolarizing channels $\mathcal{E}: S(\mathbb{C}^2) \rightarrow S(\mathbb{C}^2)$ have Kraus operators \citep[p.\ 397]{chuang00}
\begin{equation*}
E_0 = \sqrt{1 - \frac{3 \epsilon}{4}} \mathbb{I}_2, \quad E_1 = \sqrt{\frac{\epsilon}{4}} \sigma_x, \quad E_2 = \sqrt{\frac{\epsilon}{4}} \sigma_y, \quad E_3 = \sqrt{\frac{\epsilon}{4}} \sigma_z.
\end{equation*}
This family of channels forms a subset of the set of Pauli channels. The family of {\it Pauli channels} $\mathcal{E}: S(\mathbb{C}^2) \rightarrow S(\mathbb{C}^2)$, indexed by the parameters $(p_0, p_1, p_2, p_3)$, with $p_j \geq 0$ and $\sum_{j=0}^3 p_j =1$, is the set of channels \citep{fujimai03}
\begin{equation}
\rho_0 \mapsto \sum_{i=0}^3 p_i \sigma_i \rho_0 \sigma_i,
\label{eq.paulichan}
\end{equation}
where $\sigma_0 = \mathbb{I}_2$ and $\sigma_1, \sigma_2$ and $\sigma_3$ are the Pauli matrices (see (\ref{eg:Paulim})).

The family of {\it generalized Pauli channels} $\mathcal{E}: S(\mathbb{C}^d) \rightarrow S(\mathbb{C}^d)$, indexed by the parameters $(p_0, p_1, p_2,\dots, p_{d^2-1})$, with $p_j \geq 0$ and $\sum_{j=0}^{d^2-1} p_j =1$, is the set of channels  \citep{fujimai03}\index{generalized Pauli channels}
\begin{equation}
\rho_0 \mapsto \sum_{k=0}^{d^2-1} p_k U_k \rho_0 U_k^\dagger, \qquad \tr\{ U_k^\dagger U_l \} = d \delta_{kl}.
\label{eq.genpauli}
\end{equation}
The choice of unitary matrices $U_j$ is arbitrary. 

The family of {\it amplitude damping channels} $\mathcal{E}: S(\mathbb{C}^2) \rightarrow S(\mathbb{C}^2)$, indexed by the parameter $\gamma$, is the set of channels with Kraus operators \citep[p.\ 380]{chuang00}
\begin{equation}
E_0 = \left( \begin{array}{cc}
1 & 0 \\
0  & \sqrt{1-\gamma} 
\end{array} \right), \quad
E_1 = \left( \begin{array}{cc}
0 & 0 \\
0  & \sqrt{\gamma} 
\end{array} \right).
\label{eq.ampdampk}
\end{equation}
This channel describes energy dissipation: every state is brought closer to 
the pure state $| 0 \rangle \langle 0 |$. 

The family of {\it generalized damping channels}  $\mathcal{E}: S(\mathbb{C}^2) \rightarrow S(\mathbb{C}^2)$  \citep[p. 382]{chuang00}, indexed by the parameters $\gamma,p$ is the set of channels with Kraus operators 
\begin{eqnarray}
E_0 &=& \sqrt{p} \left( \begin{array}{cc}
1 & 0 \\
0  & \sqrt{1-\gamma} 
\end{array} \right), \quad
E_1 = \sqrt{p}  \left( \begin{array}{cc}
0 & 0 \\
0  & \sqrt{\gamma} 
\end{array} \right) \nonumber \\
E_2 &=& \sqrt{1-p} \left( \begin{array}{cc}
\sqrt{\gamma} & 0 \\
0  & 1 
\end{array} \right), \quad
E_3 = \sqrt{1-p}  \left( \begin{array}{cc}
0 & 0 \\
\sqrt{1-\gamma}  & 0 
\end{array} \right).
\label{eq.ampdampk1}
\end{eqnarray}
The parameter $p \in [0,1]$  represents the temperature of the environment.

\section{Fisher information}
\subsection{One-parameter case}
Given a univariate family of probability distributions with probability density functions $p(x; \theta)$, the {\it Fisher information}, introduced by \cite{fisher22}, is defined as
\begin{eqnarray}
 F_\theta &\equiv&  \int   p(x;\theta) \left( \frac{\partial \ln p(x;\theta)}{\partial \theta} \right)^2 dx \label{eq.FI00a} \\
&=&  \int  \frac{1}{p(x;\theta)} \left( \frac{\partial p(x;\theta)}{\partial \theta} \right)^2 dx.
\label{eq:FI00}
\end{eqnarray}
Intuitively, Fisher information \index{Fisher information} gives a measure of the amount of `information' about $\theta$ contained in an observation. 
If the random variable $X$ is discrete with probabilities $p(1;\theta),\dots , p(n;\theta)$, then the Fisher information can be expressed as
\begin{eqnarray*}
 F_\theta = \sum_{m=1}^{n} \frac{1}{p(m;\theta)} \left( \frac{d p (m;\theta)}{d \theta}\right)^2 .
\label{eq:FI01}
\end{eqnarray*}

\begin{proposition}
The Fisher information from $n$ {\it i.i.d.} observations $X_1, X_2,$ $\dots X_n$ is equal to $n F_\theta$ where $F_\theta$ is the Fisher information from a single observation $X_j$.
\end{proposition}

\proof
The Fisher information for a single observation $X_j$ can be written as
\begin{equation}
F_\theta = -E \left[ \frac{d^2 l(\theta;x)}{d \theta^2} \right],
\label{eq.fialt}
\end{equation}
 where $l(\theta;x) = \log L(\theta;x)$ is the log-likelihood (the natural logarithm of the likelihood function). For the case of $n$ observations, 
 \begin{equation}
 L(\theta;x_1, \dots, x_n) \equiv \prod_{i=1}^n p(x_i; \theta). 
 \end{equation}
  Thus
   \begin{equation}
l(\theta;x_1, \dots, x_n) = \sum_{i=1}^n \log p(x_i; \theta), 
 \end{equation}
 and the Fisher information from $n$ observations is equal to 
 \begin{eqnarray}
F_\theta^{(n)} &=& -E \left[ \sum_{i=1}^n \frac{d^2 l(\theta;x_i)}{d \theta^2} \right] \\
&=& n F_\theta. 
\label{eq.fialtn}
\end{eqnarray}
 
The importance of Fisher information is seen in the {\it Cram\'er--Rao inequality}. \index{Cram\'er--Rao inequality} This states that the mean square error of an unbiased estimator $t$ is greater than or equal to the reciprocal of the Fisher information, i.e.
\begin{equation}
E[(\hat \theta-\theta)^2] \geq \frac{1}{ F_\theta}.
\label{eq:CRineq}
\end{equation}
The right hand side of (\ref{eq:CRineq}) is known as the {\it Cram\'er--Rao bound}.\index{Cram\'er--Rao bound}
Under mild regularity conditions for $p(x ; \theta)$,  using a maximum likelihood estimator,  as the number of observations $n \rightarrow \infty$ \citep[p.\ 63]{vaart98}
\begin{equation}
\sqrt{n}(\hat \theta - \theta ) \leadsto \mathcal{N}(0, F_\theta^{-1}),
\end{equation}
and so, assuming the estimator is unbiased,  
\begin{equation}
n E[(\hat \theta-\theta)^2]  \rightarrow \frac{1}{ F_\theta}.
\label{eq:CRineq2}
\end{equation}
(The symbol $\leadsto$ denotes convergence in distribution.)
The larger the Fisher information, the more accurately the unknown parameter can be estimated. A standard approach to estimation of a parameter, in a known family of distributions, is to use the maximum likelihood estimator. Consequently the result (\ref{eq:CRineq2}) is of great importance: it enables the asymptotic behaviour of an estimate to be quantified.

\subsection{Multi-parameter case}
Given a $p$-parameter family of probability distribution with probability density functions $p(x; \theta^1, \dots, \theta^p)$, the {\it Fisher information}, is  the $p \times p$ matrix $F_\theta$ with entries
\begin{eqnarray*}
( F_\theta)_{jk} &\equiv&  \int   p(x;\theta) \left( \frac{\partial \ln p(x;\theta)}{\partial \theta^j} \right)    \left( \frac{\partial \ln p(x;\theta)}{\partial \theta^k} \right)  dx \\
&=&  \int  \frac{1}{p(x;\theta)} \left( \frac{\partial p(x;\theta)}{\partial \theta^j} \right) \left( \frac{\partial p(x;\theta)}{\partial \theta^k} \right)dx.
\label{eq:FI00m}
\end{eqnarray*} 
The Cram\'er--Rao inequality becomes a matrix inequality. This states that the mean square error of an unbiased estimator $t$ is greater than or equal to the inverse of the Fisher information, i.e.
\begin{equation}
E[(\hat \theta - \theta)(\hat \theta - \theta)^T] \geq F_\theta^{-1}.
\label{eq:CRineq2y}
\end{equation}
This means that the matrix $E[(\hat \theta - \theta)(\hat \theta - \theta)^T]  - F_\theta^{-1}$ is positive semi-definite, i.e.\ for all $v \in \mathbb{R}^p$,
\begin{equation*}
v^T (E[(\hat \theta - \theta)(\hat \theta - \theta)^T]  - F_\theta^{-1}) v \geq 0.
\end{equation*}
 
\section{Quantum information}  \label{subsec:SLDQFI}
\begin{definition}
A  Riemannian metric on a manifold $\mathcal{M}$ is a mathematical object that assigns smoothly to each point $x$ of $\mathcal{M}$, and each coordinate system $\theta=(\theta^1, \dots, \theta^p)$ round $x$, a positive semi-definite $p \times p$ matrix $g_\theta(x)$ such that, for another coordinate system $\phi=(\phi^1, \dots, \phi^p$),
\begin{equation}
g_\phi(x) = \left(\frac{d \theta}{d\phi} \right) g_\theta(x) \left(\frac{d \theta}{d\phi} \right)^T,
\end{equation}
or, in terms of elements of $g_\phi(x)$,
\begin{equation}
g_\phi(x)_{ij} = \sum_{k,l}g_\theta(x)_{kl}\frac{d \theta^i}{d \phi^k}\frac{d \theta^j}{d \phi^l}.
\label{mectriccy}
\end{equation}
\end{definition}

It has been shown by \cite{moro90} that, up to a constant factor, the Fisher information is the unique monotone Riemannian metric on $\Theta$.
Several types of {\it quantum information} \index{quantum information} have been suggested as quantum versions of Fisher information  \citep{petz96}, defined from a parametric family of states $\rho_\theta$.  As the Fisher information is a Riemannian metric, any quantum analogue of Fisher information should also be a Riemannian metric.  
The following properties are important when considering Riemannian metrics.
\\
{\noindent{ \bf Invariance.}} \index{invariance}
\\
Two parametric families of states $\rho_\theta$ and $\sigma_{\theta}$  are said to be {\it equivalent}  ($\rho_\theta \sim \sigma_{\theta}$) \citep{petz96} if there exist two fixed TP-CP maps $\mathcal{E}, \mathcal{F}$ such that 
\begin{equation*}
\rho_\theta= \mathcal{E}(\sigma_{\theta}), \qquad \sigma_{\theta} = \mathcal{F}(\rho_\theta).
\end{equation*}
The Riemannian metric $J$ is said to be {\it invariant} \citep{petz96} if \index{invariant}
\begin{equation*}
\rho_\theta \sim \sigma_{\theta} \quad \mathrm{implies} \quad  J(\rho_\theta) = J(\sigma_{\theta}).
\end{equation*}
{\noindent{\bf Monotonicity.}} \index{monoticity}
\\
The Riemannian metric $J$ is said to be {\it monotone} \citep{petz96} if \index{monotone}
\begin{equation*}
J(\rho_\theta) \geq J(\mathcal{E}(\rho_\theta))
\end{equation*}
for all TP-CP maps $\mathcal{E}$. 
\medskip

 A well-defined Riemannian metric must be invariant and it is desirable that it is monotone. (If a metric is monotone then it is also invariant.)
It has been shown by \cite{petz96} that there is no unique monotone quantum information quantity. The most frequently encountered monotone metrics in recent literature are the {\it Symmetric Logarithmic Derivative (SLD)}, {\it Right Logarithmic Derivative (RLD)} and {\it Kubo-Mori-Bogoliubov (KMB)} metrics, which are defined in (\ref{hoax}) -- (\ref{eq:kmbscore}) and  (\ref{eq:scoresld}) -- (\ref{hajek}).

Given a one-parameter family of states $\rho_\theta$, these quantum information quantities can be expressed  as
\begin{equation}
H^x = \tr \{ \lambda_x^\dagger \rho \lambda_x \},
\label{hoax}
\end{equation}
where the parameter $\theta$ has been suppressed, and the
quantum scores $\lambda_x$  (the quantum analogues of the logarithmic derivative $dl/d\theta$)  are defined as the solutions to the following matrix equations
\index{quantum information!SLD}\index{quantum information!RLD}\index{quantum information!KMB} \index{SLD} \index{KMB} \index{RLD}\index{SLD!quantum score} \index{KMB!quantum score} \index{RLD!quantum score} 
\begin{eqnarray}
\frac{d \rho}{d \theta} &=& \frac{1}{2}( \rho \lambda_{SLD} + \lambda_{SLD} \rho) \label{eq:sldscore} \\
 \frac{d \rho}{d \theta} &=& \rho \lambda_{RLD} \label{eq:rldscore}  \\
  \lambda_{KMB} &=& \frac{d \log \rho}{d \theta}.\label{eq:kmbscore} 
\end{eqnarray}
These all satisfy
\begin{equation}
\tr \{ \rho \lambda_x \} =0.
\end{equation}
The quantum scores $\lambda_{RLD}$ and $\lambda_{KMB}$ are defined only when $\rho_\theta$ has full rank, i.e.\ when $\rho_\theta$ is invertible. When $\rho_\theta$ does not have full rank, $ \lambda_{SLD}$ is not defined uniquely, though $H^{SLD}$ does not depend on the choice of $ \lambda_{SLD}$.

The SLD quantum information is the most commonly used quantum information quantity. It is the minimum among the set of monotone quantum information quantities \citep{petz96}.
In this thesis the SLD quantum information will be denoted simply by $H$ or $H_\theta$, and the SLD quantum score by $\lambda$ or $\lambda_\theta$.

The SLD quantum information for $n$ copies of the state $\rho_\theta$ (i.e.\ $\rho^{(n)}_\theta = \rho_\theta \otimes \rho_\theta \otimes \cdots \otimes \rho_\theta$)  is $n$ times the SLD quantum information for the state $\rho_\theta$, that is 
\begin{equation}
H( \rho^{(n)}_\theta ) =  n H_\theta( \rho_\theta ). 
\label{eq.hhbos}
\end{equation}
To see this, note that in this case 
\begin{equation}
\frac{d \rho^{(n)}}{d \theta} = \frac{d \rho}{d \theta} \otimes \rho \otimes \cdots \otimes \rho +\rho \otimes \frac{d \rho}{d \theta} \otimes \rho \otimes \cdots \otimes \rho +  \rho \otimes \cdots \otimes \rho \otimes \frac{d \rho}{d \theta}.
\end{equation}

Let $\lambda$ be a possible solution for the SLD score for $\rho$. Then 
a possible solution for the SLD score for $\rho^{(n)}$ is
\begin{equation}
\lambda^{(n)} = \lambda \otimes \mathbb{I} \otimes \cdots \otimes \mathbb{I} + \mathbb{I} \otimes \lambda \otimes \mathbb{I} \otimes \cdots \otimes \mathbb{I} + \cdots + \mathbb{I} \otimes \cdots \otimes \mathbb{I} \otimes \lambda.
\end{equation}

The SLD quantum information can be written as
\begin{eqnarray*}
H( \rho^{(n)}_\theta ) &=& \tr \left\{ \frac{d \rho^{(n)}}{d \theta} \lambda^{(n)} \right\}  \\
&=& \tr \{ \frac{d \rho}{d \theta} \lambda \otimes \rho \otimes \cdots \otimes \rho \}  \\
&+&  \tr \{ \frac{d \rho}{d \theta} \otimes \rho \lambda \otimes \cdots \otimes \rho \} \\
&\vdots& \\
&+& \tr \{ \rho \otimes \cdots \otimes \rho \otimes \frac{d \rho}{d \theta} \lambda \}.\\
\end{eqnarray*}
Since $\tr\{\rho \lambda \} =0$, the only non-zero terms are those of the form $\tr \{ \rho \otimes \cdots \otimes \rho \otimes  (d\rho/d\theta) \lambda \otimes \rho \cdots \otimes \rho \}$. As there are $n$ terms of this form, (\ref{eq.hhbos}) holds.

 \subsection{Multi-parameter models}
Given a $p$-parameter family of quantum states $\rho(\theta^1, \dots, \theta^p)$, the quantum analogues of the logarithmic derivative $(\partial l/\partial \theta^j)$ are defined as the solutions to the following matrix equations
\begin{eqnarray}
\frac{\partial \rho}{\partial \theta^j} &=& \frac{1}{2}( \rho \lambda_{SLD}^j + \lambda_{SLD}^j \rho)  \label{eq:scoresld}\\
\frac{\partial \rho}{\partial \theta^j}  &=& \rho \lambda_{RLD}^j 
\label{eq:scorerld}\\
  \lambda_{KMB} &=&\frac{\partial \log \rho}{\partial \theta^j}. 
  \label{eq:scorekmb}
\end{eqnarray}
These all satisfy
\begin{equation*}
\tr \{ \rho \lambda_x^j \} =0.
\end{equation*}
The quantum informations are $p \times p$ matrices with entries
\begin{equation}
(H^x_\theta)_{jk} = \Re \tr \{ \lambda_x^{\dagger j}\rho \lambda_x^k \}.
\label{hajek}
\end{equation}

\section{Braunstein-Caves inequality}
Consider the parametric statistical model resulting from a measurement $M = \{ M_m \}$ of a parametric family of states $\rho_\theta$. This has the probability function 
\begin{equation}
p(m; \theta) = \tr \{ \rho_\theta M_m \},
\end{equation}
which gives Fisher information $F_\theta^M$. 
 \cite{braunsteincaves94}  \index{Braunstein-Caves inequality} proved the inequality
\begin{equation}
F_\theta^M \leq H_\theta
\label{bcii}
\end{equation}
for the one-parameter case. They showed that, in the one-parameter case, the SLD quantum information gives the maximum Fisher information that can be obtained from measuring a model $\rho_\theta$.

\subsection{Multi-parameter Braunstein-Caves inequality}
\begin{theorem}
Let $F_\theta^M$ be the Fisher information given by a measurement $M$ on a parameterised quantum model
$\{ \rho_\theta: \theta = \theta^1, \dots, \theta^p \in \mathbb{R}^p \}$, with SLD quantum information $H_\theta$. Then  
\begin{equation}
F_\theta^M \leq H_\theta.
\label{bciim}
\end{equation}
This means that the matrix $H_\theta- F_\theta^M$ is non-negative, which is equivalent to
 \begin{equation}
\sum_{j,k} x_j x_k  (F_\theta^M)_{j, k} \leq \sum_{jk} x_j x_k (H_\theta)_{jk},
\end{equation}
for all vectors $x = (x_1, \dots, x_p) \in \mathbb{R}^p$. 
\label{thm:fmh}
\end{theorem}
Not only does (\ref{bciim}) give an upper bound on the Fisher information, but the proof gives necessary and sufficient conditions for equality. The following proof is similar to that in \cite[p.\ 26]{ballester05}.
\\
\proof
\\
Denote by $\Omega^+$ the set of outcomes which occur with non-zero probability. Then
\begin{eqnarray}
\sum_{j,k} x_j x_k (F_\theta^M)_{jk} &=&  \sum_{j,k} x_j x_k \sum_{m \in \Omega^+} \frac{1}{p(m;\theta)} \left( \frac{\partial p(m;\theta)}{\partial \theta^j} \right)\left( \frac{\partial p(m;\theta)}{\partial \theta^k} \right) \nonumber \\
&=& \sum_{j,k} x_j x_k \sum_{m \in \Omega^+} \frac{1}{\tr \{ \rho M_m \}} \left( \tr \left\{ \frac{\partial \rho}{\partial \theta^j} M_m \right\} \right) \left( \tr \left\{ \frac{\partial \rho}{\partial \theta^k} M_m \right\} \right) \nonumber  \\
&=& \sum_{j,k} x_j x_k \sum_{m \in \Omega^+} \frac{1}{\tr \{ \rho M_m \}} \left( \Re \tr \{ \lambda^j \rho M_m\} \right) \left( \Re \tr \{ \lambda^k \rho M_m\} \right) \nonumber \\
&\leq& \sum_{j,k} x_j x_k \sum_{m \in \Omega^+} \frac{1}{\tr \{ \rho M_m \}} | \tr \{ \lambda^j \rho M_m\}|  |  \tr \{ \lambda^k \rho M_m\}| \nonumber  \\
&=&  \sum_{m \in \Omega^+} \frac{1}{\tr \{ \rho M_m \}} | \tr \{ \lambda \rho M_m\}|^2, \quad \mathrm{where} \,  \lambda = \sum_j x_j \lambda^j,  \nonumber  \\
&=&  \sum_{m \in \Omega^+} \frac{1}{\tr \{ \rho M_m \}} | \tr \{ (M_m^{1/2} \lambda \rho^{1/2} ) (M_m^{1/2} \rho^{1/2} )^\dagger  \}|^2 \nonumber  \\
&\leq&  \sum_{m \in \Omega^+} \frac{1}{\tr \{ \rho M_m \}} \tr \{ \rho M_m \} \tr \{  M_m \lambda \rho \lambda \} \nonumber  \\
&=&  \sum_{m \in \Omega^+} \tr \{  M_m \lambda \rho \lambda \} \nonumber  \\
&\leq&   \tr \{  \lambda \rho \lambda \} \label{fi13}  \\
&=&   \sum_{j, k} x_j x_k \tr \{  \lambda^j \rho \lambda^k \} \\
&=& \sum_{j, k} x_j x_k (H_\theta)_{jk}.   \nonumber
\end{eqnarray}
Inequality  (\ref{fi13}) follows from the fact that $\sum_{m \in \Omega^+} M_m \leq \mathbb{I}$,  since  $\sum_{m \in \Omega} M_m = \mathbb{I}$ and  $M_m \geq 0$.  

  \index{Braunstein-Caves inequality!equality} 
  \begin{theorem}
  Equality holds in (\ref{bciim}) if and only if 
  \begin{equation}
 M_m^{1/2} \lambda^j \rho^{1/2} = \xi_m^j  M_m^{1/2}  \rho^{1/2},\quad \xi_m^j \in \mathbb{R} \quad \forall j,m.
 \label{bcii2e}
\end{equation}
\label{thm:eqfmh}
  \end{theorem}
  The following proof is similar to that in \cite[p.\ 27]{ballester05}.
\\
  \proof
  \\
  Equality holds in (\ref{bciim}) if and only if the following three conditions are met
\begin{eqnarray}
\Im \tr \{ \lambda^j \rho M_m\} &=& 0 \quad \forall j, m, \label{bcii1}  \\
M_m^{1/2} \lambda \rho^{1/2} &=& z_m M_m^{1/2} \rho^{1/2}, \quad \mathrm{some} \, z_m \in \mathbb{C},  \quad \forall m, \lambda, \label{bcii2}  \\
  \sum_{m \in \Omega^+} \tr \{  M_m \lambda \rho \lambda \} &=&  \tr \{  \lambda \rho \lambda \}, \quad \forall m,\lambda.
   \label{bcii3}
  \end{eqnarray}
Since
 \begin{equation*}
 \tr \{  \lambda \rho \lambda \} =  \sum_{m \in \Omega^+} \tr \{  M_m \lambda \rho \lambda \} + \sum_{m \in \Omega \setminus \Omega^+} \tr \{  M_m \lambda \rho \lambda \}, 
  \end{equation*}
 equality holds in (\ref{bciim}) if and only if the following three conditions are met
\begin{eqnarray}
\Im \tr \{ \lambda^j \rho M_m\} &=& 0 \quad \forall j,m, \label{bcii1b}  \\
M_m^{1/2} \lambda^j \rho^{1/2} &=& z_m^j M_m^{1/2} \rho^{1/2}, \quad \mathrm{some} \, z_m^j \in \mathbb{C}, \quad \forall j,m, \label{bcii2b}  \\
  \sum_{m \in \Omega \setminus \Omega^+} \tr \{  M_m \lambda^j \rho \lambda^k \} &=& 0, \quad \forall j, k,m. \label{bcii3b} 
  \end{eqnarray}
  Theorem \ref{thm:eqfmh} will be proved by showing that 
  \begin{enumerate}
  \item[(i)] if (\ref{bcii2e}) holds, then (\ref{bcii1b}), (\ref{bcii2b})  and (\ref{bcii3b}) hold, and thus equality holds in  (\ref{bciim}) (consequently (\ref{bcii2e}) is a sufficient condition);
  \item[(ii)] the conditions  (\ref{bcii1b}) and (\ref{bcii2b}) both hold only if (\ref{bcii2e}) holds (thus (\ref{bcii2e}) is a necessary condition).
  \end{enumerate}
(i) If (\ref{bcii2e}) holds, then  (\ref{bcii2b}) obviously holds. Pre-multiplying (\ref{bcii2e}) by $M_m^{1/2}$, post-multiplying by $\rho^{1/2}$ and taking the trace shows that (\ref{bcii1b})  also holds.
 Note that
\begin{equation}
\tr \{ A^\dagger A \} = 0 \quad  \mathrm{if \, and \, only \, if} \quad  A = 0.
\label{eq.aat}
\end{equation}
For $m \in \Omega \setminus \Omega^+$, $p_m =0$, and so, since $p_m = \tr \{ ( M_m^{1/2} \rho^{1/2} )^\dagger  ( M_m^{1/2} \rho^{1/2} ) \}$, by (\ref{eq.aat}) it follows that  $M_m^{1/2} \rho^{1/2} =0$.
If (\ref{bcii2e})  holds, then  for $m \in \Omega \setminus \Omega^+$, $M_m^{1/2} \lambda^j \rho^{1/2} =0$ and so (\ref{bcii3b}) holds.
\\
\\
(ii) First, it will be assumed that  (\ref{bcii2b}) holds. Pre-multiplying (\ref{bcii2b}) by $M_m^{1/2}$, post-multiplying by $\rho^{1/2}$ and taking the trace gives
\begin{equation}
\tr \{ M_m \lambda^j \rho \} = z_m^j  p_m ,\quad \forall j,m.
\end{equation}
For condition (\ref{bcii1b}) to hold, $z_m^j$ must be real. Thus (\ref{bcii1b}) and (\ref{bcii2b}) both hold only if (\ref{bcii2e}) holds.

\subsection{Equality} \label{quasi2}
For one-parameter models, equality holds in (\ref{bcii}) if and only if
\begin{equation}
 M_m^{1/2} \lambda \rho^{1/2} = \xi_m  M_m^{1/2}  \rho^{1/2},\quad \forall m,  \quad \xi_m \in \mathbb{R}.
 \label{bcii2e1}
\end{equation}
As $\lambda$ is self-adjoint, it can be written as
\begin{equation*}
\lambda = \sum_i \mu_i | e_i \rangle \langle e_i |.
\end{equation*}
The POVM $M = \{ M_i = | e_i \rangle \langle e_i | \}$ satisfies (\ref{bcii2e1}) and so, using this POVM, equality holds in 
(\ref{bcii}).
It has been shown by \cite{barndorffnielsengill00} that, in general, the optimal POVM will depend on the unknown parameter. To get around this, an adaptive measurement scheme can be used, as described in Section \ref{ssec:adaptm}.
There are a few families of states for which the optimal POVM does not depend on the parameter, such as the set of states corresponding to the `equator' of the Bloch ball, given by the set of density matrices
\begin{equation*}
\rho_\theta = 1/2 \left( \begin{array}{cc}
1 & e^{-i 2\pi\theta} \\
e^{i 2\pi\theta}  & 1 
\end{array} \right), \quad \theta \in [0,1),
\end{equation*}
and sets of quasi-classical states. {\it Quasi-classical states} are defined as sets of states for which the eigenvectors $\{ | w_i \rangle \}$ are known, i.e.\ families of states of the form
\begin{equation*}
\rho_\theta = \sum_{i=1}^d p_i(\theta) | w_i \rangle \langle w_i |.
\end{equation*}

In the multi-dimensional case there exist sets of states for which the bound (\ref{bciim}) is not attainable  even using an adaptive scheme \citep{barndorffnielsengill00}. In Section \ref{sec:nrps}, it is shown that for any  non-degenerate set of pure states (these can be parameterised by a maximum of $2(d-1)$ parameters), equality holds in (\ref{bciim}) only if the number of parameters $p \leq d -1$.

The fact that the SLD quantum information is not in general attainable means that it cannot in general be used to find the optimal estimation method for quantum states. The problem of optimally estimating $n$ identical quantum states has recently been solved by \cite{gutakahn07,gutakahn09}.
The solutions presented in these papers are based on {\em quantum local asymptotic normality}: given $n$ copies of a state, as $n \rightarrow \infty$ the joint state converges to a statistical model consisting of a classical Gaussian  distribution and a quantum Gaussian  distribution. The optimal estimation procedure for these models is known, having been solved by \cite{yuen73,holevo82}. 
Quantum local asymptotic normality was first studied in \cite{hayashi03b,hayash03} and used for estimation in \cite{haymat04}. It was later made more rigorous by \citep{gut06,gutaJa2007}.

The optimal estimation of qubits has been solved explicitly in the Bayesian set-up, in the particular case of an invariant prior in \citep{bagan06}.

\subsection{Equality in the case of pure states}  \label{subsec:matz}
Putting together (\ref{eq:CRineq2y}) and (\ref{bciim}) gives the {\it quantum Cram\'er--Rao inequality} 
\begin{equation}
E[(\hat \theta - \theta)(\hat \theta - \theta)^T] \geq  H_\theta^{-1}.
\label{eq:QCRineq}
\end{equation}
A result of \cite{mats97} will now be considered. In the case of pure states, it gives a concise necessary and sufficient condition for the existence of a POVM such that equality holds in (\ref{eq:QCRineq}).

\begin{theorem}
 Let $\{ \rho_\theta : \theta \in \Theta \}$ be a parameterised family of pure states with $\rho_\theta = | \psi_\theta \rangle \langle \psi_\theta |$. Then there exists a POVM and estimator such that equality holds in (\ref{eq:QCRineq}) at $\theta=\theta_0$, if and only if 
\begin{equation}
\Im \langle l_j (\theta_0) | l_k (\theta_0) \rangle = 0, \quad \forall j,k,
\label{eq.matzcondq}
\end{equation}
where $| l_j(\theta) \rangle = \lambda^j_\theta | \psi_\theta \rangle$ \citep{mats97,mats02,fuj01a}.
\label{thm:matzy}
\end{theorem}

In Section \ref{sec:nrps}, it is shown that condition (\ref{eq.matzcondq}) is equivalent to the simpler condition
\begin{equation}\Im \langle \psi^{(j)} (\theta_0) | \psi^{(k)} (\theta_0) \rangle = 0,   \quad \forall j,k, \quad | \psi^{(j)} \rangle = \frac{\partial  | \psi \rangle}{\partial \theta^j} .
\label{eq.matzcondq2}
\end{equation}

When (\ref{eq.matzcondq}) is satisfied, a POVM giving equality in (\ref{eq:QCRineq}) is given explicitly by
\citep{ballesterb} 
\begin{eqnarray}
M_m &=& | b_m \rangle \langle b_m |, \qquad m = 1, \dots, p+1,\nonumber \\
M_{p+2} &=& \mathbb{I} - \sum_{m=1}^{p+1} M_m, \nonumber \\
| b_m \rangle &=& \sum_{n =1}^{p+1} O_{m n} |v_n\rangle, \nonumber \\
| v_m \rangle &=& \sum_n (H^{-1/2})_{mn} |l_n \rangle, \quad | v_{p+1} \rangle = | \psi \rangle,
\label{ballPOVM}
\end{eqnarray}
with $O$ a $(p+1)\times (p+1)$ real orthogonal matrix satisfying $O_{m,p+1} \neq 0$.
That this POVM does indeed give equality in (\ref{eq:QCRineq}) can be seen from Lemma 9 of \cite{fuj01a}.

An original proof of the necessity part of Theorem \ref{thm:matzy} will now be given.
\begin{lemma}
Condition (\ref{eq.matzcondq}) is a necessary condition for equality in (\ref{eq:QCRineq}).
\end{lemma}
\proof
For equality in (\ref{eq:QCRineq}) it is necessary that equality holds in (\ref{bciim}). Equality holds in (\ref{bciim}) if and only if
\begin{equation}
 M_m^{1/2} \lambda^k \rho^{1/2} = \xi_m^k  M_m^{1/2}  \rho^{1/2},\quad \xi_m^k \in \mathbb{R} \quad \forall k,m.
\label{eq:eqfhq} 
\end{equation}
For pure states, (\ref{eq:eqfhq}) becomes
\begin{equation*}
 M_m^{1/2} | l_k  \rangle  \langle \psi | = \xi_m^k  M_m^{1/2}  | \psi \rangle \langle \psi |, \quad \forall k,m.
\label{eq:eqfhq1} 
\end{equation*}
Thus equality holds in (\ref{bciim}) if and only if
 \begin{equation}
M_m^{1/2} | l_k \rangle  = \xi_m^k M_m^{1/2} | \psi \rangle, \quad \forall k,m.
\label{eq:eqfhq2} 
\end{equation}
Taking the transpose of (\ref{eq:eqfhq2}) gives 
 \begin{equation}
\langle l_j  | M_m^{1/2}  =\xi_m^j \langle \psi  | M_m^{1/2} , \quad \forall j,m.
\label{eq:eqfhq2b} 
\end{equation}
Pre-multiplying the left hand side of (\ref{eq:eqfhq2}) by the left hand side of (\ref{eq:eqfhq2b}), and the right hand side of (\ref{eq:eqfhq2}) by the right hand side of (\ref{eq:eqfhq2b}) gives the necessary condition  
 \begin{equation*}
 \langle l_j  | M_m  | l_k  \rangle  =\xi_m^j\xi_m^k p_m, \quad \forall j,k,m.
\label{eq:eqfhq3} 
\end{equation*}
Summing over $m$, and using the result $\sum_m M_m = \mathbb{I}$, gives
 \begin{equation*}
 \langle l_j  |  l_k  \rangle  = \sum_m \xi_m^j\xi_m^k  p_m.
\label{eq:eqfhq4} 
\end{equation*}
As $\xi_m^j\xi_m^k$ and $p_m$ are all real,  it follows that $ \langle l_j  |  l_k  \rangle$ is real and (\ref{eq.matzcondq}) is a necessary condition for equality in (\ref{bciim}), and thus a necessary condition for equality in (\ref{eq:QCRineq}).  
\medskip

That (\ref{eq.matzcondq}) is a sufficient condition for equality in (\ref{eq:QCRineq}), follows from  Ballester's result that if (\ref{eq.matzcondq}) holds, then the POVM given in (\ref{ballPOVM})
gives equality in (\ref{bciim}).

\subsection{Attainable measurements - the $2$-dimensional case}
For quantum statistical models with $\mathcal{H} = \mathbb{C}^2$,  equality holds in the Braunstein-Caves inequality (\ref{bcii})  only if every element of the POVM $M = \{ M_k \}$ has rank $1$.
 This was shown for pure states by \cite{barndorffnielsengill00}, and for mixed states by \cite{luati04}.
An original proof of this result, which includes mixed and pure state models, will now be given.
A necessary condition for equality in (\ref{bcii}) is
\begin{equation*}
M_m^{1/2} \lambda \rho^{1/2} = \xi_m  M_m^{1/2}  \rho^{1/2},\quad \xi_m \in \mathbb{R}, \quad \forall m.
\label{eq:eqfhq33} 
\end{equation*}
Pre-multiplying by $M_m^{1/2}$ and post-multiplying by $\rho^{1/2}$ gives
\begin{equation*}
M_m \lambda \rho =\xi_m M_m \rho, \quad \forall m.
\label{eq:eqfhq43} 
\end{equation*}
Then
\begin{equation}
M_m A_m  = 0,\qquad \forall m,
\label{eq:mA} 
\end{equation}
where 
\begin{equation*}
A_m = \lambda \rho - \xi_m \rho.
\label{eq:rew} 
\end{equation*}
Now, from (\ref{eq:mA}), it is seen that $M_m$ is singular unless $A_m=0$ for all $\theta$. It will be assumed that $A_m=0$ for all $\theta$. If this is so then
\begin{equation}
\lambda \rho = \xi_m \rho.
\label{eq:rhoma} 
\end{equation}
 Taking the trace of (\ref{eq:rhoma}) gives
 \begin{eqnarray}
\tr \{\lambda \rho \} &=& \xi_m \tr \{\rho\} \nonumber \\
0 &=& \xi_m.
\label{eq:rhoma1} 
\end{eqnarray}
Thus from (\ref{eq:rhoma}) and (\ref{eq:rhoma1}),
\begin{equation*}
\lambda \rho = 0,
\end{equation*}
and so 
\begin{equation*}
 \lambda \rho + (  \lambda \rho )^\dagger = 2 \frac{d \rho}{d \theta } =0.
\end{equation*}
Thus, if $A_m =0$, the model does not depend on $\theta$. Assuming that the model does depend on $\theta$, it follows that $A_m \neq 0$ and so $M_m$ is singular.
A consequence of this is that in the $2$-dimensional case, the elements of attainable measurements have rank $1$.

\section{Estimation}
Many quantum information processes can be represented as quantum channels. In practice, quantum channels are not known {\em a priori} and estimating them is of great importance.

There are several ways to estimate a quantum channel. One approach is quantum process tomography, \index{tomography}\index{quantum process tomography}which is discussed in chapter $8$ of \cite{chuang00}. For this approach it is necessary to estimate how the channel acts on different bases of the Hilbert space plus linear combinations thereof.  A problem with this method is that in many practical situations it is not possible to prepare these input states in the laboratory \citep{demartinietal03}.

Another approach is to assume that the channel comes from a given parametric family of channels \citep{fuj01,fuj01a,fuj04,fujimai03,ballesterb,ballestera,saromilb06}.
(The latter approach will be followed in this thesis.)
A family of channels parametrized by a real parameter $\theta$ can be represented by Kraus operators depending on $\theta$ as
\begin{eqnarray}
 \rho_{0} \mapsto  \sum_{k}E_{k}(\theta) \rho_{0} E_{k}^{\dagger}(\theta)  .
\label{eq:1pchannela}
\end{eqnarray}
When estimating a quantum channel, there are many different factors to consider:
how should the channels be arranged, and what combination of input state, POVM and estimator is best.
The idea of finding the optimal input state was considered by \cite{acin01}.

In general, for a parametric family of channels, different input states lead to different families of output states. The input state is chosen such that the family of output states has the maximum attainable SLD quantum information. The measurement which gives equality in (\ref{bciim}) is chosen (an adaptive measurement may be needed), and the maximum likelihood estimator used.

In this thesis the performance of an estimation procedure is usually measured either by the mean square error $E[(\hat \theta - \theta)^2]$ or for unitary channels, where $U_{\hat\theta}$ is the estimate of the unitary matrix $U_\theta$,  by
\begin{equation}
1- \langle F( U_{ \hat \theta},U_\theta) \rangle= 1 - \frac{\left \langle |\tr \{ U^{-1}_\theta  U_{ \hat \theta}|^2 \right \rangle}{d^2},
\label{eq.fhuu3}
\end{equation}
where $\langle \cdot \rangle$ denotes expectation. Often this cost function will be denoted simply by 
$1 - \langle F \rangle$.
Given a family of channels $\mathcal{E}(\theta)$, an estimate $\hat \theta$ of a parameter $\theta$ will depend on $n$, the number of times the channel $\mathcal{E}(\theta)$ is used.
Similarly, an estimate $U_{\hat \theta}$ of a unitary matrix $U_\theta$ will also depend on $n$.
It is of interest to see how rapidly $\hat \theta$ approaches $\theta$, and $ U_{\hat \theta}$ approaches $U_\theta$, as $n \rightarrow \infty$. The `big $O$' notation will be used for this purpose.
It is said that `$f(n)$ is $O(g(n))$' if there exist constants $c$ and $n_0$ such that for all $n > n_0$,
$f(n) \leq c g(n)$ \citep[p.\ 136]{chuang00}. That is, for large $n$, up to an unimportant factor, the function $g(n)$ is an upper bound on $f(n)$.

\subsection{Important developments in channel estimation} \label{sec:devel}
Here a brief review is given of the major advances in the estimation of quantum channels.

A channel $\mathcal{E}: S(\mathbb{C}^d) \mapsto S(\mathbb{C}^d)$, can be extended to a channel $\mathbb{I} \otimes \mathcal{E}: S( \mathbb{C}^{d^2}) \rightarrow S( \mathbb{C}^{d^2})$ by
 \begin{equation}
\rho_1 \mapsto (\mathbb{I}_d \otimes \mathcal{E})(\rho_1), \quad \rho_1 \in S(\mathbb{C}^{d^2}).
\label{eq.ex1}
\end{equation} 
For many channels $\mathcal{E}$, when using (\ref{eq.ex1}), a maximally entangled input state is optimal, in terms of Fisher information. Often the Fisher information is significantly greater than can be obtained from the unextended channel  $\mathcal{E}: S(\mathbb{C}^d) \mapsto S(\mathbb{C}^d)$. 
This was shown for a completely unknown unitary matrix in $SU(2)$ by \cite{fuj01a}, and $SU(d)$ (close to the identity) by \cite{ballestera}.
This has also been shown for several non-unitary channels, in particular the $2$-dimensional depolarizing channel \citep{fuj01} and, more generally, the generalized Pauli channels \citep{fujimai03}. 

Another advantage of the extended channel $\mathbb{I} \otimes \mathcal{E}$ is that, using a maximally entangled input state, the output states are in one-to-one correspondence with the channel. This means that, in contrast to quantum tomography,  the experimenter does not require many different input states: it is enough to have many copies of a maximally entangled state.

Using the extension (\ref{eq.ex1}) with a maximally entangled input state the mean square error and $1 - \langle F \rangle$ are $O(1/n)$ \citep{hayashi06}.
This rate at which $1 - \langle F \rangle$ approaches zero is known as the {\it standard quantum limit} \citep{burghbart05}, but can be surpassed \citep{hayashi06,kahn07,fujimai07}.

Another major step in estimation, when $n$ copies of a channel are available, was the idea of using the following extension with an entangled input state, so that
\begin{equation}
\rho_2 \mapsto  \mathcal{E}^{\otimes n}(\rho_2), \quad \rho_2 \in S(\mathbb{C}^{d^n}).
\label{rho2maps}
\end{equation}
One of the first clear uses of this method for estimation was by \cite{huelga97}. 

Using the experimental setup (\ref{rho2maps}), it has been shown that it is possible to estimate a unitary matrix with $1 - \langle F \rangle = O(1/n^2)$. This has been shown for estimation of an unknown unitary matrix in $SU(2)$ by \cite{hayashi06} and $SU(d)$ by \cite{kahn07}.
This rate at which $1 - \langle F \rangle$ approaches zero is known as the {\it Heisenberg limit} \citep{gio04} and cannot be surpassed \citep{kahn07}.

A {\it reference frame} is a specific coordinate system. \index{reference frame}
Estimation of a unitary matrix in $SU(2)$ is equivalent to the problem of transmiting a $3$-dimensional reference frame from Alice to Bob. Alice encodes information about her reference frame in quantum particles, and then sends these to Bob. Bob measures the quantum particles, and from his results estimates Alice's reference frame.
It has been shown that it is possible to do this with $1 - \langle F \rangle = O(1/n^2)$ \citep{baganetal04a,baganetal04b,chiribella04}. 

For most channels it is not possible to surpass the standard quantum limit. 
This has been shown for generalized Pauli channels by \cite{fujimai03}. 
Recently it has been shown \citep{fujimai08} that for most channels, given $n$ copies and using the setup (\ref{rho2maps}),  the SLD quantum information is $O(n)$.  A consequence of this is that, from the quantum Cram\'er-Rao inequality (\ref{eq:QCRineq}), for these channels, the mean square error is $O(1/n)$.

It is also possible to use a channel repeatedly on the same input state, i.e.
 \begin{equation}
\rho_0 \mapsto  \mathcal{E}^{ n}(\rho_0).
\end{equation}
\cite{kit95} suggested an $l$-stage iterative estimation scheme for the unitary matrix (\ref{eq.kitu}).
 At the $k$th stage $U_\theta$ acts $2^{k-1}$ times on the same input state. At each stage, several measurements are made. 
Using this information, an estimate $\hat \theta$ of $\theta$ is obtained satisfying 
$\mathrm{Pr} (| \hat \theta - \theta|_1 \leq 1/2^{l+2} ) \geq 1 -\epsilon$. The value of $\epsilon$ can be made arbitrarily small by doing more measurements at each stage.

 For a similar estimation scheme,  \cite{rudolph03} showed that, by choosing $\epsilon = 1/{2^{2l}}$,  $1 - \langle F \rangle = O((\log n/n)^2)$. The advantage of these estimation schemes is their simplicity:  they require no entanglement and only a single copy of $\mathcal{E}$. In spite of this, $1- \langle F \rangle$ is still close to the Heisenberg limit.
 
This thesis contains, as far as the author is aware, the first complete method for iterative estimation similar to that of \cite{kit95}. It is also shown that an extension similar to (\ref{rho2maps}) can be used to estimate $n$ non-identical channels, with an entangled input state. This results in an increase in the rate at which the mean square error decreases, relative to using a separable state.

\chapter{Attainability of the information bound of Sarovar and Milburn}
\label{ch:SM1}

\section{Introduction} \label{sec:introSM1}
The problem of estimating non-unitary quantum channels is more difficult than that of estimating unitary channels. The output states of non-unitary channels are mixed, and the SLD quantum information is generally more cumbersome to compute. Also, for multi-parameter families of mixed states, there is no known analogue of Matsumoto's condition (\ref{eq.matzcondq}) for equality in the Quantum Cram\'er--Rao inequality (\ref{eq:QCRineq}); neither is there a known method for computing the optimal POVM. 

\cite{saromilb06} introduced an upper bound on the Fisher information obtained from measuring the output states of a parameterised family of channels.
They also gave necessary and sufficient conditions for equality.
Their bound depends on the Kraus operators of the channel and not on the set of output states. 
In this chapter it is shown that this bound is not generally attainable, and consequently does not generally give the optimal POVM. 
Thus the attempt of Sarovar and Milburn to find the optimal estimation strategy for 
non-unitary quantum channels is not succesful. (The work in this chapter has been published in \cite{oloan07}.)

The problem of how to express the SLD quantum information of a noisy channel in terms of its Kraus operators has recently been solved by \cite{fujimai08} for the extended channel $\mathbb{I}_d \otimes \mathcal{E}: S(\mathbb{C}^{d^2}) \mapsto  S(\mathbb{C}^{d^2})$. 
This puts an upper bound on the SLD quantum information for the unextended channel  $\mathcal{E}: S(\mathbb{C}^{d}) \mapsto  S(\mathbb{C}^{d})$, but this bound will not, in general, be attainable. 

\subsection{The approach of Sarovar and Milburn} \label{sec:introSM1b}
 
Sarovar and Milburn looked at estimating one-parameter quantum channels of the form
\begin{eqnarray}
 \rho_{0} \mapsto  \sum_{k}E_{k}(\theta) \rho_{0} E_{k}^{\dagger}(\theta),
 \label{eq:1pchannel}
\end{eqnarray}
(see (\ref{eq.krausdec})).
The input state $\rho_0$ is a known pure state, and is chosen such that the output state is in one-to-one correspondence with the channel. Since a specific value of $\theta$ corresponds to a specific channel, estimation of the channel reduces to a parameter estimation problem. 
Sarovar and Milburn were interested in finding the maximal Fisher information that can be obtained by measuring the output states of the set of channels (\ref{eq:1pchannel}). They were also interested in finding POVMs that attain this bound. First, Sarovar and Milburn derived the inequality
 \begin{eqnarray}
F_\theta^M \leq C_{E}(\theta).
 \label{eq:smbound}
 \end{eqnarray}
where $E$ denotes a set of Kraus operators $\{ E_k \}$ and 
\begin{eqnarray}
C_{E}(\theta)  = 4 \sum_{k} \tr \{   E_{k}'(\theta)  \rho_{0} E_{k}'^{\dagger}(\theta) \}, \qquad E_{k}'(\theta) = \frac{d }{d \theta}E_{k}(\theta).
\label{ce}
 \end{eqnarray}
However, it was noted that $C_{E}(\theta)$  depends on the Kraus representation $E$ \citep{saromilb06}. 
For any channel $\mathcal{E}$, the Kraus representation is not unique. 
Given a unitary matrix $U = [u_{jk}]$ then the set of operators $\{ F_j \}$ given by
\begin{equation*}
F_j = \sum_{k} u_{jk} E_k,
\end{equation*}
lead to the same quantum channel  \cite[p.\ 372]{chuang00}.
That is, for all $\rho_0$,
\begin{equation*}
\sum_k E_k \rho_0 E_k^\dagger = \sum_j F_j \rho_0 F_j^\dagger.
\end{equation*}
To obtain a bound which depends only on the channel and not on the Kraus representation,  Sarovar and Milburn chose the bound given by the canonical Kraus operators. {\it Canonical Kraus operators}  $\{ \Upsilon_{k}(\theta) \}$ are  defined as Kraus operators satisfying
\begin{eqnarray}
\tr \{ \Upsilon_{k}(\theta)  \rho_{0}  \Upsilon_{j}^{\dagger }(\theta)  \} = \delta_{jk} p_{k}(\theta), \qquad \forall j,k.
\label{eq:condkco}
\end{eqnarray}
From (\ref{eq:smbound}) it follows that
\begin{eqnarray}
F_\theta^M \leq C_{\Upsilon}(\theta). 
 \label{eq:smboundups}
 \end{eqnarray}
 \begin{remark}
The canonical Kraus operators are unique only up to a choice of phase (see p.\ 267 of \cite{beng06}).
In Chapter \ref{ch:smqi} it is shown that this leads to ambiguity in the bound $C_\Upsilon(\theta)$.
 However, this does not affect the results in this chapter.
\end{remark}
Throughout the rest of this chapter the right hand side of (\ref{eq:smboundups}) will be referred to as the {\it SM bound}. \index{SM bound} The bound (\ref{eq:smboundups}) is said to be {\it uniformly attainable} if, for all $\theta$ in $\Theta$, there exists a POVM $M$, possibly depending on $\theta$, such that $F_\theta^M = C_{\Upsilon}(\theta)$.
If this bound is not uniformly attainable, then no bound of the form (\ref{ce}) is uniformly attainable \citep{saromilb06}. To achieve equality in (\ref{eq:smboundups}) the POVM $\{ M_m \}$ must satisfy
\begin{equation}
M_{m}^{1/2}  \Upsilon_{k}'(\theta) \rho_{0}^{1/2} = \xi_{m}(\theta)  M_{m}^{1/2}\Upsilon_{k}(\theta)  \rho_{0}^{1/2}, \qquad \forall m, k,
\label{eq:eupsrho}
\end{equation}
for some real $\xi_m(\theta)$. (This condition is analogous to (\ref{bcii2e}).)
For channels with quasi-classical output states (see Section \ref{quasi2}), it was shown in \cite{saromilb06} that this bound is attainable. Channels of this type will be called {\it quasi-classical channels}.
Sarovar and Milburn asked 
\begin{enumerate}
\item[(i)] whether their bound (\ref{eq:smboundups}) is attainable more generally,  
\item[(ii)] whether explicit expressions for optimal POVMs can be derived from the attainability conditions (\ref{eq:eupsrho}). 
\end{enumerate}
It is very important for an upper bound on Fisher information to be attainable, otherwise it gives an unrealistic view of how well a parameter can be estimated.  

\section{One-parameter channels}
In this Chapter the extended channel will be considered, i.e.
\begin{equation}
\rho_0 \mapsto \mathbb{I}_d \otimes \mathcal{E} (\rho_0), \qquad \rho_0 \in \mathcal{S}(\mathbb{C}^{d^2}).
\end{equation}
The canonical Kruas operators $\{ \Upsilon_k(\theta) \}$ are $d^2 \times d^2$ Kraus operators satisfying (\ref{eq:condkco}).

When the input state is pure, with $\rho_{0} = | \psi_{0} \rangle \langle \psi_{0} |$, condition (\ref{eq:condkco}) for the canonical Kraus decomposition is equivalent to the condition 
\begin{eqnarray}
\langle v_{j}(\theta)  | v_{k}(\theta) \rangle  = \delta_{jk} p_{k}(\theta), \qquad  \mathrm{where}   \quad  | v_{k}(\theta)  \rangle  =  \Upsilon_{k} (\theta) | \psi_{0}  \rangle .
\label{eq.defspecd}
\end{eqnarray}
The output state is 
\begin{eqnarray*}
\rho_\theta = \sum_{k} | v_{k}(\theta)  \rangle  \langle v_{k}(\theta) |.
\end{eqnarray*}
This can be rewritten as
\begin{eqnarray}
\rho_\theta = \sum_{k} p_{k}(\theta) | w_{k}(\theta) \rangle  \langle w_{k}(\theta) | , \qquad | w_{k}(\theta)   \rangle  =\frac{1}{\sqrt{p_{k}(\theta)}}| v_{k}(\theta)  \rangle.
\label{eq:outstate}
\end{eqnarray}
Thus the canonical decomposition leads to the spectral decomposition of the output state \citep{saromilb06}. 

\begin{proposition}
The SM bound, $C_{\Upsilon}(\theta)$, can be expressed as (omitting $\theta$)
\begin{eqnarray}
 C_{\Upsilon} &=& \sum_{k, p_k \neq 0} \frac{p_{k}'^{2}}{p_{k}} +  \sum_{ j < k, p_j + p_k > 0 } 4 (p_{j} + p_{k}) |\langle w_{j}' | w_{k}  \rangle |^2  \nonumber \\
 &+&  4    \sum_{k, p_k \neq 0} p_{k} |\langle w_{k}' | w_{k}  \rangle |^2 .
\label{eq:samr}
\end{eqnarray}
\label{prop29}
\end{proposition}

\proof
For simplicity, it is assumed that for all $p_j(\theta)$  either 
\begin{enumerate}
\item[(i)] $p_j(\theta) > 0$ for all $\theta$, 
\item[(ii)] $p_j(\theta) = 0$ for all $\theta$.
\end{enumerate}
When $p_j(\theta) = 0$ for all $\theta$, it follows from (\ref{eq.defspecd}) and (\ref{eq:outstate}) that 
 \begin{eqnarray*}
 \Upsilon_{j}| \psi_{0} \rangle &=& \sqrt{p_{j}} | w_{j}  \rangle = 0,\\
\Upsilon_{j}' | \psi_{0} \rangle &=&  0, \\
  \tr \{  \Upsilon_{j}^{'}  \rho_{0} \Upsilon_{j}^{\dagger '}  \} &=& \langle \psi_{0} | \Upsilon_{j}^{\dagger '} \Upsilon_{j}' | \psi_{0}  \rangle = 0.
  \end{eqnarray*} 
 When $p_j(\theta) > 0$, for all $\theta$,
\begin{eqnarray*}
 \Upsilon_{j}| \psi_{0} \rangle &=& \sqrt{p_{j}} | w_{j}  \rangle,\\
\Upsilon_{j}' | \psi_{0} \rangle &=& \frac{p_{j}'}{2 \sqrt{p_{j}}}  | w_{j}  \rangle +   \sqrt{p_{j}} | w_{j} ' \rangle .\\
\end{eqnarray*}
Then
\begin{eqnarray}
\langle \psi_{0} | \Upsilon_{j}^{\dagger '} \Upsilon_{j}' | \psi_{0}  \rangle &=& \left(\frac{p_{j}'}{2 \sqrt{p_{j}}}  \langle w_{j} | +   \sqrt{p_{j}} \langle w_{j} '| \right) \left(\frac{p_{j}'}{2 \sqrt{p_{j}}}  | w_{j}  \rangle +   \sqrt{p_{j}} | w_{j} ' \rangle \right),\nonumber \\
&=& \frac{p_{j}'^2}{4p_{j}} + \frac{p_{j}'}{2} \left( \langle w_{j}' | w_{j}  \rangle + \langle w_{j} | w_{j}'  \rangle \right) + p_{j} \langle w_{j}' | w_{j}'  \rangle. \label{eqpjpj}
\end{eqnarray}
The right hand side of (\ref{eqpjpj}) can be simplified, because
\begin{eqnarray}
\langle w_{j}' | w_{j}  \rangle + \langle w_{j} | w_{j}'  \rangle = \frac{\partial}{\partial \theta}  \tr \{\rho_{j} \}  = 0 , \qquad  \rho_{j} =  | w_{j} \rangle  \langle w_{j} |.
\label{eq:diffjj}
\end{eqnarray}
(It follows from (\ref{eq:diffjj}) that $\langle w_{j}' | w_{j}  \rangle$ is purely imaginary.)
Thus 
\begin{eqnarray*}
C_{\Upsilon}&=& 4 \sum_{j, p_j \neq 0}\left( \frac{p_{j}'^2}{4p_{j}} + p_{j}\langle w_{j}' | w_{j}'  \rangle \right). 
\end{eqnarray*}
Inserting the identity $\mathbb{I}_d = \sum_{k=1}^{d} | w_{k}  \rangle  \langle   w_{k} |$ into $\langle w_{j}' | w_{j}'  \rangle $ gives
\begin{eqnarray}
C_{\Upsilon} &=& \sum_{j, p_j \neq 0}  \frac{p_{j}'^2}{p_{j}} + \sum_{j, k, p_j \neq 0}  4 p_{j}\langle w_{j}' | w_{k}  \rangle  \langle   w_{k} |  w_{j}'  \rangle, \nonumber \\
&=&  \sum_{j, p_j \neq 0}  \frac{p_{j}'^2}{p_{j}} +  \sum_{j, k, p_j \neq 0}  4 p_{j} |\langle w_{j}' | w_{k}  \rangle |^2.
\label{labelthis}
\end{eqnarray}
The right hand side of (\ref{labelthis}) will be re-written in (\ref{21312}). Since
\begin{equation*}
\langle w_{j} | w_{k} \rangle = \delta_{jk}, 
\end{equation*}
it follows that
\begin{eqnarray}
\frac{\partial}{\partial \theta} \langle w_{j} | w_{k}  \rangle &=& \langle w_{j}' | w_{k}  \rangle + \langle w_{j} | w_{k}'  \rangle = 0,\nonumber \\
\langle w_{j}' | w_{k}  \rangle &=& - \langle w_{j} | w_{k}'  \rangle, \label{eq:wjwk}  \\
|\langle w_{j}' | w_{k}  \rangle |^2 &=& \langle w_{j}' | w_{k}  \rangle  \langle w_{k} | w_{j}'  \rangle \nonumber \\
&=& (- \langle w_{j} | w_{k}'  \rangle) (- \langle w_{k}' | w_{j}  \rangle) =  |\langle w_{k}' | w_{j}  \rangle |^2. \label{eq:diffwjwk}
\end{eqnarray}
Now,
\begin{eqnarray}
\sum_{j, k, p_j \neq 0}  4 p_{j} |\langle w_{j}' | w_{k}  \rangle |^2 &=& \sum_{j < k, p_j \neq 0}  4 p_{j} |\langle w_{j}' | w_{k}  \rangle |^2 + \sum_{ k < j , p_j \neq 0}  4 p_{j} |\langle w_{j}' | w_{k}  \rangle |^2  \nonumber \\
&+& \sum_{j = k, p_j \neq 0}  4 p_{j} |\langle w_{j}' | w_{k}  \rangle |^2. \label{lblbl}
\end{eqnarray}
Swapping the indices $j$ and $k$ in the second term and using (\ref{eq:diffwjwk}) simplifies (\ref{lblbl}) further to 
\begin{equation}
 \sum_{j, k, p_j \neq 0}  4 p_{j} |\langle w_{j}' | w_{k}  \rangle |^2 = \sum_{j < k, p_j + p_k \neq 0}  4( p_{j} + p_k ) |\langle w_{j}' | w_{k}  \rangle |^2 + \sum_{j , p_j \neq 0}  4 p_{j} |\langle w_{j}' | w_{j}  \rangle |^2.
\label{bingb}
\end{equation}
Thus, from (\ref{labelthis}) and (\ref{bingb}), the SM bound $C_{\Upsilon}(\theta)$ can be rewritten as
\begin{eqnarray}
C_{\Upsilon} &=& \sum_{j,  p_j \neq 0}  \frac{p_{j}'^2}{p_{j}} + \sum_{j < k,  p_j+p_k > 0}  4 ( p_{j} + p_{k} ) |\langle w_{j}' | w_{k}  \rangle |^2 \nonumber \\
&+&   4\sum_{ p_k \neq 0}  p_{k} |\langle w_{k}' | w_{k}  \rangle |^2. \label{21312}
\end{eqnarray}

\begin{remark}
It can be seen that $C_\Upsilon(\theta)$ can be described solely in terms of the family of output states.
The SM bound was originally derived as an upper bound on the Fisher information for a one-parameter family of quantum channels. Since any  parametric family of quantum states can be written in the form
\begin{eqnarray*}
\rho_\theta = \sum_k p_{k}(\theta) | w_{k}(\theta) \rangle  \langle w_{k}(\theta) |,
\label{eq:outstate2}
\end{eqnarray*}
$C_\Upsilon(\theta)$ can be extended to an upper bound on the Fisher information for one-parameter families of states. 

It can be seen from the form of (\ref{21312}) that $C_\Upsilon(\theta)$ is a Riemannian metric on a $1$-dimensional manifold (see Section \ref{subsec:SLDQFI}).
\end{remark}

\begin{proposition}
The SLD quantum information can be written as (omitting $\theta$)
\begin{equation}
H = \sum_{k, p_k \neq 0} \frac{p_{k} '^{2}}{p_{k}} +   \sum_{  j < k, p_j + p_k > 0  } 4\frac{ ( p_{j} - p_{k})^2}{p_{j} + p_{k}} |\langle w_{j}' | w_{k}  \rangle|^2.
\label{eq:Hsldr}\\
\end{equation}
\label{prop:sld}
\end{proposition}

\proof
 The SLD is defined as any self-adjoint solution $\lambda$ of the matrix equation 
\begin{equation}
\frac{d \rho}{d\theta} = \frac{1}{2}\left(\rho \lambda+\lambda \rho \right).
\label{sld:def}
 \end{equation}
The SLD quantum information is defined as
\begin{equation*}
H = \tr \{ \rho \lambda^2 \} .
\label{Hsld:def2}
\end{equation*}
Substituting (\ref{eq:outstate}) into (\ref{sld:def}) gives
\begin{eqnarray}
\sum_{i=1} \left\{ p_{i}' | w_i \rangle \langle  w_i | + p_{i} (  | w_{i}'  \rangle \langle w_i | + | w_i  \rangle \langle w_{i} ' | ) \right\} \nonumber \\ =   \frac{1}{2} \left( \sum_{l} p_{l}| w_l  \rangle \langle w_l | \lambda + \lambda \sum_{m}  p_{m} | w_m  \rangle \langle w_m | \right) . 
\label{sld:def2}
 \end{eqnarray}
From (\ref{sld:def2}) the components of the SLD are calculated. First,  the diagonal elements $\lambda_{jj}$ are considered.  Pre-multiplying (\ref{sld:def2}) by $\langle w_j |$
 and post-multiplying  $| w_j \rangle$ gives, on the left hand side,
 \begin{equation*} 
 p_{j}' +   p_{j} ( \langle w_j  | w_{j}'  \rangle + \langle w_{j}'  | w_{j}  \rangle ) = p_{j}' 
 \end{equation*}
by (\ref{eq:diffjj}), and on the right hand side 
 \begin{equation*} 
p_{j} \langle w_j | \lambda | w_j \rangle.
  \end{equation*}
Hence, provided that $p_j > 0$, 
 \begin{equation*} 
\lambda_{jj} = \frac{p_{j}'}{p_j}. 
  \end{equation*}
The diagonal elements   $\lambda_{jj}$  are not defined when $p_j = 0$. In this case,  a particular solution of $\lambda$ is chosen for which $\lambda_{jj} =0$.  Next, the off-diagonal components $\lambda_{jk}$ are considered. Pre-multiplying (\ref{sld:def2})  by $\langle w_j |$ and post-multiplying by $| w_k \rangle$ gives, on the left hand side
 \begin{equation*} 
0 + p_{k}  \langle w_j | w_{k}' \rangle + p_{j}  \langle w_{j}' | w_k \rangle = (p_{j} - p_{k}) \langle w_{j}' | w_k \rangle,
  \end{equation*}
by (\ref{eq:wjwk}), and on the right hand side 
\begin{equation*} 
\frac{1}{2} (p_{j} + p_{k})  \langle w_j | \lambda | w_k \rangle.
  \end{equation*}
Thus, provided that $p_j + p_k > 0$, 
\begin{equation*} 
\lambda_{jk} = \frac{ 2 (p_{j} - p_{k}) \langle w_{j}' | w_k \rangle }{p_{j} + p_{k}}.
  \end{equation*}
The entries  $\lambda_{jk}$ are not defined when $p_j + p_k = 0$. Again a particular solution of $\lambda$ is chosen for which $\lambda_{jk} =0$, when $p_j + p_k = 0$.   This gives the following particular solution  of the SLD 
 \begin{equation}
\tilde \lambda =  \sum_{k, p_k \neq 0} \frac{ p_k '}{ p_k} | w_k \rangle \langle w_k |  + 
 \sum_{j \neq k, p_j+p_k > 0}  2 \frac{  p_j - p_k}{ p_j + p_k} \langle w_j' | w_k \rangle | w_j \rangle \langle  w_k |.
   \label{eq:Lsldr}
\end{equation}
Denote by $\tilde \lambda^{2*}$ the part of $\tilde \lambda^2$ which makes a non-zero contribution to $\tr \{\rho \tilde \lambda^2 \}$.
Only terms of the form $z_k | w_k \rangle \langle w_k |$, with $z_k \in \mathbb{C}$, in $\tilde \lambda^{2}$ will contribute to $\tr \{\rho \tilde \lambda^2 \}$. Thus, 
\begin{eqnarray*}
\tilde \lambda^{2*} &=&  \sum_{k, p_k \neq 0} \left(\frac{ p_k '}{ p_k}\right)^2 | w_k \rangle \langle w_k |  +  \sum_{j \neq k, p_j+p_k > 0}  4 \frac{  p_j - p_k}{ p_j + p_k}  \frac{  p_k - p_j}{ p_k + p_j} \langle w_j' | w_k \rangle \langle w_k' | w_j \rangle | w_j \rangle \langle  w_j |
   \label{eq:Lsldr2}
   \\
   &=&  \sum_{k, p_k \neq 0} \left(\frac{ p_k '}{ p_k}\right)^2 | w_k \rangle \langle w_k |  + 
  \sum_{j \neq k, p_j+p_k > 0}  4 \left(\frac{  p_j - p_k}{ p_j + p_k}\right)^2 |\langle w_j' | w_k \rangle|^2 | w_j \rangle \langle  w_j |,
   \label{eq:Lsldr2b}
\end{eqnarray*}
using (\ref{eq:wjwk}).
This gives
\begin{equation}
H = \sum_{k, p_k \neq 0}\frac{ p_k^{'2}}{ p_k}  + 
  \sum_{j \neq k, p_j+p_k > 0} 4 p_j  \left(\frac{  p_j - p_k}{ p_j + p_k}\right)^2 |\langle w_j' | w_k \rangle|^2. 
   \label{eq:Lsldr2b2}
\end{equation}
The second term on the right hand side of  (\ref{eq:Lsldr2b2}) can be rewritten as
\begin{eqnarray*}
  \sum_{j \neq k, p_j+p_k > 0}  4 p_j  \left(\frac{  p_j - p_k}{ p_j + p_k}\right)^2 |\langle w_j' | w_k \rangle|^2
 &=&   \sum_{j <  k, p_j+p_k > 0}  4p_j  \left(\frac{  p_j - p_k}{ p_j + p_k}\right)^2 |\langle w_j' | w_k \rangle|^2 \nonumber \\
&+&  \sum_{k <  j , p_j+p_k > 0} 4 p_j  \left(\frac{  p_j - p_k}{ p_j + p_k}\right)^2 |\langle w_j' | w_k \rangle|^2. \qquad 
\label{poppie1}
\end{eqnarray*}
Swapping the indices, $j$ and $k$, in the second term on the right hand side of the above equation and using (\ref{eq:diffwjwk}) gives
\begin{equation}
  \sum_{j \neq k, p_j+p_k > 0}  4 p_j  \left(\frac{  p_j - p_k}{ p_j + p_k}\right)^2 |\langle w_j' | w_k \rangle|^2
 =  \sum_{j <  k, p_j+p_k > 0}  4 \frac{(  p_j - p_k)^2}{ p_j + p_k} |\langle w_j' | w_k \rangle|^2.
 \label{eq221}
 \end{equation}
 The required result (\ref{eq:Hsldr}) follows from (\ref{eq:Lsldr2b2}) and (\ref{eq221}).

\begin{theorem}
\begin{equation}
 H_\theta \leq C_{\Upsilon}(\theta),
\label{eq:ineqcs2}
\end{equation}
with equality if and only if 
\begin{equation}
\langle w_{j}' | w_{k}  \rangle = 0, \quad \forall j,k \quad \mathrm{with} \quad  p_j, p_k > 0.
\label{wjwk0}
\end{equation}
\label{eq:ineqcs}
\end{theorem}

\proof
The first terms in $H_\theta$, given in (\ref{eq:Hsldr}), and $C_\Upsilon(\theta)$, given in (\ref{eq:samr}), are identical. Thus
\begin{equation*}
C_{\Upsilon}(\theta) - H_\theta =  A_{C}(\theta) - A_H(\theta) +  B_{C}(\theta),
\end{equation*}
where (omitting $\theta$)
\begin{eqnarray*}
A_{H} &=&  \sum_{  j < k, p_j + p_k > 0 } 4\frac{( p_{j} - p_{k})^2}{p_{j} + p_{k}} |\langle w_{j}' | w_{k}  \rangle|^2, \\
A_{C} &=&  \sum_{  j < k, p_j + p_k > 0 } 4(p_{j} + p_{k}) |\langle w_{j}' | w_{k}  \rangle |^2 ,\\
B_{C} &=& 4 \sum_{k, p_k \neq 0} p_{k} |\langle w_{k}' | w_{k}  \rangle |^2 .
\label{eq:equality1}
\end{eqnarray*}
The terms $A_C$ and $A_H$ are symmetric in $j$ and $k$ due to (\ref{eq:diffwjwk}). Now
\begin{eqnarray*}
A_C - A_H &=& 2 \sum_{  j \neq k, p_j + p_k > 0 } \frac{( p_{j} + p_{k})^2 - ( p_{j} - p_{k})^2}{p_{j} + p_{k}} |\langle w_{j}' | w_{k}  \rangle|^2, \\
&=& 8 \sum_{  j \neq k, p_j + p_k > 0 } \frac{p_{j} p_{k}}{p_{j} + p_{k}} |\langle w_{j}' | w_{k}  \rangle|^2.
\end{eqnarray*}
Changing the range of the summation  to $j \neq k$ where $p_j,p_k > 0$, and adding $B_C$ gives
\begin{equation}
C_{\Upsilon}- H = 8 \sum_{  j, k, p_j ,p_k > 0 } \frac{p_{j} p_{k}}{p_{j} + p_{k}} |\langle w_{j}' | w_{k}  \rangle|^2.
\label{cmh}
\end{equation}
Since the right hand side of (\ref{cmh}) is non-negative,  (\ref{eq:ineqcs2}) follows.

Equality holds in (\ref{eq:ineqcs2}) if and only if the right hand side of (\ref{cmh}) is zero, which holds if and only if (\ref{wjwk0}) holds.
  
  \begin{lemma}
For channels, with output states, for which $p_j(\theta) > 0$ for all $j$ and $\theta$, the bound (\ref{eq:ineqcs2}) is achievable if and only if the channel is quasi-classical.
\label{lemc}
\end{lemma}
\proof 
Equality holds in (\ref{eq:ineqcs2}) if and only if (\ref{wjwk0}) is satisfied. When $p_j(\theta) > 0$ for all $j$ and $\theta$, 
condition (\ref{wjwk0}) is satisfied if and only if  $|w_{j}'  \rangle$ has zero components along every vector $|w_{k} \rangle$. This is possible only if $|w_{j}'  \rangle = 0$ and hence the channel is quasi-classical.

\begin{lemma}
For unitary channels, the bound (\ref{eq:ineqcs2}) is achievable if and only if
\begin{equation}
\tr \{ U_\theta \rho_{0} U_\theta'^{\dagger}  \} = 0.
\label{eq:unitconds}
\end{equation}
\label{lemd}
\end{lemma}
\proof 
Equality holds in  (\ref{eq:ineqcs2}) if and only if (\ref{wjwk0}) is satisfied. For unitary channels there is only one non-zero $p_j$ and $| w_j \rangle = U_\theta | \psi_0 \rangle$, where $\rho_0 =  | \psi_0 \rangle \langle \psi_0 |$.  Condition (\ref{wjwk0}) is satisfied if and only if $\langle w_j ' | w_j \rangle = 0$. This is equivalent to (\ref{eq:unitconds}).

\begin{remark}
Note that, for the most common unitary channels -- those of the form $\exp(i\theta H)$, with $H$ a self-adjoint matrix -- condition (\ref{eq:unitconds}) is satisfied.
\end{remark}

\begin{example}
There exist channels which are neither quasi-classical or unitary for which equality holds in (\ref{eq:ineqcs2}). The channel with an arbitray pure input state and output states
\begin{equation*}
\rho_\theta = \theta^2 | w_1(\theta) \rangle \langle w_1(\theta)  | + ( 1 - \theta^2 ) | w_2(\theta)  \rangle \langle w_2 (\theta) |, \quad 0 < \theta < 1,\\
\end{equation*}
where
\begin{equation*}
| w_1(\theta)  \rangle = ( \theta, \sqrt{ 1 - \theta^2}, 0 )^T, \quad  | w_2(\theta)  \rangle = ( 0, 0, 1 )^T,
\end{equation*}
satisfies (\ref{wjwk0}), and so equality holds in (\ref{eq:ineqcs2}).
 \end{example}
 
  \begin{theorem}
  \begin{equation}
  F_\theta^M \leq C_\Upsilon(\theta),
  \label{fupsc2}
  \end{equation}
with equality if and only if 
 \begin{equation}
\langle w_{j}' | w_{k}  \rangle = 0, \quad \forall j,k \quad \mathrm{with} \quad  p_j, p_k > 0.
\label{fupsec}
\end{equation}
   \end{theorem}
 \proof
Inequality (\ref{fupsc2}) follows from (\ref{bcii}) and (\ref{eq:ineqcs2}). Equality holds in (\ref{fupsc2}) if and only if there is equality in both (\ref{bcii}) and (\ref{eq:ineqcs2}). For one-parameter families of states it is always possible to find a POVM $M_{\theta}$, depending on $\theta$, which achieves equality in (\ref{bcii})  \citep{braunsteincaves94}. However, equality holds in  (\ref{eq:ineqcs2}) if and only if  (\ref{fupsec}) is satisfied.

 \begin{theorem}
 \begin{equation}
 H_\theta \leq C_{E}(\theta),
   \label{hlessce2}
 \end{equation}
 with equality if and only if the set of output states satisfies (\ref{wjwk0}), and a fixed unitary matrix $U = [u_{jk} ]$ exists such that the Kraus operators $E_j$ are related to the canonical Kraus operators $\Upsilon_k$ by
\begin{equation*}
E_j(\theta) = \sum_k u_{j k} \Upsilon_k(\theta).
\end{equation*}
  \label{hlessce}
   \end{theorem}
  \proof
Inequality (\ref{hlessce2}) will be proved by considering two cases:
 \begin{enumerate} 
\item[(i)] When equality is attainable in (\ref{eq:smbound}), it is attainable also in (\ref{eq:smboundups}) \citep{saromilb06}. In this case,  $C_{\Upsilon}(\theta) \leq C_{E}(\theta)$ for all other sets of Kraus operators $E = \{ E_j \}$ \citep{saromilb06}.  Inequality (\ref{eq:ineqcs2}) gives $H_\theta \leq C_{E}(\theta)$.

\item[(ii)] When (\ref{eq:smbound}) is not attainable, $F_\theta^M <  C_{E}(\theta)$ for all $M$.
 For one-parameter families of states there always exists a measurement $M_{\theta}$ such that $F_{\theta}^{M_\theta} = H_\theta$.  Thus  $H_\theta = F_{\theta}^{M_\theta} <  C_{E}(\theta)$.
 \end{enumerate}  

Equality holds in (\ref{hlessce2}) only if the bound given by the canonical Kraus operators $C_{\Upsilon}$ is attainable.  The bound  $C_{\Upsilon}$ is attainable if and only if the set of ouput states satisfies (\ref{wjwk0}). It has been shown \citep[p.\ 372]{chuang00} that if two sets of Kraus operators $\{ E_j \}$ and $\{ F_k \}$  lead to the same quantum channel then they must be related by
\begin{equation}
E_j = \sum_k u_{j k} F_k,
\label{EUF}
\end{equation}
where $U = [u_{jk} ]$ is a unitary matrix. When  $C_{\Upsilon}$ is attainable \citep{saromilb06},
\begin{equation*}
C_E = C_{\Upsilon} + 4 \sum_{jk} p_j | u_{jk}'|^2.
\end{equation*}
Thus for equality in (\ref{hlessce2}) it is further required that $\sum_{jk} p_j | u_{jk}'|^2 = 0$. This is satisfied if and only if a unitary matrix $U= [u_{jk} ]$ exists satisfying (\ref{EUF}) that does not depend on $\theta$.

\begin{remark}
Condition (\ref{eq:eupsrho}) cannot be used generally to test for optimality of POVMs. 
Condition  (\ref{eq:eupsrho}) is a necessary and sufficient condition for equality between the Fisher information and the SM bound. Since it is not generally possible to achieve equality between the Fisher information and the SM bound, condition (\ref{eq:eupsrho}) cannot be achieved for general models. Thus it cannot be used generally to test for POVMs giving maximal Fisher information.
\end{remark}
  
   \section{Multi-parameter channels}
\subsection{The multi-parameter SM bound}
 The SM bound for a multi-parameter family of channels will be defined as the matrix $C_{\Upsilon}(\theta)$ with entries
    \begin{eqnarray}
C_{\Upsilon}(\theta)_{jk} = 4 \sum_{l} \Re \tr \left\{ \Upsilon_{l}(\theta)^{(j)} \rho_{0}  \Upsilon_{l}(\theta)^{(k) \dagger}   \right\}, \qquad \Upsilon_{l}(\theta)^{(k)} = \frac{\partial }{\partial \theta^k}\Upsilon_{l}(\theta).
\label{multipcupsdef}
\end{eqnarray}

\begin{proposition}
For $\theta$ and $ v$ in $\mathbb{R}^p$, and $t \rightarrow 0$,
\begin{eqnarray}
\frac{d}{dt}  \Upsilon_{k} (\theta + t  v  ) &=& \sum_l  \Upsilon_{k} (\theta)^{(l)} v^l + O(t),
\label{eq:multip1}\\
\tilde  \lambda_t  &=&  \sum_l \tilde \lambda_\theta^{(l)} v^l + O(t),\label{eq:multip3}
\end{eqnarray}
where $\tilde \lambda_t$ is defined by (\ref{eq:Lsldr}) with respect to the parameter $t$, $\tilde \lambda_\theta^{(l)}$ is defined by (\ref{eq:Lsldr}) with respect to the parameter $\theta^l$ and $v^l$ is the $l$th component of the vector $v$.  
\label{prop:444}
\end{proposition}

\proof \\
Put $\phi(t) = \theta + t v $,  with components $\phi^l(t) = \theta^l + t  v ^l $. Using the chain rule to differentiate $\Upsilon_{k} (\phi(t) )$ gives
\begin{equation}
\frac{d}{d t}  \Upsilon_{k} (\phi(t)) = \sum_l  \frac{ \partial \Upsilon_{k} (\phi)}{\partial \phi^l} \frac{ \partial \phi^l}{\partial t} .
\label{eq:appca8}
\end{equation}
Now,
\begin{eqnarray*}
  \frac{ \partial \Upsilon_{k} (\phi)}{\partial \phi^l}  &=&  \left. \frac{ \partial \Upsilon_{k} (\phi)}{\partial \phi^l}  \right|_{t=0}+ O(t) = \frac{ \partial \Upsilon_{k} (\theta)}{\partial \theta^l} + O(t) ,\\
  \frac{ \partial \phi^l}{\partial t}  &=& v^l.
  \end{eqnarray*}
Substituting these back into (\ref{eq:appca8}) gives (\ref{eq:multip1}). Similarly, for $t \rightarrow 0$, 
 \begin{eqnarray}
p_k(\theta+tv) &=& p_k (\theta) + O(t),\label{O1} \\
 \frac{d p_k(\theta+tv)}{d t } &=& \sum_l  p_k^{(l)}  v^{l} + O(t) , \quad p_k^{(l)} = \frac{\partial p_k(\theta)}{ \partial \theta^{l}}  \label{O2}\\
 \frac{d | w_k(\theta+tv) \rangle }{d t } &=&  \sum_l  | w_k^{(l)} \rangle  v^l + O(t), \quad| w_k^{(l)} \rangle  = \frac{\partial | w_k(\theta) \rangle}{ \partial \theta^l}.
  \label{O3}
 \end{eqnarray}
 Substituting (\ref{O1}) -- (\ref{O3}) into (\ref{eq:Lsldr}) gives
 \begin{eqnarray*}
\tilde \lambda_t &=&  \sum_{k, p_k \neq 0}  \frac{ \sum_l  p_k^{(l)}  v^{l} + O(t)}{p_k + O(t)} | w_k \rangle \langle w_k |   \\
&+& \sum_{  j \neq k, p_j + p_k > 0} 2 \frac{  p_j  - p_k+ O(t)}{p_j +  p_k + O(t)} \left( \sum_l v^l  \left\langle w_j^{(l)} | w_{k} \right\rangle   + O(t)\right)| w_j \rangle \langle  w_k | \\
&=& \sum_l v^l \left( \sum_{k, p_k \neq 0} \frac{ p_k^{(l)}}{ p_k} | w_k \rangle \langle w_k |  + 
 \sum_{j \neq k, p_j+p_k > 0}  2 \frac{  p_j - p_k}{ p_j + p_k} \left\langle w_j^{(l)} | w_k \right\rangle | w_j \rangle \langle  w_k | \right) + O(t).
\end{eqnarray*}
Thus $\tilde \lambda_t$ has the form (\ref{eq:multip3}).
\begin{theorem}
For multi-parameter channels,
\begin{equation}
 H_\theta \leq C_{\Upsilon}(\theta),
\label{mulithc}
\end{equation}
with equality if and only if 
\begin{eqnarray}
\bigg\langle w_j^{(l)} \bigg| w_k \bigg\rangle &=& 0, \quad \forall j,k,l   \quad \mathrm{with} \quad   p_j,p_k > 0, \label{eq:unitcond3a} \\
\mathrm{where} \quad  \bigg| w_j^{(l)} \bigg\rangle &=& \frac{\partial}{\partial \theta^l}    \bigg| w_j  \bigg \rangle.
\nonumber
\end{eqnarray}
\label{lm:4a}
\end{theorem}
\proof
Equation (\ref{mulithc}) is equivalent to 
\begin{equation}
 v^T   H_\theta  v \leq  v^T  C_{\Upsilon}(\theta) v ,\quad \mathrm{for \, all} \, v \in \mathbb{R}^p.
\label{eq:cCca}
\end{equation}
 For given $\theta$ and $ v$ in $\mathbb{R}^p$, consider the set of one-parameter channels
\begin{equation}
 \rho_{0} \mapsto  \sum_{k}  \Upsilon_{k} (\theta + t  v ) \rho_{0} \Upsilon_{k}^{\dagger}(\theta + t v  ) ,\quad t \in \mathbb{R}.
\label{eq:multiproof1}
\end{equation}
 From Theorem \ref{eq:ineqcs} it is known that $  H_t \leq C_{\Upsilon}(t)$, i.e.
\begin{equation*}
 \tr \left\{ \tilde \lambda_t \rho_{\theta + t v} \tilde \lambda_t \right\} \leq 4  \sum_{l=1}^{d} \tr \left\{ \frac{d}{d t}  \Upsilon_{l} (\theta + t v ) \rho_0  \frac{d}{d t}  \Upsilon_{l} (\theta + t  v  )^{\dagger} \right\}.
 \label{eq:34a}
\end{equation*}
Using (\ref{eq:multip1}) and (\ref{eq:multip3}) and evaluating at $t=0$  gives
\begin{equation*}
\sum_{m,n} v^m v^n \tr \left\{ \tilde \lambda_\theta^{(m)} \rho_\theta \tilde \lambda_\theta^{(n)} \right\} \leq 4 \sum_{m,n,l} v^m v^n \tr \left\{  \Upsilon_{l}(\theta)^{(m)} \rho_0  \Upsilon_{l}(\theta  )^{(n) \dagger} \right\}.
\label{eq:34ab}
\end{equation*}
This is equivalent to (\ref{eq:cCca}). Since this holds for all $v \in \mathbb{R}^p$,  (\ref{mulithc}) holds. 
\\
\\
Equality in (\ref{mulithc}) is equivalent to 
\begin{equation}
 v^T   H_\theta  v =  v^T  C_{\Upsilon}(\theta) v ,
\label{eq:cCc2a}
\end{equation}
for all $v \in \mathbb{R}^p$.  It follows that  (\ref{eq:cCc2a}) holds for all $v \in \mathbb{R}^p$ if and only if, for one-parameter channels of the form (\ref{eq:multiproof1}) for given $\theta$ and $v \in \mathbb{R}^p$,  $H_t |_{t=0} =  C_{\Upsilon}(t) |_{t=0}$. From Theorem \ref{eq:ineqcs}, this holds if and only if the channel (\ref{eq:multiproof1}) satisfies (\ref{wjwk0}) at the point $t=0$. This condition is equal to
\begin{equation*}
\left. \left( \frac{d}{dt} \langle w_j | \right) | w_k \rangle\right|_{t=0} = 0, \quad \forall j,k  \quad \mathrm{with} \quad   p_j,p_k > 0. 
\end{equation*}
Using (\ref{O3}), this condition can be rewritten as 
  \begin{eqnarray}
\sum_{l=1}^{m} v^l \bigg\langle w_j^{(l)} \bigg| w_k \bigg\rangle = 0 \quad \forall j,k  \quad \mathrm{with} \quad   p_j,p_k > 0.
\label{eq:unitcond2a}
\end{eqnarray}
Condition (\ref{eq:unitcond2a}) holds for all $v$ if and only if (\ref{eq:unitcond3a}) is satisfied.

\begin{lemma}
For channels, with output states for which $p_j(\theta) > 0$ for all $j$ and $\theta$, equality holds in (\ref{mulithc}) if and only if the channel is quasi-classical.
\end{lemma}
\proof
This follows from (\ref{eq:unitcond3a}) and the same analysis as in Lemma \ref{lemc}.

  \begin{lemma}
For unitary channels, equality holds in (\ref{mulithc}) if and only if
\begin{eqnarray*}
\tr \left\{ U_\theta \rho_{0} \frac{\partial U_\theta}{\partial \theta^l}^{\dagger}  \right\} = 0, \quad \forall l.
\label{eq:unitcond}
\end{eqnarray*}
\end{lemma}
\proof
This follows from (\ref{eq:unitcond3a}) and the same analysis as in Lemma \ref{lemd}.

\begin{example}
There exist channels which are neither quasi-classical or unitary for which equality holds in  (\ref{mulithc}). The channel with an arbitrary pure input state and output states 
\begin{equation*}
\rho_\theta = f(\theta)^2 | w_1(\theta) \rangle \langle w_1(\theta)  | + ( 1 - f(\theta)^2 ) | w_2(\theta)  \rangle \langle w_2 (\theta)|,
\end{equation*}
where $f(\theta)$ and $g(\theta)$ are real functions of $\theta$ with $0 \leq f(\theta), g(\theta) \leq 1$  and 
\begin{equation*}
| w_1(\theta)  \rangle = ( g(\theta), \sqrt{ 1 - g(\theta)^2}, 0 )^T, \quad  | w_2(\theta)  \rangle = ( 0, 0, 1 )^T,
\end{equation*}
\end{example}
satisfies (\ref{eq:unitcond3a}) and hence achieves equality in (\ref{mulithc}).

\begin{theorem}
For multi-parameter channels, 
\begin{equation}
F_\theta^M \leq C_{\Upsilon}(\theta),
\label{mulithc1}
\end{equation}
with equality if and only if (\ref{eq:unitcond3a}) holds and there exists a POVM satisfying 
  \begin{equation}
 M_m^{1/2} \lambda^j \rho^{1/2} = \xi_m^j  M_m^{1/2}  \rho^{1/2},  \quad \xi_m^j \in \mathbb{R}, \quad \forall j,m.
 \label{bcii2ebb}
\end{equation}
\end{theorem}
\proof
This follows from  Theorems  \ref{thm:fmh}, \ref{thm:eqfmh} and \ref{lm:4a}. 

\begin{theorem}
For multi-parameter channels,
\begin{equation}
H_\theta \leq C_{E}(\theta),
\label{multihlessce2}
\end{equation}
with equality if and only if the set of output states satisfies (\ref{eq:unitcond3a}) and a fixed unitary matrix $U = [ u_{jk} ]$ exists such that the Kraus operators $E_j$ are related to the canonical Kraus operators $\Upsilon_k$ by
 \begin{equation}
E_j(\theta) = \sum_k u_{j k} \Upsilon_k(\theta).
\label{EUF2}
\end{equation}
\label{multihlessce}
\end{theorem}

\proof
Inequality (\ref{multihlessce2}) follows from  (\ref{hlessce2}) and the same analysis used in the proof of Theorem \ref{lm:4a} with $\Upsilon_k$ replaced by $E_k$.
\\
\\
Equality holds in (\ref{multihlessce2}) if and only if, for the set of channels (\ref{eq:multiproof1}), $\left. H_t \right|_{t=0} = \left. C_E (t)\right|_{t=0}$ for all $v$.
From Theorem \ref{hlessce} this is satisfied if and only if the output states of the channel satisfy (\ref{wjwk0}) at $t=0$ and the Kraus operators $E_j$ are related to the canonical Kraus operators $\Upsilon_k$ by
\begin{equation*}
\left. E_j(\theta+tv) \right|_{t=0} =  \sum_k u_{j k}(\theta+tv) \Upsilon_k(\theta+tv)\left. \right|_{t=0},
\end{equation*}
where
\begin{equation}
 \sum_{jk} p_j \left| \left. \frac{d u_{jk}}{d t}\right|_{t=0} \right|^2= 0.
\label{unitzar}
\end{equation}
From the proof of Theorem \ref{lm:4a} it can be seen that for channels of the form (\ref{eq:multiproof1}),  satisfying  (\ref{wjwk0}) at $t=0$ is equivalent to satisfying (\ref{eq:unitcond3a}).
Condition (\ref{unitzar}) can be rewritten as
\begin{equation*}
 \sum_{jk}  p_j  \left|  \sum_l \frac{\partial u_{jk}}{\partial \theta^l} v^l \right|^2 =  0.
\end{equation*}
This is satisfied for all $v$ if and only if a unitary matrix $U= [u_{jk} ]$ exists satisfying (\ref{EUF2}) that does not depend on $\theta$.

\begin{theorem}
For multi-parameter channels,
\begin{eqnarray}
  F_\theta^M \leq C_{E}(\theta),
  \label{multiflessce}
\end{eqnarray}
with equality if and only if the set of output states satisfies (\ref{eq:unitcond3a}), a fixed unitary matrix $U = [ u_{jk} ]$ exists such that the Kraus operators $E_j$ are related to the canonical Kraus operators $\Upsilon_k$ by (\ref{EUF2}) and there exists a POVM satisfying (\ref{bcii2ebb}).
\end{theorem}
  \proof
 This follows from Theorems \ref{thm:fmh}, \ref{thm:eqfmh} and \ref{multihlessce}.

\chapter{The bound of Sarovar and Milburn as a metric on the space of quantum states} \label{ch:smqi}

\section{Introduction} \label{sec:introSMQI}
Various statistical notions can be expressed in differential-geometric terms
\citep{amari00}. This area is sometimes known as `information geometry'.
Of special importance is Fisher information, which is the unique monotone metric on the space of probability measures \citep{moro90}. However, there is no unique monotone metric on the space of quantum states  \citep{petz96}.
(Definitions of monotonicity and invariance were given below (\ref{mectriccy}).)

The following theorem of \cite{moro90}  is of great interest.
\begin{theorem}
A Riemannian metric is invariant if and only if at every density matrix 
\begin{equation*}
\rho = \sum_j p_j | j \rangle \langle j|,
\end{equation*}
 the squared length of any tangent vector $A$  is of the form
\begin{equation}
C \sum_i \frac{1}{p_i} | A_{ii} |^2 + 2 \sum_{j<k} c(p_j, p_k)  | A_{jk}|^2, \quad A_{jk} = \langle j |A | k \rangle,
\label{eq:chemor1}
\end{equation}
where $C$ is a constant, $c(\alpha x,  \alpha y) = \alpha^{-1} c(x, y)$ and $c(x,y) = c(y,x)$.
\end{theorem}
This result was augmented by the following theorem of \cite{petz96}.  
\begin{theorem}
A Riemannian metric on the space of quantum states is monotone if and only if  at every density matrix 
\begin{equation*}
\rho = \sum_j p_j | j \rangle \langle j|,
\end{equation*}
the squared length of any tangent vector $A$  is of the form (\ref{eq:chemor1}) and the function $f(t) = 1/c(t,1)$ is operator monotone. (A function $f(t)$ is operator monotone if for self-adjoint $n \times n$ matrices $A$ and $B$,  with $A \leq B$, $f(A) \leq f(B)$, \citep[Section 12.1]{beng06}.)
\end{theorem}

For parametric families of states, put $A = d \rho/d\theta$.
In this case (\ref{eq:chemor1}) becomes
\begin{equation}
C \sum_i \frac{1}{p_i} \left| \left( \frac{d \rho}{d \theta} \right)_{ii} \right|^2 + 2 \sum_{j<k} c(p_j, p_k)  \left| \left( \frac{d \rho}{d \theta} \right)_{jk}\right|^2.
\label{eq:chemor}
\end{equation}

 For the SLD, KMB and RLD quantum informations \citep{petz96}, $C =1$ and
\begin{eqnarray*}
c_{SLD}(x,y) &=& \frac{2}{x + y} \\
c_{KMB}(x,y)  &=& \frac{\ln x - \ln y}{x - y} \\
c_{RLD}(x,y)  &=& \frac{1}{2}\bigg(\frac{1}{x} + \frac{1}{y}\bigg).
\end{eqnarray*}
For a more thorough background to the theory of metrics on the space of quantum states see \cite[Chapter 14]{beng06}. 

The  {\it Symmetric Logarithmic Derivative (SLD)}, {\it Kubo-Mori Bogoliubov (KMB)} and {\it Right Logarithmic Derivative (RLD)} metrics (see Section \ref{subsec:SLDQFI}) are the most frequently encountered monotone metrics in recent literature. 
The SLD quantum information is the minimum monotone metric on the space of quantum states \citep{petz96}.
It has been used widely in the estimation of states \citep{helstrom67b,helstrom76,holevo82,hayashi05} and quantum channels \citep{fuj01,fuj01a,fuj04,fujimai03,ballesterb,ballestera}. For one-parameter families of states, the SLD quantum information is equal to the maximum attainable  Fisher information \citep{braunsteincaves94}.
The SLD quantum information is related to the bures distance,
\begin{equation}
b^2(\rho,\sigma) = 1 - \mathrm{tr}\{\sqrt{\rho^{1/2} \sigma \rho^{1/2} }\},
\end{equation}
in the following way \citep[(6.23)]{hayashi06b}
\begin{equation}
H_\theta^S = 8  \lim_{\epsilon \rightarrow 0} \frac{b^2(\rho_\theta,\rho_{\theta+\epsilon})}{\epsilon^2}.
\label{hsbrd}
\end{equation}
The bures distance is a quantum analogue of the Hellinger distance 
\begin{equation}
d_2^2(p \| q) = 1 - \sum_{i=1}^k \sqrt{p_i} \sqrt{q_i},
\end{equation}
where $p=(p_1, \dots, p_k)$ and $q=(q_1, \dots, q_k)$.
The result (\ref{hsbrd}) is interesting since, given a probability distribution $p_\theta = \{ p_i(\theta) \}$, the `classical' Fisher information is related to the Hellinger distance by
\begin{equation}
F_\theta = 8  \lim_{\epsilon \rightarrow 0} \frac{d_2^2(p_\theta \| p_{\theta+\epsilon})}{\epsilon^2}.
\end{equation}

The KMB quantum information is equal to the limit of the quantum relative entropy $D(\rho \| \sigma ) = \tr ( \rho ( \ln \rho - \ln \sigma))$ \citep{hayashi02}. That is,
\begin{equation}
H^K_\theta = \lim_{\epsilon \rightarrow 0} \frac{2 D(\rho_{\theta} \| \rho_{\theta + \epsilon} )}{\epsilon^2}.
\end{equation}
This is analogous to the fact that the `classical' Fisher information is the limit of the `classical' relative entropy $D(p \| q) = \sum_{i=1}^k p_i \ln (p_i/q_i)$, where $p=(p_1, \dots, p_k)$, $q=(q_1, \dots, q_k)$. That is, given a probability distribution $p_\theta = \{ p_i(\theta) \}$,  
\begin{equation}
F_\theta= \lim_{\epsilon \rightarrow 0} \frac{2 D(p_{\theta} \| p_{\theta + \epsilon} )}{\epsilon^2}.
\end{equation}

 The RLD quantum information is the maximal monotone metric on the space of quantum states  \citep{petz96}. It has also been used in estimation theory \citep{fuji94}.

In Chapter \ref{ch:SM1} it was shown that Sarovar and Milburn's bound $C_\Upsilon(\theta)$ for one-parameter channels could be generalized to a Riemannian metric on $\Theta$. In this chapter $C_\Upsilon(\theta)$ will be referred to as the {\it SM quantum information}.
 It seems natural to look at the properties of $C_\Upsilon(\theta)$.
Is it is well-defined?  Is it useful?

In this chapter it is shown that the SM quantum information is not a well-defined metric, since different choices of phase of the eigenvectors lead to different metrics. A new metric $C_L$ is defined from $C_\Upsilon$. Properties of $C_L$ are investigated and it is seen that it is invariant but not monotone.

\section{Analysis of the SM quantum information}
  The SM quantum information for the family of states
 \begin{equation}
 \rho_\theta = \sum_{k=1}^d p_k(\theta) | w_k (\theta) \rangle \langle w_k(\theta) |
\label{eq.statesw}
 \end{equation}
was shown in Proposition \ref{prop29} to be equal to
  \begin{eqnarray}
C_{\Upsilon} &=&  \sum_i \frac{1}{p_i} \bigg(\frac{d p_i}{d \theta} \bigg)^2 + 4 \sum_{j<k} (p_j + p_k) | \langle w_j' | w_k\rangle |^2 \nonumber  \\
  &+& 4 \sum_i p_i | \langle w_i' | w_i \rangle |^2.
   \label{eq.origsm}
 \end{eqnarray}
This can be rewritten as 
    \begin{eqnarray}
C_{\Upsilon} &=&    \sum_i \frac{1}{p_i} \bigg(\frac{d p_i}{d \theta} \bigg)^2 + 4 \sum_{j<k} \frac{p_j + p_k}{(p_j - p_k)^2}  \bigg| \bigg\langle w_j \bigg| \frac{d \rho}{d \theta} \bigg| w_k \bigg\rangle \bigg|^2 \nonumber  \\
  &+& 4 \sum_i p_i(\theta) | \langle w_i' | w_i \rangle |^2. 
\label{eq:smche}
\end{eqnarray}
It can be seen that $C_{\Upsilon}(\theta)$ is not of the form (\ref{eq:chemor}), and hence is neither invariant nor monotone. The SM quantum information $C_{\Upsilon}(\theta)$ for a family of states is defined in terms of its eigenvectors and eigenvalues by (\ref{eq.origsm}). The eigenvectors of a state are unique up to a change of phase. It turns out that different choices of phase for the eigenvectors lead to different metrics.

\begin{example}
Consider the set of 2-dimensional states
 \begin{equation}
\rho_{r,\theta,\phi} =  \frac{1}{2} \left( \begin{array}{cc}
\ 1 + r \cos\theta & r \sin\theta e^{ - i \phi} \\
\ r \sin\theta e^{ i \phi}  & 1 -r \cos\theta   \end{array} \right),
\label{eq.qubit}
\end{equation}
with $0 \leq r \leq1$, $0 \leq \theta \leq \pi$ and $0 \leq \phi \leq 2\pi$. 
Any qubit, mixed or pure, can be written in the form (\ref{eq.qubit}) with specific values of $r$, $\theta$ and $\phi$.
Each state  (\ref{eq.qubit}) has spectral decomposition
\begin{eqnarray*}
\rho_{r,\theta,\phi}  &=&\frac{1+r}{2} \bigg| w_1 (\theta,\phi) \bigg \rangle \bigg \langle w_1 (\theta,\phi) \bigg | + \frac{1-r}{2} \bigg | w_2 (\theta,\phi) \bigg \rangle \bigg \langle w_2 (\theta,\phi) \bigg | ,\\
| w_1 (\theta,\phi)\rangle &=& ( \cos(\theta/2) e^{-i \phi/2}, \sin(\theta/2 ) e^{i \phi/2})^T,\\
| w_2 (\theta,\phi)\rangle &=& ( \sin(\theta/2) e^{-i \phi/2}, - \cos(\theta/2) e^{i \phi/2})^T.
\end{eqnarray*}
 The SM quantum information for the family of states $\rho_\theta$ calculated from the above eigenvalues and eigenvectors is 
  \begin{equation*}
 C_{\Upsilon}(r,\theta,\phi) = \left( \begin{array}{ccc}
\ \displaystyle \frac{1}{1-r^2} & 0 & 0\\
\ 0 & 1 & 0 \\
\ 0 & 0 &1  \end{array} \right).
\end{equation*}
 Changing the eigenvectors by the phase shift $e^{-i \phi/2}$, i.e.\ $| w_k(\theta,\phi) \rangle \mapsto e^{-i \phi/2}| w_k(\theta,\phi) \rangle$, leaves the density matrix unchanged but the SM quantum information calculated from the eigenvalues and shifted eigenvectors  becomes
   \begin{equation*}
 C_{\Upsilon}(r,\theta,\phi) = \left( \begin{array}{ccc}
\ \displaystyle \frac{1}{1-r^2} & 0 & 0\\
\ 0 & 1 & 0 \\
\ 0 & 0 &2 + 2r \cos\theta  \end{array} \right).
\end{equation*}
\label{ex:111}
\end{example}

Hence the SM quantum information is not a well-defined metric.

\section{A new metric}
The  {\it $C_L$ quantum information} for the family of states (\ref{eq.statesw}) will be defined as
\begin{equation}
C_L = C_{\Upsilon} - 4 \sum_i p_i | \langle w_i' | w_i \rangle |^2. \label{eq.dercl}
\end{equation}
Thus
\begin{eqnarray}
C_L&=& \sum_i \frac{1}{p_i} \bigg(\frac{d p_i}{d \theta} \bigg)^2 + 4 \sum_{j<k} (p_j + p_k)  | \langle w_j' | w_k \rangle|^2  \\
&=& \sum_i \frac{1}{p_i} \bigg(\frac{d p_i}{d \theta} \bigg)^2 + 4 \sum_{j<k} \frac{p_j + p_k}{(p_j - p_k)^2}  \bigg| \bigg\langle w_j \bigg| \frac{d \rho}{d \theta} \bigg| w_k \bigg\rangle \bigg|^2. 
  \label{calebmetric}
\end{eqnarray}
\begin{remark}
Unlike the RLD and KMB quantum informations, the $C_L$ quantum information can be defined for families of pure states. For pure states, $C_L(\rho_\theta) = H(\rho_\theta)$ (see  (\ref{eq.clminush})).
\end{remark}
The $C_L$ quantum information is of the form (\ref{eq:chemor}) with $C=1$ and
\begin{equation}
c_{L}(p_j,p_k) = 2 \frac{p_j + p_k}{(p_j - p_k)^2}.
\end{equation}
This function is symmetric and  $c_L(\alpha x,  \alpha y) = \alpha^{-1} c_L(x, y)$. Hence, $C_L$ is invariant. Thus it does not suffer the same defect as $C_\Upsilon$. The $C_L$ quantum information provides each parameterized family $\{ \rho_\theta : \theta \in \Theta \}$ with a unique Riemannian metric on $\Theta$. 

For a metric to be monotone, it must be of the form (\ref{eq:chemor1}) and the function $f(t)$ associated with the metric must be monotone and satisfy $f(t) = tf(t^{-1})$.
  The functions associated with the SLD, KMB and RLD quantum informations are
  \begin{eqnarray*}
  f_{SLD}(t) &=& \frac{1+t}{2}\\
  f_{KMB}(t) &=& \frac{t-1}{\log t}\\
  f_{RLD}(t) &=& \frac{2t}{1+t}.
  \end{eqnarray*}
Calculation shows that the function associated with $C_L$ is
 \begin{equation}
 f_{C_L}(t) = \frac{(t -1)^2}{2(1+t)}.
 \end{equation}
If $f$ is a monotone function then $f(0) \leq f(t_1) \leq f(t_2)$ whenever $0 \leq t_1 \leq t_2$. The function $f_{C_L}(t)$ satisfies $f_{C_L}(t) = tf_{c_L}(t^{-1})$ but is not monotone, as $ f_{C_L}(0) > f_{C_L}(1)$. Hence, $C_L$ is an invariant but not monotone Riemannian metric.
 
\begin{example}
The depolarizing channel,(\ref{eq.depngd}), acts on $3$-dimensional states in the following way
\begin{equation}
\rho_0 \mapsto (1-\epsilon) \rho_0 + \frac{\epsilon}{3} \mathbb{I}_3, \qquad 0 \leq \epsilon \leq 1.
\end{equation}
Consider the one-parameter set of $3$-dimensional mixed states 
  \begin{eqnarray*}
\rho_\theta &=& (1-2 \delta) | w_1 \rangle \langle w_1 | + \delta | w_2 \rangle \langle w_2 | + \delta | w_3 \rangle \langle w_3 | ,\\
| w_1 \rangle &=& ( 1, 0, 0 )^T,\\
| w_2 \rangle &=& ( 0, \cos\theta, \sin\theta)^T,\\
| w_3 \rangle &=& ( 0, -\sin\theta, \cos\theta)^T,
\end{eqnarray*}
where $\delta$ is fixed. The $C_L(\theta)$ quantum information of this family of states is $8 \delta$.
Under the action of the depolarizing channel the set of output states is  
\begin{eqnarray*}
\mathcal{E}(\rho_\theta) &=& \bigg( (1-\epsilon)(1-2\delta) + \frac{\epsilon}{3} \bigg) | w_1 \rangle \langle w_1 | \\
&+& \bigg(  (1-\epsilon)\delta + \frac{\epsilon}{3}  \bigg) | w_2 \rangle \langle w_2 | + \bigg( (1-\epsilon)\delta + \frac{\epsilon}{3}  \bigg) | w_3 \rangle \langle w_3 | 
\end{eqnarray*}
with $| w_i \rangle$ unchanged. The $C_L$ quantum information for the family of states $\mathcal{E}(\rho_\theta)$ is $8\delta+ 8 \epsilon(1/3 - \delta)$. Now
\begin{equation}
C_{L}(\mathcal{E}(\rho_\theta)) - C_{L}(\rho_\theta) = 8 \epsilon(1/3 - \delta).
\end{equation}
For $\epsilon > 0 $ and $\delta < 1/3$,  $C_L$ has increased under the action of a TP-CP map, thus demonstrating the non-monotonicity of $C_L$.
\end{example} 

\section{Ordering of $C_L$, $C_\Upsilon$ and $H$}
\begin{theorem}
Given a parameterised quantum model $\{\rho_\theta = \sum_{k=1}^d p_k(\theta) | w_k (\theta) \rangle \langle w_k(\theta) |: \theta  \in \mathbb{R}^p \}$,
\begin{equation}
H_\theta \leq C_L(\theta) \leq C_\Upsilon(\theta),
\end{equation} 
where the multi-parameter versions of $H_\theta$, $C_L(\theta)$ and $C_\Upsilon(\theta)$ are defined in (\ref{eq.multihups}), (\ref{cla1}) and (\ref{eq.multicups}) respectively.
Equality holds in $H_\theta \leq C_L(\theta)$  for families of states $\rho_\theta$ if and only if 
\begin{equation}
 \bigg\langle w_j^{(m)}  \bigg| w_k  \bigg\rangle = 0, \quad \forall m, \quad j \neq k, p_j,p_k > 0.
\label{eq:unitcond3b}
\end{equation}
Equality holds in $C_L(\theta) \leq C_\Upsilon(\theta)$ for families of states $\rho_\theta$ if and only if 
\begin{equation}
  \bigg\langle w_i^{(m)}  \bigg| w_i  \bigg\rangle  = 0, \quad \forall m,i, \quad   p_i > 0.
\label{eq:unitcond3}
\end{equation}
\label{thm:hclcu}
\end{theorem}
A proof of Theorem \ref{thm:hclcu} will be given first for the one-parameter case and then for the general case. 

\subsection{One-parameter case }
\begin{lemma}
\begin{eqnarray}
C_L(\theta) \leq C_{\Upsilon}(\theta),
\label{eq.clecups}
\end{eqnarray}
with equality if and only if
\begin{eqnarray}
 \langle w_i' | w_i \rangle  = 0, \quad \forall i, \quad   p_i > 0.
 \label{eq.eq1pclcu}
 \end{eqnarray}
\label{lem.clleqcups1}
\end{lemma}
\proof
This follows from the definition of $C_L$, (\ref{eq.dercl}), and the fact that $| \langle w_i' | w_i \rangle |$  is non-negative.

\begin{lemma}
\begin{equation}
H_\theta \leq C_L(\theta),
\label{eq.1phleqcl}
\end{equation}
with equality if and only if
 \begin{equation}
\langle w_{j}' | w_{k}  \rangle = 0, \quad \forall j \neq k, p_j, p_k  >0.
\label{eq.eq1phcl2}
\end{equation}
\label{eq.eq1phcl}
 \end{lemma} 
\proof
Proposition \ref{prop:sld} showed that
\begin{equation*}
H = \sum_{k} \frac{1}{p_k} \bigg(\frac{d p_k}{d \theta} \bigg)^2 +   \sum_{  j < k} 4\frac{ ( p_{j} - p_{k})^2}{p_{j} + p_{k}} |\langle w_{j}' | w_{k}  \rangle|^2,
\label{eq:Hsldra}
\end{equation*}
and hence
\begin{eqnarray}
C_L - H &=& 16 \sum_{  j < k } \frac{p_{j} p_{k}}{p_{j} + p_{k}} |\langle w_{j}' | w_{k}  \rangle|^2 \nonumber \\
 &=&  8 \sum_{  j \neq k } \frac{p_{j} p_{k}}{p_{j} + p_{k}} |\langle w_{j}' | w_{k}  \rangle|^2, \quad 
\label{eq.clminush}
\end{eqnarray}
since $|\langle w_{j}' | w_{k}  \rangle|^2$ is symmetric with respect to $j$ and $k$ (\ref{eq:diffwjwk}).
The right hand side of (\ref{eq.clminush}) is non-negative, and equal to zero if and only if (\ref{eq.eq1phcl2}) holds.

\subsection{The multi-parameter case}
\begin{proposition}
In the multi-parameter case the SM quantum information is the matrix with entries
\begin{eqnarray}
(C_{\Upsilon})_{kl} &=& \sum_i \frac{1}{p_i} \bigg(\frac{\partial p_i}{\partial \theta^k} \bigg)\bigg(\frac{\partial p_i}{\partial \theta^l} \bigg)  + 4 \Re \sum_{i < j} (p_i + p_j ) \bigg\langle w_i^{(k)} \bigg| w_j \bigg\rangle \bigg\langle w_j \bigg| w_i^{(l)} \bigg\rangle \nonumber \\
&+& 4  \sum_i p_i \bigg\langle w_i^{(k)} \bigg| w_i \bigg\rangle \bigg\langle w_i \bigg| w_i^{(l)} \bigg\rangle.
\label{eq.multicups}
\end{eqnarray}
\end{proposition}

\proof
The multi-parameter version of $C_\Upsilon$ was defined, (\ref{multipcupsdef}), as the matrix with entries 
  \begin{equation*}
(C_{\Upsilon})_{kl} = 4 \sum_{i} \Re \tr \left\{ \Upsilon_{i}^{(k)} \rho_{0}  \Upsilon_{i}^{(l) \dagger}   \right\}, \qquad \Upsilon_{i}^{(k)} = \frac{\partial }{\partial \theta^k}\Upsilon_{i}.
\end{equation*}
Using (\ref{eq.defspecd}) and (\ref{eq:outstate}) 
\begin{eqnarray}
\Re \tr \left\{ \Upsilon_i^{(k)} \rho_0  \Upsilon_i^{(l) \dagger}   \right\} &=& \Re \bigg[ \frac{1}{4 p_i} \frac{\partial p_i}{\partial \theta^k } \frac{\partial p_i}{\partial \theta^l }  + p_i \bigg\langle w_i^{(l)} \bigg| w_i^{(k)} \bigg\rangle \nonumber \\
&+& \frac{1}{2}\left(  \frac{\partial p_i}{\partial \theta^l } \bigg\langle w_i \bigg| w_i^{(k)} \bigg\rangle + \frac{\partial p_i}{\partial \theta^k } \bigg\langle w_i^{(l)}  \bigg| w_i \bigg\rangle \right) \bigg].
\label{cmultrs1}
\end{eqnarray}
The contributions of the final two terms on the right hand side of (\ref{cmultrs1}) are zero since they are purely imaginary (see below (\ref{eq:diffjj})).
Thus,
 \begin{equation}
(C_{\Upsilon})_{kl} =  \sum_i \frac{1}{p_i} \bigg(\frac{\partial p_i}{\partial \theta^k} \bigg)\bigg(\frac{\partial p_i}{\partial \theta^l} \bigg)  + 4 \Re \sum_i  p_i \bigg\langle w_i^{(l)} \bigg| w_i^{(k)} \bigg\rangle.
\label{almost}
\end{equation}
Inserting the identity $\mathbb{I}_d = \sum_j | w_j \rangle \langle w_j|$  into the second term on the right hand side of (\ref{cmultrs1}) gives
\begin{eqnarray}
\sum_i p_i \bigg\langle w_i^{(l)} \bigg| w_i^{(k)} \bigg\rangle &=& \sum_{i \neq j} p_i \bigg\langle w_i^{(l)} \bigg| w_j \bigg\rangle \bigg\langle w_j \bigg|  w_i^{(k)} \bigg\rangle \nonumber \\
&+& \sum_{i} p_i \bigg\langle w_i^{(l)} \bigg| w_i \bigg\rangle \bigg\langle w_i \bigg|  w_i^{(k)} \bigg\rangle.
\label{exe1}
\end{eqnarray}
The first term on the right hand side of (\ref{exe1}) can be written as
\begin{eqnarray}
\sum_{i \neq j} p_i \bigg\langle w_i^{(l)} \bigg| w_j \bigg\rangle \bigg\langle w_j \bigg|  w_i^{(k)} \bigg\rangle &=& \sum_{i < j} p_i \bigg\langle w_i^{(l)} \bigg| w_j \bigg\rangle \bigg\langle w_j \bigg|  w_i^{(k)} \bigg\rangle 
\nonumber \\
&+& \sum_{i > j} p_i \bigg\langle w_i^{(l)} \bigg| w_j \bigg\rangle \bigg\langle w_j \bigg|  w_i^{(k)} \bigg\rangle.
\label{exe2}
\end{eqnarray}
Swapping the indices, $i$ and $j$, on the second term on the right hand side of (\ref{exe2}) gives 
\begin{eqnarray}
\Re \sum_{i > j} p_i \bigg\langle w_i^{(l)} \bigg| w_j \bigg\rangle \bigg\langle w_j \bigg|  w_i^{(k)} \bigg\rangle &=& \Re \sum_{i < j} p_j \bigg\langle w_j^{(l)} \bigg| w_i \bigg\rangle \bigg\langle w_i \bigg|  w_j^{(k)} \bigg\rangle 
\nonumber \\
 &=& \Re \sum_{i < j} p_j \bigg\langle w_j \bigg| w_i^{(l)} \bigg\rangle \bigg\langle w_i^{(k)} \bigg|  w_j \bigg\rangle \label{exe2b},
  \end{eqnarray}
using (\ref{eq:wjwk}).
From (\ref{exe1}), (\ref{exe2}) and (\ref{exe2b}) it follows that 
\begin{eqnarray}
\Re \sum_i p_i \bigg\langle w_i^{(k)} \bigg| w_i^{(l)} \bigg\rangle &=& \Re \sum_{i < j} (p_i + p_j) \bigg\langle w_i^{(k)} \bigg| w_j \bigg\rangle \bigg\langle w_j \bigg| w_i^{(l)} \bigg\rangle \nonumber \\
&+& \Re \sum_{i} p_i \bigg\langle w_i^{(k)} \bigg| w_i \bigg\rangle \bigg\langle w_i \bigg|  w_i^{(l)} \bigg\rangle.
\label{nearlythere}
\end{eqnarray}
The required result follows from (\ref{almost}) and (\ref{nearlythere}). 
\\
\\
The multivariate version of $C_L$ will be defined as the matrix with entries
\begin{eqnarray}
(C_L)_{kl} &=& (C_{\Upsilon})_{kl}  - 4 \Re \sum_i p_i \bigg\langle w_i^{(k)} \bigg| w_i \bigg\rangle \bigg\langle w_i \bigg| w_i^{(l)} \bigg\rangle \label{cla1} \\
&=& \sum_i \frac{1}{p_i} \bigg(\frac{\partial p_i}{\partial \theta^j} \bigg)\bigg(\frac{\partial p_i}{\partial \theta^k} \bigg) + 4 \Re \sum_{i < j} (p_i + p_j ) \bigg\langle w_i^{(k)} \bigg| w_j \bigg\rangle \bigg\langle w_j \bigg| w_i^{(l)} \bigg\rangle. \nonumber 
\end{eqnarray} 

 \begin{lemma}
\begin{equation}
C_L(\theta) \leq C_{\Upsilon}(\theta).
\label{mulithc2}
\end{equation}
with equality if and only if (\ref{eq:unitcond3}) holds. 
\label{lm:4}
\end{lemma}

\proof
Equation (\ref{mulithc2}) is equivalent to 
\begin{equation}
 v^T   C_L(\theta)  v \leq  v^T  C_{\Upsilon}(\theta) v ,
\label{eq:cCc}
\end{equation}
for all $v \in \mathbb{R}^p$.  For given $\theta$ and $ v$ in $\mathbb{R}^p$, consider the set of one-parameter states 
\begin{equation*}
 \rho_{\theta + t v} = \sum_{k=1}^d p_k(\theta + t v)| w_k(\theta + t v) \rangle \langle w_k(\theta + t v) |, \qquad t \in \mathbb{R}.
\end{equation*}
It was shown in the proof of Proposition \ref{prop:444} that
\begin{eqnarray}
\frac{d}{d t}  p_{k} (\theta + t  v  ) &=& \sum_l  \frac{\partial p_{k} (\theta)}{\partial \theta^l} v^l + O(t), \quad t \rightarrow 0,
\label{eq:multip1b}\\
\frac{d}{d t}  | w_{k} (\theta + t  v  ) \rangle &=& \sum_l   \bigg|w_k(\theta)^{(l)} \bigg\rangle v^l + O(t),  \quad t \rightarrow 0,
\label{eq:multip2} \\
 \bigg|w_k(\theta)^{(l)} \bigg\rangle &=& \frac{\partial}{\partial \theta^l}   | w_k(\theta) \rangle, 
\nonumber
\end{eqnarray}
where $ v^l$ is the $l$th component of the vector $v$.
From Lemma \ref{lem.clleqcups1} it is known that $  C_L(t) \leq C_{\Upsilon}(t)$, i.e.
\begin{eqnarray}
\sum_i \frac{1}{p_i(\theta +tv)} \bigg(\frac{d p_i}{d t} \bigg)^2 &+& 
4 \sum_{j<k} (p_j(\theta + tv) + p_k(\theta + tv))  \bigg| \bigg\langle \frac{d w_j}{dt} \bigg| w_k \bigg\rangle \bigg|^2 \nonumber \\
 \leq \sum_i \frac{1}{p_i(\theta +tv)} \bigg(\frac{d p_i}{d t} \bigg)^2 &+& 
4 \sum_{j<k} (p_j(\theta + tv) + p_k(\theta + tv))  \bigg| \bigg\langle \frac{d w_j}{dt}\bigg| w_k \bigg\rangle \bigg|^2\nonumber \\
&+& 4 \sum_i p_i(\theta) \bigg| \bigg\langle \frac{d w_i}{dt} \bigg| w_i \bigg\rangle \bigg|^2 .
\label{eq:34}
\end{eqnarray}
Using (\ref{eq:multip1b}) and (\ref{eq:multip2}) and evaluating at $t=0$  gives
\begin{eqnarray*}
\sum_{m,n} v^m v^n \left( \sum_i \frac{1}{p_i} \bigg(\frac{\partial p_i}{\partial \theta^m} \bigg)\bigg(\frac{\partial p_i}{\partial \theta^n} \bigg)  + 4 \sum_{i < j} (p_i + p_j ) \bigg\langle w_i^{(m)}  \bigg| w_j  \bigg\rangle  \bigg\langle w_j  \bigg| w_i^{(n)}  \bigg\rangle \nonumber \right) \\
\leq \sum_{r,s} v^r v^s \left( \sum_i \frac{1}{p_i} \bigg(\frac{\partial p_i}{\partial \theta^r} \bigg)\bigg(\frac{\partial p_i}{\partial \theta^s} \bigg)  + 4 \sum_{i < j} (p_i + p_j )  \bigg\langle w_i^{(r)}  \bigg| w_j  \bigg\rangle  \bigg\langle w_j  \bigg| w_i^{(s)}  \bigg\rangle \right) \nonumber \\
+ \left. 4  \sum_i p_i  \bigg\langle w_i^{(m)}  \bigg| w_i  \bigg\rangle  \bigg\langle w_i  \bigg| w_i^{(n)}  \bigg\rangle \right) .
\label{eq:34b}
\end{eqnarray*}
This can be rewritten as
\begin{equation*}
\sum_{m,n} v^m v^n C_L(\theta)_{mn} \leq \sum_{r,s} v^r v^s C_\Upsilon (\theta)_{rs}. 
\end{equation*}
This is equivalent to (\ref{eq:cCc}). Since this holds for all $v$ in $\mathbb{R}^p$, (\ref{mulithc2}) holds. 
\\
\\
Equality in (\ref{mulithc2}) is equivalent to 
\begin{equation}
 v^T   C_L(\theta)  v =  v^T  C_{\Upsilon}(\theta) v ,
\label{eq:cCc2}
\end{equation}
for all $v \in \mathbb{R}^p$.  From the proof of Lemma \ref{lm:4} it is seen that (\ref{eq:cCc2}) is satisfied for all $v \in \mathbb{R}^p$ if and only if, for one-parameter families of states  $\rho_{\theta + tv}$,  $C_L(t) |_{t=0} =  C_{\Upsilon}(t) |_{t=0}$. From Lemma \ref{lem.clleqcups1} this is possible if and only if the channel satisfies (\ref{eq.eq1pclcu}) at the point $t=0$. This condition is equal to
\begin{equation}
\left. \left. \sum_i p_i(t) \left| \left\langle \frac{d w_i}{dt} \right| w_i \right\rangle \right|^2 \right|_{t=0} = 0. 
\end{equation}
Using (\ref{eq:multip2}), this condition can be rewritten  as 
  \begin{eqnarray}
\sum_{l=1}^{m} v^m v^n   \sum_i p_i  \bigg\langle w_i^{(m)}  \bigg| w_i  \bigg\rangle  \bigg\langle w_i  \bigg| w_i^{(n)}  \bigg\rangle = 0, \quad \forall m,n.
\label{eq:unitcond2}
\end{eqnarray}
Condition (\ref{eq:unitcond2}) holds for all $v$ if and only if (\ref{eq:unitcond3}) is satisfied.

\begin{proposition}
In the multi-parameter case the SLD quantum information is the matrix with entries
\begin{equation}
(H)_{kl}  = \sum_i \frac{1}{p_i} \bigg(\frac{\partial p_i}{\partial \theta^k} \bigg)\bigg(\frac{\partial p_i}{\partial \theta^l} \bigg)  + 4 \Re \sum_{i < j} \frac{(p_i - p_j)^2}{p_i + p_j} \bigg\langle w_i^{(k)} \bigg| w_j \bigg\rangle \bigg\langle w_j \bigg| w_i^{(l)} \bigg\rangle.
\label{eq.multihups}
\end{equation}
\label{pro39}
\end{proposition}

\proof
In the multi-parameter case a particular choice of SLD with respect to the parameter $\theta^k$ is
 \begin{equation}
\tilde \lambda^k =  \sum_{i, p_i \neq 0} \frac{ 1}{ p_i}\frac{ \partial p_i}{ \partial \theta^k}  \bigg| w_i \bigg\rangle \bigg\langle w_i \bigg|  + 
 \sum_{i \neq j, p_i+p_j > 0}  2 \frac{  p_i - p_j}{ p_i + p_j} \bigg\langle w_i^{(k)} \bigg| w_j \bigg\rangle \bigg| w_i \bigg\rangle \bigg\langle  w_j \bigg|.
   \label{eq:Lsldrm}
\end{equation}
Proposition \ref{pro39} follows almost identically to the one-parameter case (see proof of Proposition \ref{prop:sld}).

 \begin{lemma}
\begin{equation}
H_\theta \leq C_L(\theta),
\label{mulithc2hcl}
\end{equation}
\label{lm:5hc}
\end{lemma}
with equality if and only if (\ref{eq:unitcond3b}) holds.

\proof
This follows from Lemma \ref{eq.eq1phcl} in  the same way as Lemma \ref{lm:4} follows from Lemma \ref{lem.clleqcups1}.

\section{$C_L$ as the minimum of $C_\Upsilon$}
Example \ref{ex:111} showed that for $C_\Upsilon$, different choices of eigenvectors of $\rho_\theta$ can result in completely different metrics. Since $C_\Upsilon$ is an upper bound on Fisher information, 
it seems sensible to choose the minimum among possible values of $C_\Upsilon$.
It will now be investigated whether there exists a choice of eigenvectors such that $C_\Upsilon = C_L$.

\subsection{One-parameter case}
Consider a family of states $\rho_\theta = \sum_i p_i (\theta) | w_i(\theta) \rangle \langle w_i(\theta) |$.
A phase change of the eigenvectors $| w_1(\theta) \rangle, \dots, | w_d(\theta) \rangle$ sends these vectors to $| v_1(\theta) \rangle, \dots, | v_d(\theta) \rangle$, where $| v_j (\theta) \rangle = e^{i \alpha_j(\theta)} | w_j (\theta) \rangle$ for some real-valued functions $\alpha_1, \dots, \alpha_d$. The density matrix $\rho_\theta$ is unchanged. Now
\begin{equation*}
\frac{d}{d \theta} | v_k (\theta) \rangle = i \frac{d \alpha_k}{d \theta} e^{i \alpha_k(\theta)} | w_k (\theta) \rangle +  e^{i \alpha_k(\theta)} \frac{d}{d \theta} | w_k(\theta) \rangle
\end{equation*}
and hence 
\begin{equation*}
 \langle v_k' | v_k \rangle  =  -i \frac{d \alpha_k}{d \theta} +  \langle w_k' | w_k \rangle.
\end{equation*}
Choosing
\begin{equation*}
\alpha_k(\theta) = - i  \int_{\theta_0}^\theta \langle w_k'(\phi) | w_k(\phi)  \rangle d\phi,
\end{equation*}
 (\ref{eq.eq1pclcu}) is satisfied.  (Since $ \langle w_k' | w_k \rangle$ is purely imaginary,  $\alpha_k$ is real.)  
Thus in the one-parameter case $C_L$ is the minimum among $C_\Upsilon$. 

\subsection{Multi-parameter case}
 A phase change of the eigenvectors $| w_1(\theta) \rangle, \dots, | w_d(\theta) \rangle$ sends these vectors to $| v_1(\theta) \rangle, \dots, | v_d(\theta) \rangle$, where $| v_j (\theta) \rangle = e^{i \alpha_j(\theta)} | w_j (\theta) \rangle$ for some real-valued functions $\alpha_1, \dots, \alpha_d$. In this case $\theta = (\theta^1, \dots, \theta^p)$.
Equality holds in (\ref{mulithc2}) if and only if (\ref{eq:unitcond3}) holds. Now,
\begin{equation*}
\frac{\partial}{\partial \theta^m} | v_j (\theta) \rangle = i \frac{\partial \alpha_j}{\partial \theta^m} e^{i \alpha_j(\theta)} | w_j (\theta) \rangle +  e^{i \alpha_j(\theta)} \frac{\partial}{\partial \theta^m} | w_k(\theta) \rangle
\end{equation*}
and hence 
\begin{equation*}
  \bigg\langle v_j^{(m)}  \bigg| v_j  \bigg\rangle   = -  i \frac{\partial \alpha_j}{\partial \theta^m} +   \bigg\langle w_j^{(m)}  \bigg| w_j  \bigg\rangle.
\end{equation*}
This is zero if and only if
\begin{equation*}
i \frac{\partial \alpha_j}{\partial \theta^m} = \bigg\langle \frac{\partial w_j}{\partial \theta^m}\bigg| w_j \bigg\rangle \in i \mathbb{R} \qquad \forall j,m.
\end{equation*}
This is solvable for $\alpha_1, \dots, \alpha_d$ if and only if
\begin{equation*}
\frac{\partial^2 \alpha_j}{\partial \theta^k \partial \theta^l} = \frac{\partial^2 \alpha_j}{\partial \theta^l \partial \theta^k} \quad \forall j,k,l. 
\end{equation*}
This is equivalent to
\begin{equation*}
\frac{\partial}{\partial \theta^k} \bigg\langle \frac{\partial w_j}{\partial \theta^l}\bigg| w_j \bigg\rangle = \frac{\partial}{\partial \theta^l} \bigg\langle \frac{\partial w_j}{\partial \theta^k}\bigg| w_j \bigg\rangle \quad \forall j,k,l ,
\end{equation*}
which is equivalent to 
\begin{equation*}
 \bigg\langle \frac{\partial^2 w_j}{\partial \theta^k \partial \theta^l}\bigg| w_j \bigg\rangle +  \bigg\langle \frac{\partial w_j}{\partial  \theta^l}\bigg| \frac{ \partial w_j}{\partial \theta^k} \bigg\rangle= \bigg\langle \frac{\partial^2 w_j}{\partial \theta^l \partial \theta^k} \bigg| w_j \bigg\rangle + \bigg\langle \frac{\partial w_j}{\partial \theta^k}\bigg| \frac{\partial w_j}{\partial \theta^l} \bigg\rangle \quad \forall j,k,l. 
\end{equation*}
Since $| w_j \rangle$ is assumed to be continuously differentiable, 
\begin{equation*}
 \bigg\langle \frac{\partial^2 w_j}{\partial \theta^k \partial \theta^l}\bigg| w_j \bigg\rangle = \bigg\langle \frac{\partial^2 w_j}{\partial \theta^l \partial \theta^k} \bigg| w_j \bigg\rangle \quad \forall j,k,l ,
 \end{equation*}
 and hence it is required that
  \begin{equation*}
 \bigg\langle \frac{\partial w_j}{\partial  \theta^l}\bigg| \frac{ \partial w_j}{\partial \theta^k} \bigg\rangle= \bigg\langle \frac{\partial w_j}{\partial \theta^k}\bigg| \frac{\partial w_j}{\partial \theta^l} \bigg\rangle \qquad \forall j,k,l.
\end{equation*}
This is satisfied if and only if
\begin{equation}
 \bigg\langle \frac{\partial w_j}{\partial  \theta^l}\bigg| \frac{ \partial w_j}{\partial \theta^k} \bigg\rangle \in \mathbb{R} \qquad \forall j,k,l,
 \label{eq:condclcum}
\end{equation}
which, in general, does not hold.
Hence, for multi-parameter families of states, $C_L$ is not generally the minimum among $C_\Upsilon$.

\begin{example}
For the family of states given in Example \ref{ex:111}, 
\begin{eqnarray*}
 \bigg\langle \frac{\partial w_1}{\partial  \theta}\bigg| \frac{ \partial w_1}{\partial \phi} \bigg\rangle &=& \frac{i}{2} \sin(\theta/2) \cos(\theta/2),\\
  \bigg\langle \frac{\partial w_2}{\partial  \theta}\bigg| \frac{ \partial w_2}{\partial \phi} \bigg\rangle &=& \frac{-i}{2} \sin(\theta/2) \cos(\theta/2).
\end{eqnarray*}
Since (\ref{eq:condclcum}) is not satisfied,  $C_L$ is not the minimum among $C_\Upsilon$, for this family of states. 
\end{example}
 
 \section{Relationship between $C_L$ and SLD information of mixtures}
For a general family of states $\rho_\theta = \sum_i p_i(\theta) | w_i(\theta) \rangle \langle w_i(\theta) |$, $C_L$ was defined in (\ref{cla1}) as the matrix with entries
\begin{eqnarray}
(C_L)_{kl} &=& (C_\Upsilon)_{kl} -  4 \sum_i p_i \bigg\langle w_i^{(k)} \bigg| w_i \bigg\rangle \bigg\langle w_i \bigg| w_i^{(l)} \bigg\rangle \nonumber \\
&=&  \sum_i \frac{1}{p_i} \bigg(\frac{\partial p_i}{\partial \theta^k} \bigg)\bigg(\frac{\partial p_i}{\partial \theta^l} \bigg) \nonumber\\
&+& 4 \sum_i p_i \bigg( \Re \bigg\langle w_i^{(k)} \bigg| w_i^{(l)} \bigg\rangle -
\bigg\langle w_i^{(k)} \bigg| w_i \bigg\rangle \bigg\langle w_i \bigg| w_i^{(l)} \bigg\rangle \bigg)
\label{eq.nnum} 
\end{eqnarray}
by (\ref{almost}). It is not difficult to show that the SLD quantum information for  $\rho_i(\theta) = |w_i(\theta) \rangle \langle w_i(\theta)|$ is the matrix with entries 
\begin{equation}
(H(\rho_i))_{kl} = 4 \Re \bigg\langle w_i^{(k)} \bigg| w_i^{(l)} \bigg\rangle -
\bigg\langle w_i^{(k)} \bigg| w_i \bigg\rangle \bigg\langle w_i \bigg| w_i^{(l)} \bigg\rangle.
\end{equation}
Thus, 
\begin{equation}
C_L\left(\sum_i p_i \rho_i\right) =  F_\theta (p) +  \sum_i p_i H(\rho_i), \label{eq.339}
\end{equation}
where $F_\theta (p)$ is the Fisher information matrix for $p = (p_1, \dots, p_d)$, which has entries 
\begin{equation}
(F_\theta (p))_{kl} = \sum_i \frac{1}{p_i} \bigg(\frac{\partial p_i}{\partial \theta^k} \bigg)\bigg(\frac{\partial p_i}{\partial \theta^l} \bigg).
\label{clamaar}
\end{equation}
The result (\ref{eq.339})  states that the $C_L$ quantum information is equal to the classical Fisher information of the probability distribution $\{ p_1,\dots,p_d \}$ plus a weighted sum of the SLD quantum informations of the pure states $\rho_i(\theta)$ of which the state $\rho_\theta$ is a convex mixture. 
From (\ref{eq.nnum}) it can be seen that for pure states, for which there is only one non-zero $p_i$,  $C_L = H$, and so
\begin{equation}
C_L\left(\sum_i p_i \rho_i\right) =  F_\theta (p) +  \sum_i p_i C_L(\rho_i). \label{eq.339b}
\end{equation}
Note that (\ref{eq.339}) and (\ref{eq.339b}) are analogous to (7.4) of \cite{amari82}:
Given random variables $X$ and $Y$ depending on $\theta$ with
\begin{equation}
f(x,y;\theta) = g(x;\theta) h (y| x;\theta),
\end{equation}
\begin{equation}
F_\theta^{X,Y} = F_\theta^X + E_X [ F_\theta^Y| x ],
\end{equation}
where
\begin{eqnarray*}
F_\theta^{X,Y} &=& \int \int f(x,y;\theta) \left(\frac{d \log f(x,y;\theta)}{d \theta} \right)^2 dx dy\\
F_\theta^{X} &=& \int  g(x;\theta) \left(\frac{d \log g(x;\theta)}{d \theta} \right)^2 dx \\
E_X [ F_\theta^Y| x ] &=& \int g(x;\theta)\left( \int h (y| x;\theta) \left(\frac{d \log h(y| x;\theta)}{d \theta} \right)^2 dy \right) dx .
\end{eqnarray*}

\chapter{Simultaneous estimation of several commuting quantum unitary channels} \label{ch:niprl}

\section{Introduction}
The situation in which there are  $n$ non-identical commuting channels which are `dependent' (having the same parameter but different forms) will be considered. This chapter introduces the idea of estimation of different commuting unitary channels simultaneously, as opposed to estimating them separately.
Using the SLD quantum information as a measure of performance, it will be shown that this can give considerable improvement over estimating the channels individually.

\subsection{Estimation of unitary channels}
Estimation of an unknown or partially unknown unitary channel has received a lot of attention recently, see \cite{rudolph03,zheng06,acin01,baganetal04a,baganetal04b,ballesterb,ballestera,demartinietal03,fuj01a,hayashi06}.  Almost every quantum information protocol assumes perfect knowledge of a quantum channel. In practice, knowledge will be imperfect; hence estimation of quantum channels has to precede most other quantum information schemes, and its optimization is of fundamental importance.  

It will be assumed that the unitary channel comes from a parametric family of channels.
When estimating a parameter $\theta$ in a one-parameter model, the SLD quantum information $H_\theta$ will be used as a measure of performance. When $H_\theta$ is attainable, i.e.\ there exists an $M$ such that $F_\theta^M = H_\theta$, the following result is of importance: As the number of observations $N \rightarrow \infty$, using the POVM $M$ and an unbiased maximum likelihood estimator,
\begin{equation}
NE[(\hat\theta-\theta)^2] \rightarrow \frac{1}{H_\theta}
\label{eq:CRineq2b}
\end{equation}
\citep[p. 63]{vaart98}. When $\mathrm{dim} \, \theta > 1$, the performance of estimation will be quantified using the trace of the SLD quantum information ($\tr \{H_\theta \}$). When $\mathrm{dim} \, \theta > 1$, the SLD quantum informations for different parametric families of states may be incomparable. That is, given two families of states $\rho_\theta^{(1)}$ and $\rho_\theta^{(2)}$, with SLD quantum informations $H^{(1)}_\theta$ and $H^{(2)}_\theta$, it may be that $H^{(1)}_\theta \not\geq H^{(2)}_\theta$ and $H^{(1)}_\theta \not\leq H^{(2)}_\theta$. The quantity $\tr \{H_\theta \}$  is useful since \citep{ballesterb} 
\begin{enumerate}
\item[(i)] it treats the parameters $\theta^1, \dots, \theta^p$ with equal importance, 
\item[(ii)] if $\tr \{H^{(1)}_\theta \} \geq \tr \{H^{(2)}_\theta \}$ then $H^{(1)}_\theta \not< H^{(2)}_\theta$.
\end{enumerate}
The output state will be measured using POVMs which satisfy (\ref{bcii2e}) (possibly using an adaptive
measurement), and an estimate of $\theta$ and hence $U_\theta$ will be obtained using the maximum likelihood estimator.

Previous work in estimation (see Section \ref{sec:devel}) has looked at the case where there are $n$ copies of some $U_\theta$.
In this chapter a more general problem is considered:  given $n$ channels which are not identical,
is it better to estimate each of them individually or is it possible to improve on this by using the channels in parallel, as in the case of $n$ identical channels?
It may be that in practice, more commonly, there are $n$ channels which are different (but functionally dependent) than $n$ channels which are identical.

In this chapter the performance of estimation will be considered as a function of $N$, the number of times each of the $n$ channels is used.
It will be assumed that each channel can be used only once on each input state. 

\section{Simplifying Matsumoto's equality condition} \label{sec:nrps}
The following result will simplify later calculations.

It was mentioned in Theorem \ref{thm:matzy} that for pure state models $\rho_\theta = | \psi_\theta \rangle \langle \psi_\theta |$, there exists a POVM and estimator such that equality holds in the quantum Cram\'er-Rao inequality, (\ref{eq:QCRineq}), at $\theta=\theta_0$ if and only if 
\begin{equation}
\Im \langle l_j (\theta) | l_k (\theta) \rangle = 0, \quad \forall j,k,
\label{eq.matzcond}
\end{equation}
where $| l_j(\theta) \rangle = \lambda^j_\theta | \psi_\theta \rangle$ \citep{mats97,fuj01a,mats02}.
An equivalent condition which is simpler to check, and will be used in this chapter, is given in the following lemma.  
\begin{lemma}
For families of pure states  $\rho_\theta = | \psi_\theta \rangle \langle \psi_\theta |$, equality holds in the quantum Cram\'er-Rao inequality, (\ref{eq:QCRineq}), at $\theta=\theta_0$ if and only if
\begin{equation}
\Im \langle \psi_\theta^{(j)}  | \psi_\theta^{(k)} \rangle =0 \quad \forall j,k,
\label{eq.calebcond}
\end{equation}
where $| \psi_\theta^{(j)}\rangle = \partial | \psi_\theta\rangle/\partial \theta^j$.\label{lm:caleblemz}\end{lemma}
\proof 
For pure states, equality holds in (\ref{eq:QCRineq}) at $\theta=\theta_0$ if and only if (\ref{eq.matzcond}) is satisfied. Now $| l_j(\theta) \rangle = \lambda^j_\theta | \psi_\theta \rangle$ is independent of the choice of $\lambda^j_\theta$ \cite[Appendix A, before (7)]{fuj01a}. A possible choice is 
\begin{equation}
\lambda^j_\theta = 2 \partial \rho_\theta/\partial \theta^j = 2(|\psi_\theta^{(j)}\rangle \langle \psi_\theta|+|\psi_\theta\rangle \langle \psi_\theta^{(j)}|).
\end{equation}
A little algebra gives 
\begin{equation}
\langle l_j (\theta) | l_k (\theta)\rangle = 4(\langle \psi_\theta^{(j)} | \psi_\theta^{(k)} \rangle + \langle \psi_\theta^{(j)} | \psi_\theta \rangle \langle \psi_\theta^{(k)} | \psi_\theta \rangle).
\end{equation}
The second term is always real, since $\langle \psi^{(l)}_\theta | \psi \rangle $ is purely imaginary for all $l$ (see below (\ref{eq:diffjj})). Thus condition (\ref{eq.matzcond}) is equivalent to condition (\ref{eq.calebcond}).
\begin{remark}
Although $\langle \psi_\theta^{(j)} | \psi_\theta^{(k)}\rangle$ depends on the choice of phase of $| \psi_\theta \rangle$, $\Im  \langle \psi_\theta^{(j)} | \psi_\theta^{(k)}\rangle$ does not.
 \end{remark}

\begin{lemma}
If $| x_1 \rangle, \dots, | x_n \rangle \in \mathbb{C}^d$ such that 
\begin{enumerate}
\item[(i)] $| x_1 \rangle, \dots, | x_n \rangle$ are $\mathbb{R}$-linearly independent, 
\item[(ii)] $\langle x_j | x_k \rangle \in \mathbb{R}$ for all $j,k=1, \dots,n,$
\end{enumerate}
then  $n \leq d$.
\label{pejslem}
\end{lemma}
\proof
Suppose that  $\exists$ $\alpha_1, \dots, \alpha_n \in \mathbb{C}$ such that
 \begin{equation*}
\sum_{j=1}^n \alpha_j | x_j \rangle = 0.
\end{equation*}
Putting $\alpha_j = a_j +i b_j$, where $a_j, b_j \in \mathbb{R}$ for $j=1, \dots,n$, gives
\begin{equation*}
\sum_{j=1}^n (a_j +i b_j)| x_j \rangle = 0,
\end{equation*}
and so
\begin{equation}
\sum_{j=1}^n (a_j +i b_j)\langle x_k | x_j \rangle = 0, \quad \mathrm{for \ all}\  k.
\label{eq.pej}
\end{equation}
From condition (ii),
\begin{equation*}
\sum_{j=1}^n a_j \langle x_k | x_j \rangle = 0, \ \mathrm{for \ all}\  k.
\end{equation*}
Thus
 \begin{equation*}
 \sum_{j,k=1}^n a_j a_k \langle x_k | x_j \rangle = 0, 
 \end{equation*}
 and so 
 \begin{equation*}
 \sum_{j=1}^n a_j  | x_j \rangle = 0.
  \end{equation*}
  Since by (i)  $| x_1 \rangle, \dots, | x_n \rangle$ are $\mathbb{R}$-linearly independent, $a_j =0$ for $j=1, \dots,n$. Similarly (\ref{eq.pej}) gives $b_j =0$ for $j=1, \dots,n$. Thus $\alpha_j =0$ for $ j=1, \dots,n$. Consequently,  $| x_1 \rangle, \dots, | x_n \rangle$  are $\mathbb{C}$-linearly independent. Therefore, if $| x_1 \rangle, \dots, | x_n \rangle \in \mathbb{C}^d$ satisfy (i) and (ii), then $n \leq d$.
\begin{theorem} 
For a $d$-dimensional non-degenerate family of pure states $\rho_\theta = | \psi_\theta \rangle \langle \psi_\theta |$,  $\theta = (\theta^1, \dots, \theta^p)$, $H_\theta$ is attainable only if $p \leq d-1$.
\label{thm:calebsthm1}
\end{theorem}
\proof
The vectors $\{ | l_j(\theta) \rangle \}$, where $| l_j (\theta) \rangle = \lambda^j_\theta | \psi_\theta \rangle$ are $\mathbb{R}$-linearly independent (due to the nondegeneracy of the parameterization $\theta \mapsto \rho_\theta)$ \cite[Appendix A]{fuj01a}. Since $\langle l_j(\theta) | \psi_\theta \rangle = \tr \{ \rho_\theta \lambda^j_\theta \} =0$ for all $j$, the vectors $\{ |\psi_\theta \rangle, | l_1 (\theta)\rangle, \dots, | l_p (\theta)\rangle \}$ are also $\mathbb{R}$-linearly independent. 
From (\ref{eq.matzcond}) it is seen that $H_\theta$ is attainable if and only if the set of vectors $\{ |\psi_\theta \rangle, | l_1 (\theta)\rangle, \dots, | l_p (\theta)\rangle \}$ satisfy conditions  (i) and (ii) in Lemma \ref{pejslem}.
It follows from Lemma \ref{pejslem} that $H_\theta$ is attainable only if $p  \leq d-1$.

\begin{remark}
As any unitary channel can be specified  by $d^2-1$ parameters, Theorem \ref{thm:calebsthm1} shows the importance of enlarging the Hilbert space to estimate a completely unknown $U \in SU(d)$, i.e.\ letting $\mathbb{I}_d \otimes U$ act on a state  $| \phi \rangle \in \mathbb{C}^{d^2}$. In this case it is possible to have a maximum of $d^2-1$ parameters such that $H_\theta$ is attainable.
\end{remark}

For the channels considered in this chapter an extension of the form $\mathbb{I}_d \otimes  \mathcal{E}$ does not increase the maximum attainable Fisher information.

\section{A $2$-dimensional family of non-identical channels}  \label{sec:dep2}
Consider the following set of $2$-dimensional channels,  which are all functions of the parameter $\theta$, 
\begin{equation}
U^{1}_\theta = \left( \begin{array}{cc}
1 & 0 \\
0 & e^{i f_1(\theta)} \end{array} \right),
\dots, 
\quad U^{n}_\theta = \left( \begin{array}{cc}
1 & 0 \\
0 & e^{i f_n(\theta)}
 \end{array} \right),
 \label{eq.u12un}
 \end{equation}
 where $0 \leq \theta \leq q$, for some $q$, and $f_j: \mathbb{R} \rightarrow  \mathbb{R}$.  The following conditions are imposed on the functions $f_j$:  
 \begin{enumerate}
\item[(a)]  $\displaystyle \frac{df_j(\theta)}{d\theta} > 0$, 
\item[(b)] $0 \leq \sum_j f_j(\theta) \leq  \pi$, 
\end{enumerate}
for all $j$ and $\theta$.
\begin{remark}
Throughout this chapter similar restrictions will be given on the unitary matrices to be estimated. Condition (a) means that as $\theta$ is increased the angle through which states are rotated is also increased, though the amount by which the phase increases varies from unitary to unitary; condition (b) can be thought of as having some prior information about the phases to be estimated, possibly through a knowledge of the experimental arrangements.
\end{remark}
The SLD quantum informations of the schemes
\begin{enumerate}
\item[(i)] letting each of the $n$ channels act on identical copies of $| \psi_x \rangle = 1/\sqrt{2}(| 0 \rangle + |1\rangle)$,
i.e.
\begin{equation*}
| \psi_x \rangle \mapsto U_\theta^j | \psi_x \rangle,
\end{equation*}
 \item[(ii)]  arranging all $n$ of the channels in parallel and using the  entangled input state $| \psi \rangle = 1/\sqrt{2}(| 0 0 \cdots 0 \rangle + |11 \cdots 1\rangle) \in \mathbb{C}^{2^n}$, i.e.\
 \begin{equation}
| \psi \rangle \mapsto  (U_\theta^1 \otimes \cdots  \otimes U_\theta^n) | \psi \rangle,
 \end{equation}
 \end{enumerate}
 will be compared. If $U^j_\theta$ acts on the state $| \psi_x \rangle$, the output state is $ 1/\sqrt{2} (|0 \rangle + e^{i f_j(\theta)}| 1 \rangle)$. This gives $H_\theta^j=(d f_j(\theta)/d\theta)^2$, which is attainable by measuring in $x$, i.e.\ using the POVM $M^x = \{ M_0 =| \psi_{x} \rangle\langle \psi_{x} |, \mathbb{I} - M_0 \}$. Thus for the $n$ channnels 
\begin{equation}
H_\theta^{(i)} = \sum_{j=1}^n \left(\displaystyle \frac{d f_j(\theta)}{d\theta}\right)^2
\label{eq.mse1}
\end{equation}
 and is attainable. An estimate $\hat \theta^{(i)}$ is obtained using the maximum likelihood estimator.

Now the $n$-partite input state $| \psi \rangle$ will be considered. The output state is $1/\sqrt{2}(| 00 \cdots 0 \rangle + e^{i \sum_{j=1}^n f_j(\theta)}| 11 \cdots 1\rangle)$. Computation gives 
\begin{equation}
H_\theta^{(ii)} = \left(\sum_{j=1}^n \frac{\displaystyle d f_j(\theta)}{d\theta}\right)^2,
\label{eq.mse2}
\end{equation}
which is attainable using the POVM $M = \{ M_0 =| \psi \rangle \langle \psi|,  \mathbb{I} - M_0 \}$. Because of conditions (a) and (b), $\theta$ can be identified. An estimate $\hat \theta^{(ii)}$ is obtained using the maximum likelihood estimator.

The SLD quantum informations (\ref{eq.mse1}) and  (\ref{eq.mse2}) may look similar, but they are not. The position of the bracket makes a considerable difference. From condition (a) on the functions $f_j$, (\ref{eq.mse2}) is considerably larger than (\ref{eq.mse1}).
For example, in the case when $f_j(\theta) = \theta$ for all $j$, the SLD quantum informations are $Nn$ and $Nn^2$, respectively.

A consequence of this is that the asymptotic limit of the mean square error is considerably smaller using approach (ii). The asymptotic limits of the mean square errors for approaches (i) and (ii) are, respectively,
\begin{eqnarray}
N E[(\hat \theta^{(i)} - \theta)^2] \rightarrow \frac{1}{\displaystyle \sum_{j=1}^n \left( \frac{d f_j(\theta)}{d\theta}\right)^2},\\
N E[(\hat \theta^{(ii)} - \theta)^2] \rightarrow  \frac{1}{\left(\displaystyle \sum_{j=1}^n  \frac{d f_j(\theta)}{d\theta}\right)^2}.
\label{mse22}
\end{eqnarray}

\subsection{Sequential method}
Here it will be shown that, without using entanglement, it is possible to obtain the same SLD quantum information for the set of non-identical channels (\ref{eq.u12un}), as was obtained in approach (ii).
A third scheme for estimating the set of channels (\ref{eq.u12un}) will be introduced, which will be referred to as the {\it sequential scheme}. The sequential scheme makes no use of entanglement. 

\begin{enumerate}
\item[(iii)] The channels (\ref{eq.u12un}) are each used once on the same separable input state $| \psi_x \rangle$, i.e.
\begin{eqnarray}
  | \psi_x \rangle  &\mapsto& U_\theta^n \cdots U_\theta^2 U_\theta^1 | \psi_x \rangle  \label{eq.seqsch}\\
&=& \frac{1}{\sqrt{2}}(|0 \rangle + e^{i \sum_{j=1}^n f_j(\theta)}|1\rangle) \nonumber.
\end{eqnarray}
\end{enumerate}
Calculation gives
\begin{equation}
H_\theta^{(iii)} = \left(\sum_j \frac{d f_j(\theta)}{d\theta} \right)^2
\end{equation}
 and is attainable by measuring in $x$.  An estimate $\hat \theta^{(iii)}$ is obtained using the maximum likelihood estimator.
The SLD quantum information obtained in approach (iii) is equal to that obtained in approach (ii), thus will have the same asymptotic limit on the mean square error, (\ref{mse22}).

\section{A more general family of one-parameter channels}  \label{sec:depH}
Often physicists are interested in unitary channels parameterised as $V_\theta = \exp(i \theta H)$, where $H$ is an observable related to the energy in a system, known as the {\it Hamiltonian}. This seemingly simple channel has many examples in interferometry and measurement of small forces.  \citep[For more on channels of this type see][and the references therein.]{gio06} 
Consider $n$ $d$-dimensional unitary channels parameterised as 

\begin{equation}
U^{j}_\theta = \exp (i f_j(\theta) H ), \quad 1 \leq j \leq n,
\label{genchanh}
\end{equation}
where $0 \leq \theta \leq q$, for some $q$, $f_j: \mathbb{R} \rightarrow  \mathbb{R}$ for all $j$. The following conditions are imposed on the functions $f_j$: 
\begin{enumerate}
\item[(a)]  $\displaystyle \frac{df_j(\theta)}{d\theta} > 0$, 
\item[(b)] $0 \leq \sum_j f_j(\theta) \leq  \pi$, 
\end{enumerate}
 for all $j$ and $\theta$.
The problem of finding the optimal input state will not be considered.
The SLD quantum informations for 
\begin{enumerate}
\item[(i)]  letting each of the $n$ channels act on identical copies of some $| \psi_{0} \rangle$, i.e.\
\begin{equation*}
| \psi_{0} \rangle \mapsto U^j_\theta | \psi_{0} \rangle,
\end{equation*} 
\item[(ii)]  letting each of the $n$ channels act on the same separable state $| \psi_{0} \rangle$, i.e.\
\begin{equation*}
| \psi_{0} \rangle \mapsto U^n_\theta \cdots U^2_\theta U^1_\theta | \psi_{0} \rangle
\end{equation*}
\end{enumerate}
will be compared.
The SLD quantum informations for (i) and (ii) are, respectively,
\begin{eqnarray}
H_\theta^{(i)} = 4  \sum_{j=1}^n \left( \frac{d f_j(\theta)}{d\theta} \right)^2 [ \langle \phi_0 | H^2 | \phi_0 \rangle - \langle \phi_0 | H | \phi_0 \rangle^2 ], 
\label{eih1}\\
H_\theta^{(ii)} =  4 \left( \sum_{j=1}^n  \frac{d f_j(\theta)}{d\theta} \right)^2  [ \langle \phi_0 | H^2 | \phi_0 \rangle - \langle \phi_0 | H | \phi_0 \rangle^2 ].
\label{eih2} 
\end{eqnarray}
Because of condition (a) the SLD quantum information of (ii), given by (\ref{eih2}), is considerably larger than that of (i), given by (\ref{eih1}). These results hold for all choices of input state $| \phi_0 \rangle$.

 \section{A $d$-dimensional family of non-identical channels } \label{depd}
The situtation of having $n$ `dependent' $d$-dimensional commuting channels will be considered.
These will be parameterised in a similar way to that used by  \cite{ballesterb}.
 \cite{ballesterb} looked at commuting unitary channels. \index{commuting unitary channels} Any commuting unitary channel can be specified using $d-1$ parameters, i.e.\ by a parameter $\theta = (\theta_1, \dots, \theta_{d-1})$.
Given a set of $d \times d$ matrices $t_k$, $k=1, \dots, d-1$, satisfying 
\begin{enumerate}\item[(i)] $t_k = t_k^\dagger$, 
\item[(ii)] $\tr\{t_k \} = 0$, 
\item[(iii)] $\tr\{t_k t_l \} = \delta_{kl}$, 
\item[(iv)] $t_k t_l = t_l t_k$,
\end{enumerate}Ballester parameterised the set of commuting unitary channels as
\begin{equation}
U_{\theta} = \exp\left(i \sum_{k=1}^{d-1} \theta_k t_k\right).
\label{eq:uform1}
\end{equation}
Since, from (iv), $t_k$ and $t_l$ commute, they share a basis $\{ | w_k \rangle \}$, which is assumed to be known. Consequently, any $t_m$ can be written as
\begin{equation}
t_m = \sum_{i=1}^d c_{m i} | w_i \rangle \langle w_i |.
\label{cm1}
\end{equation}
From condition (i) it follows that $c_{mi} \in \mathbb{R}$ for all $m,i$. Conditions (ii) and (iii) give  \begin{eqnarray}
 \sum_{i=1}^d c_{mi} &=& 0, \label{cm2}\\  \sum_{i=1}^d c_{mi} c_{ni} &=&  \delta_{mn}.
 \label{cm3}
 \end{eqnarray}
Ballester showed that there is no advantage in extending $U_\theta$ and using a maximally entangled input state.
The maximum value of $\tr\{H_\theta \}$ can be attained using the separable state 
\begin{equation}| \psi_{sep} \rangle = \frac{1}{\sqrt{d}} \sum_{k=1}^d  | w_k \rangle.\label{eq.sepstate}\end{equation} 
Consider the set of channels
 \begin{equation}
 U^{j}_\theta = \exp\left(i\sum_{k=1}^{d-1} f_j(\theta_k) t_k\right), \quad 1 \leq j \leq n,
 \label{genchan}
 \end{equation}
  where $0 \leq \theta \leq q$, for some $q$, $f_j: \mathbb{R} \rightarrow  \mathbb{R}$ and $f_j(\theta_0) =0$ for all $j$. All $n$ channels depend on the parameter $\theta = (\theta_1, \dots, \theta_{d-1})$, and each channel depends on every component of $\theta$. The following conditions are imposed on the functions $f_j$: 
\begin{enumerate}
\item[(a)]  $\displaystyle \frac{df_j(\theta)}{d\theta} > 0$ 
\item[(b)] $0 \leq \sum_j f_j(\theta_k) \leq  \pi$, 
\end{enumerate}
 for all $j, k$ and $\theta$.
The traces of the SLD quantum information for 
\begin{enumerate}
\item[(i)]  letting each of the $n$ channels act on identical copies of $| \psi_{sep} \rangle$ given in (\ref{eq.sepstate}), i.e.\
\begin{equation*}
| \psi_{sep} \rangle \mapsto U^j_\theta | \psi_{sep} \rangle,
\end{equation*} 
\item[(ii)]  letting each of the $n$ channels act on the same separable state $| \psi_{sep} \rangle$, i.e.\
\begin{equation*}
| \psi_{sep} \rangle \mapsto U^n_\theta \cdots U^2_\theta U^1_\theta | \psi_{sep} \rangle
\end{equation*}
\end{enumerate}
will be compared.

\begin{proposition}
The traces of the SLD quantum informations for 
(i) and (ii) are, respectively,
\begin{eqnarray}
\tr \{ H_\theta^{(i)} \} =   \frac{4}{d} \sum_{i=1}^{d-1} \sum_{j=1}^n \left( \frac{\partial f_j(\theta_i)}{\partial \theta_i} \right)^2, \label{sldsep} \\ 
\tr \{ H_\theta^{(ii)} \} = \frac{4}{d} \sum_{i=1}^{d-1} \left( \sum_{j=1}^n  \frac{\partial f_j(\theta_i)}{\partial \theta_i} \right)^2. \label{sldseq}
  \end{eqnarray}
  \end{proposition}
From condition (a), the trace of the SLD quantum information of (ii), given by (\ref{sldseq}), is considerably larger than that of (i), given by (\ref{sldsep}). 
\\
\\
\proof
A proof will be given for (\ref{sldseq}); the proof for  (\ref{sldsep}) is very similar. The $j$th unitary channel will be denoted by $U_\theta^j$. As the $U_\theta^j$ commute,
\begin{equation}
\prod_{j=1}^n U_\theta^j = \exp\left\{i\sum_{k=1}^{d-1} g_k(\theta_k) t_k \right\}, \quad g_k(\theta_k) = \sum_{j=1}^n  f_j(\theta_k).
\end{equation}
Using each of the $n$ channels on the single input state (\ref{eq.sepstate}) gives the output state
\begin{equation}
| \psi_\theta \rangle = \left(\prod_{j=1}^n U_\theta^j \right) | \psi_{sep} \rangle =  \exp\left\{i\sum_{k=1}^{d-1} g_k(\theta_k) t_k \right\} | \psi_{sep} \rangle.
\label{eq.ddepout}
\end{equation}
An arbitrary diagonal element of $H_\theta$ is equal to
\begin{eqnarray*}
(H_\theta^{(ii)})_{mm} &=& 4 \left[\langle \psi_\theta^{(m)} | \psi_\theta^{(m)} \rangle - | \langle \psi_\theta^{(m)} | \psi_\theta \rangle |^2 \right], \quad | \psi_\theta^{(m)}  \rangle = \partial| \psi_\theta \rangle/ \partial \theta^m, \\
&=& 4\left( \frac{\partial g_m(\theta_m)}{\partial \theta_m } \right)^2 \left[ \langle \psi_\theta | t_m t_m | \psi_\theta \rangle - | \langle \psi_\theta |t_m | \psi_\theta \rangle |^2 \right] \\
&=& 4\left( \frac{\partial g_m(\theta_m)}{\partial \theta_m } \right)^2 \left[ \langle \psi_{sep} | t_m t_m | \psi_{sep} \rangle - | \langle \psi_{sep} |t_m | \psi_{sep} \rangle |^2 \right] \\
&=& 4\left( \frac{\partial g_m(\theta_m)}{\partial \theta_m } \right)^2 \left[ \frac{1}{d} \sum_{k=1}^d c_{mk}^2 - \left|\frac{1}{d} \sum_{k=1}^d c_{mk} \right|^2 \right] \, \mathrm{by} \, (\ref{cm1})\\
&=& \frac{4}{d} \left( \frac{\partial g_m(\theta_m)}{\partial \theta_m } \right)^2 \, \mathrm{by} \, (\ref{cm2}) \, \mathrm{and}  \, (\ref{cm3}).
\end{eqnarray*}
Thus
\begin{eqnarray*}
\tr \{ H_\theta^{(ii)} \} &=& \frac{4}{d} \sum_{m=1}^{d-1}  \left( \frac{\partial g_m(\theta_m)}{\partial \theta_m } \right)^2\\
&=& \frac{4}{d} \sum_{m=1}^{d-1}  \left( \sum_{j=1}^N \frac{\partial f_j(\theta_m)}{\partial \theta_m }\right)^2.
\end{eqnarray*}
\begin{proposition}
The SLD quantum information (\ref{sldseq}) is attainable.
\end{proposition}
\proof
The set of output states is given by (\ref{eq.ddepout}). Now 
\begin{eqnarray*}
\langle \psi_\theta^{(m)} | \psi_\theta^{(n)}  \rangle &=& \left( \frac{\partial g_m(\theta_m)}{\partial \theta_m } \right)\left( \frac{\partial g_n(\theta_n)}{\partial \theta_n } \right)\langle \psi_\theta| t_m t_n |\psi_\theta \rangle,\quad | \psi_\theta^{(m)} \rangle = \partial| \psi_\theta \rangle/ \partial \theta^m, \\
&=& \left( \frac{\partial g_m(\theta_m)}{\partial \theta_m } \right) \left( \frac{\partial g_n(\theta_n)}{\partial \theta_n } \right)  \langle \psi_{sep}| t_m t_n |\psi_{sep} \rangle\\
&=& \frac{\delta_{mn}}{d} \left( \frac{\partial g_m(\theta_m)}{\partial \theta_m } \right)^2.
\end{eqnarray*}
which is always real. Thus (\ref{eq.calebcond}) is satisfied, and consequently $H_\theta$ is attainable.

\chapter{An iterative phase estimation algorithm} \label{ch:ipe}

\section{Introduction}
This chapter considers phase estimation, which is of fundamental importance to quantum information and quantum computation. Phase estimation is related to some very important problems such as estimating eigenvalues \citep{nori04,guzik05,wang08,wang09}, precision measurement of length and optical properties, and clock synchronization \citep{burghbart05}. (The
work in this chapter has been published in \cite{oloan10}.)

Consider a unitary matrix $U_\theta$ depending on an unknown parameter $\theta$ for which one of its eigenvectors $| u \rangle$ is completely known; 
furthermore $U_\theta$ acts on $| u \rangle$ by $U_\theta| u \rangle = e^{i 2 \pi\theta} | u \rangle$, where $\theta \in [0,1)$.
The task of phase estimation is to estimate the eigenvalue $e^{i 2 \pi\theta}$, and consequently 
$\theta$, as accurately as possible.
This chapter considers phase estimation of a unitary matrix with known eigenvectors, which acts on a $2$-dimensional Hilbert space. In particular,  unitary matrices of the form
\begin{equation}
U_\theta = \left(  \begin{array} {cc}
1 & 0 \\
0 & e^{i 2 \pi \theta}  \ \end{array} \right),
\label{U}
\end{equation}
are considered, where $\theta \in [0,1)$.  The angle $\theta$ will be thought of as a point on a circle of unit circumference, and confidence intervals for $\theta$ as arcs on a circle of unit circumference, known as confidence arcs.
The distance between the point $\theta$ and an estimate $\hat \theta$, will be defined as
\begin{equation}
| \hat \theta - \theta |_{1} =\mathrm{min} \left( (\hat \theta - \theta )_{\mathrm{mod \, 1}}, (\theta - \hat \theta  )_{\mathrm{mod \, 1}} \right).
\end{equation}

The performance of phase estimation schemes will be quantified in terms of the expected fidelity $\langle F( U_{\hat \theta},U_\theta) \rangle$.
The cost function
\begin{equation}
1- \langle F( U_{\hat \theta},U_\theta) \rangle = 1 - \frac{\left \langle |\tr \{ U_{\hat \theta}^{-1}  U_{\theta} \}|^2\right \rangle}{d^2} 
\end{equation}
will be used, and its asymptotic scaling analysed as a function of $n$ --- the number of times that $U_\theta$ is used.

For a simple phase estimation approach where $U_\theta$ is used once on $n$ identical copies of some input state (see Section \ref{eq.simple}), $1- \langle  F \rangle  = O(1/n)$. This rate at which $1- \langle  F \rangle$ approaches zero is known as the {\it standard quantum limit} \citep{burghbart05}.

However,  it has been shown  \citep{hayashi06,kahn07,fujimai07} that it is possible to obtain $1- \langle  F \rangle = O(1/n^2)$. This rate at which $1- \langle  F \rangle$ approaches zero is known as the {\it Heisenberg limit} \citep{gio04}, and cannot be beaten  \citep{kahn07}.  These methods require $n$ copies of $U_\theta$ and entangled states.

It has further been shown that it is possible to achieve the Heisenberg limit without entanglement, and with only a single copy of $U_\theta$ (see Section \ref{eq.bits}).
Estimation schemes of this type require a rotation gate capable of performing arbitrary rotations to perfect precision.

\cite{kit95}  sketched an iterative phase estimation method which requires only a single copy of $U_\theta$ and basic measurements: no extra rotation gate is needed.
For this method $1- \langle  F \rangle = O((\log  n/n)^2$, which is within a logarithmic factor of the Heisenberg limit.
However, as will be shown in this chapter, attempts to give a detailed account for such a scheme have been unsuccessful.
This chapter seeks to give a correct detailed phase estimation scheme similar to that of \cite{kit95}, which requires only a single copy of $U_\theta$ and basic measurements.

A selection of different phase estimation schemes will now be given.

\subsection{Simple approach} \label{eq.simple}
A very simple method of phase estimation is to let $U_\theta$ act on the input state 
$| \psi_x \rangle = 1/\sqrt{2}(|0\rangle +|1\rangle)$; the output state is $| \psi_{\theta} \rangle = 1/\sqrt{2}(|0\rangle +e^{i2 \pi \theta}|1\rangle)$.
After measuring in $x$, outcome $0$ is observed with probability $p(0; \theta) = (1+ \cos(2\pi \theta))/2$.
Performing $N$ measurements gives an estimate $ \cos(2 \pi \hat \theta) = 2 N_{x=0}/N -1$ of $\cos(2\pi\theta)$, where $N_{x=0}$ is the number of times outcome $0$ is observed.  
After measuring in $y$, outcome $0$ is observed with probability $p(0; \theta) = (1+ \sin(2\pi \theta))/2$.
Performing $N$ measurements gives an estimate $ \sin(2 \pi \hat \theta) = 2 N_{y=0}/N -1$ of $\sin(2\pi\theta)$, where $N_{y=0}$ is the number of times outcome $0$ is observed.  
From estimates of $\cos(2\pi \theta)$ and  $\sin(2\pi \theta)$ an estimate of $\theta$ can be obtained.
  
\subsection{Kitaev's procedure}
The first $l$-stage iterative phase estimation procedure was given by \cite{kit95}. (The number of stages $l$ is chosen beforehand, and will be a compromise between the precision desired and experimental resources and limitations.)
At the $k$th stage of Kitaev's procedure, $U_\theta$ acts $2^{k-1}$ times on a qubit, which is then measured.
The experimenter performs some multiple of $\log(l/\epsilon)$ measurements of $(2^{k-1}\theta)_{\mathrm{mod \, 1}}$. This ensures that it is possible to `localize each of the numbers $2^{k-1}\theta$ in one of the 8 intervals $[(s-1)/8,(s+1)/8] \, (s=0, \dots,7)$ with error probability $\leq \epsilon/l$'. Using this information, an algorithm --- which is not given --- gives an estimate $\hat \theta$ satisfying
\begin{equation}
\mathrm{Pr}\left(\left(\hat \theta -  1/2^{l+2}, \hat \theta +  1/2^{l+2} \right)  \ni \theta \right) \geq 1 - \epsilon.
\label{kitty}
\end{equation}

\subsection{The scheme of Rudolph and Grover}
 \cite{rudolph03} looked at the problem of transmitting a reference frame from Alice to Bob, which is linked to estimation of  an unknown $U \in SU(2)$, parametrized by three parameters $\alpha, \theta, \phi$. The scheme of Rudolph and Grover involves estimating the parameters $\alpha, \theta, \phi$ individually using the following $l$-stage iterative procedure.
The parameter $\theta \in [0,1)$ is thought of in terms of an infinite binary expansion $\theta=w_1 w_2 \dots w_l \dots$.
At the $k$th stage a qubit is sent back and forth between Alice and Bob in such a way that, when Bob finally measures it, he observes outcome $0$ with probability $p_k(0;\theta) = (1+ \cos(2^{k}\pi \theta))/2$.

This is repeated a minimum of $N = 32 \log_2 (2l / \epsilon)$  times \citep{rudolph03}, which ensures that Bob's estimate $\hat p_k(1;\theta)$ of $ p_k(1;\theta)$ satisfies
\begin{equation}
\mathrm{Pr}\left( \left( \hat  p_k - 1/4, \hat p_k +1/4 \right) \ni p_k \right) \geq 1 - \epsilon/l.
\label{RG270}
\end{equation}
It is assumed that if $|\hat p_k - p_k| \leq 1/4$, then Bob can estimate the $k$th bit of $\theta$ correctly.
If this is so, then from (\ref{RG270}), the probability that Bob estimates the $k$th bit of $\theta$ correctly is at least $1 - \epsilon/l$, and the probability that he estimates all of the binary digits of $\theta$ correctly is at least $1-\epsilon$. 
After $l$ stages, an estimate $\hat \theta =  \hat w_1 \hat w_2 \dots \hat w_l$ is obtained,  satisfying
\begin{equation}
\mathrm{Pr}\left( \left( \hat  \theta - 1/2^{l}, \hat \theta +1/2^{l} \right) \ni \theta \right) \geq 1 - \epsilon.
\end{equation}
A similar scheme is then used to estimate the parameters $\alpha$ and $\phi$. The method of Rudolph and Grover has been used by \cite{burghbart05} for the problem of clock synchronization. 

\subsection{The procedure of Ji {\it et al}.}
 \citet{zheng06} highlighted two errors with the method of Rudolph and Grover:
 \begin{enumerate}
\item[(i)] knowing $| \hat \theta - \theta|_1 \leq 1/2^m$ does not give the first $m$ bits of the binary expansion of $\theta$ --  
consider $\theta = 0.49$, $\hat\theta=0.5$ and $m=1$, 
\item[(ii)] the method is problematic (in the sense explained in section \ref{secz2}) for $\theta$ close to $1/2$.
\end{enumerate}
\citeauthor{zheng06}  gave the following $l$-stage procedure. In the first stage, the experimenter lets $U_\theta$ act on $| \psi_x \rangle$ and then measures in $x$;  outcome $0$ is observed with probability $p(0;\theta) = (1+ \cos(2\pi\theta))/2$. The state  $U_\theta | \psi_x \rangle$ is measured $N$ times ($N$ is some multiple  of $\log(l/\epsilon)$), which gives an estimate $\hat \theta$ satisfying
\begin{equation}
\mathrm{Pr}\left( \left( \hat \theta - 1/12, \hat \theta +1/12 \right) \ni \theta \right) \geq 1 - \epsilon/l.
\label{1/12}
\end{equation}
Having obtained an estimate $\hat \theta$,
\begin{enumerate}
\item[1)] if $\hat \theta \in [0, 5/12)$, define $r_1=2$ and $\nu_1 =0$,
\item[  2)] if $\hat \theta \in [5/12, 7/12)$, define $r_1=3$ and $\nu_1 =1$,
\item[3)] if $\hat \theta \in [7/12, 1]$, define $r_1=2$ and $\nu_1 =1$.
\end{enumerate}
At the $k$th stage the experimenter lets  $U_\theta$ act $r_1 r_2 \dots r_{k-1}$ times on $| \psi_x \rangle$. After measuring $U_\theta^{r_1 r_2 \dots r_{k-1}}|\psi_x \rangle$ $N$ times,  $(r_1 r_2 \dots r_{k-1} \theta)_{\mathrm{mod \, 1}}$ is estimated and  $r_k$ and $\nu_k$ are obtained in a similar way to  $r_1$ and $\nu_1$. 
After $l$ stages, values are obtained for $(r_1, \dots,r_l, \nu_1, \dots, \nu_l)$.
The final estimate of $\theta$ is
\begin{equation}
\hat \theta = \sum_{i=1}^l \frac{\nu_i}{\prod_{j=1}^i r_j}.
\label{eq.zhen1}
\end{equation}

\subsection{The method of \citeauthor{dobi06} } \label{eq.bits}

A popular iterative estimation method is to take $\theta$ to have a binary expansion of given length $l$ plus some small remainder, that is $\theta = w_1 w_2 \dots w_l + \Delta$. The binary digits $w_1, \dots, w_l$ are estimated one at a time with a single measurement.
This has been done by \cite{childs00,dobi06,knill07}.
The method will be reviewed as described by  \cite{dobi06}. 

At the $k$th stage the experimenter lets $U_\theta^{2^{l-k+1}}$ act on one of two qubits. The other qubit is acted on by a  Z-rotation gate $e^{i \alpha_k \sigma_z}$ before being measured ---  where  $\alpha_0 =0$ and $\alpha_k$ for $k=2, \dots, l$ depend on the results from the previous $k-1$ stages.
From this measurement, an estimate $\hat w_{l-k+1}$  is obtained of the $(l-k+1)$th binary digit. After $l$ stages an estimate $\hat \theta = \hat w_1 \hat w_2 \dots \hat w_l$ is obtained of $\theta$ which satisfies
\begin{equation}
\mathrm{Pr}\left( \left(\hat \theta - 1/2^{l+1} ,\hat \theta + 1/2^{l+1} \right) \ni \theta \right) \geq  0.81.
\end{equation}

The probability that the final interval contains $\theta$ can be increased to $1-\epsilon$ by either (a) increasing the number of rounds to $l' = l + \log(2+1/(2\epsilon))$  or (b) using  $O(\log^2(1/\epsilon))$  extra measurements of the first few binary digits \citep{dobi06}. The method of
\citeauthor{dobi06} has recently been carried out on experimental data by \cite{liu07}.
Similar work has also been done by \cite{higgins07}.

\section{Problems} \label{secz2}
There is nothing wrong with Kitaev's method of iterative estimation. However, he does not give 
an algorithm for 
\begin{enumerate}
\item[(i)] choosing which of the intervals contains $(2^{k-1}\theta)_{\mathrm{mod \, 1}} $ with probability $1 - \epsilon/l$,
\item[(ii)] reconstructing $\theta$ given confidence intervals for $(2^{k-1}\theta)_{\mathrm{mod \, 1}} $.
 \end{enumerate}
As will be seen in this section, there are gaps in the methods of \citeauthor{rudolph03}, and  Ji {\it et al}.\ for (i).
There are two main gaps in the method of \citeauthor{rudolph03}, which will now be explained.
Firstly, $p_k(0;\theta) = (1+ \cos(2^{k}\pi \theta))/2$ is a multimodal function of $\theta$. For example, $\theta =3/4$ and $\theta=1/4$ give the same value of $p_1(0;\theta)$, even though they differ in the first binary digit. To overcome this, an estimate of $\sin(2\pi \theta)$ is needed as well.
This however is a trivial point and is easily overcome.

Secondly, if $\theta = 1/2 \pm \delta$, where $\delta$ is small,  a large number of measurements is required to determine the first bit of $\theta$ correctly with high probability.
If a mistake is made then,  for the final estimate $\hat \theta$,  $| \hat \theta - \theta|_1 \geq \delta$. 
This problem, which occurs for $\theta$ close to $1/2$, was pointed out by \cite{zheng06}.

A similar problem also occurs for $\theta= 0 \pm \delta$.
Because of this, difficulties will be encountered in estimating the $k$th bit of $\theta$ whenever
$(2^{k-1}\theta)_{\mathrm{mod \, 1}}  \approx 0$, $(2^{k-1}\theta)_{\mathrm{mod \, 1}}  \approx 1$ or $(2^{k-1}\theta)_{\mathrm{mod \, 1}}  \approx 1/2$. 
However, it may also be possible to overcome this problem using extra rotation gates in these cases.

There are also gaps in the method of  \cite{zheng06}. 
Firstly, like Rudolph and Grover, they overlook the fact that $p_1(0;\theta) = (1+ \cos(2\pi\theta))/2$ is bimodal.
Secondly, the accuracy of their final estimate relies on the assumption that if $| \hat \theta - \theta|_1 \leq 1/12$ and $\hat \theta \in [0, 5/12)$, then $\theta \in [0, 1/2)$. This is not true -- consider $\theta = -1/12 \not \in [0, 1/2)$.
Similarly, they assume that  if $| \hat \theta - \theta|_1 \leq 1/12$ and $\hat \theta \in [7/12, 1)$, then $ \theta \in [1/2, 1)$, which again is not true -- consider $\theta = 1/12  \not \in [1/2, 1)$.
Again problems will be encountered at the $k$th stage if $(r_1 \cdots r_{k-1} \theta)_{\mathrm{mod \, 1}}    \approx 0$ or $(r_1 \cdots r_{k-1} \theta)_{\mathrm{mod \, 1}} \approx 1$.

\section{An iterative estimation algorithm} 
This section contains a new method of phase estimation. Firstly, an iterative algorithm is given for going from confidence arcs for $\theta, (2\theta)_{\mathrm{mod \, 1}}, (4\theta)_{\mathrm{mod \, 1}}, \dots,$  $(2^{l-1} \theta)_{\mathrm{mod \, 1}}$, of length $1/3$ and coverage probability at least $1-\epsilon/l$, to a
confidence arc for $\theta$ of length $1/(2^{l-1} \times 3)$ and  coverage probability at least $1-\epsilon$.
 Secondly, a method is given for obtaining a confidence arc for $(2^{k-1} \theta)_{\mathrm{mod \, 1}}$, of length $1/3$ and coverage probability at least $1-\epsilon/l$.
 Thirdly, one of Bernstein's inequalities is used to calculate the number of measurements needed at each stage.
 Finally, it is shown that it is possible to choose a value of $\epsilon$ such that $1- \langle F(U_{\hat \theta}, U_\theta) \rangle = O((\log n/n)^2)$.

 \subsection{The iterative algorithm} \label{32}
  First an intuitive approach is given using examples.
For computational simplicity, confidence arcs of length $0.3$ and coverage probability $1$ will be considered.
$L_k$ and $J_k$ will denote confidence arcs  for $(2^{k-1}\theta)_{\mathrm{mod \, 1}}$ and  $2^{k-1}\theta$ respectively, of length $0.3$ and coverage probability $1$.  
(In the more general algorithm $L_k$ and $J_k$ will have length $1/3$ and coverage probability at least $1-\epsilon/l$.) For the examples, $l=3$. 
\\
\\
{\bf Example 1}
\\ 
 Suppose that after doing some measurements of $U_\theta$, $U_\theta^2$ and $U_\theta^4$  it is found that
\begin{eqnarray}
L_1 &=& [0.6, 0.9 ]  \ni \theta \label{ex1l1} \\
L_2 &=& [0.3, 0.6 ] \ni (2\theta)_{\mathrm{mod \, 1}} \label{ex1l2} \\
L_3 &=& [0.8, 1.1 ] \ni (4\theta)_{\mathrm{mod \, 1}}. \label{ex1l3}
\end{eqnarray}
It follows from (\ref{ex1l1}) that 
\begin{equation}
2L_1= [1.2, 1.8] \ni 2\theta.
\label{2ex1l1}
\end{equation}
Using (\ref{ex1l2}) and  (\ref{2ex1l1}), it follows that
\begin{equation}
J_2 =  [1.3, 1.6]  \ni 2\theta.
\label{defj21}
\end{equation}
From (\ref{defj21}) it is known that
\begin{equation}
2J_2= [2.6, 3.2]  \ni 4\theta.
\label{2j21}
\end{equation}
Using (\ref{ex1l3}) and (\ref{2j21}) gives
\begin{equation}
J_3 = [2.8, 3.1]  \ni 4\theta.
\label{defj3}
\end{equation}

Using confidence arcs (\ref{ex1l1}), (\ref{ex1l2}) and (\ref{ex1l3}) for $\theta, (2\theta)_{\mathrm{mod \, 1}}$ and $ (4\theta)_{\mathrm{mod \, 1}}$ respectively, of length $0.3$ and coverage probability $1$,
a confidence arc (\ref{defj3}) has been derived for $4\theta$ of length $0.3$ and coverage probability $1$.
This gives a confidence arc for $\theta$ of length $0.3/2^{3-1}=0.075$ 
and coverage probability $1$, namely
\begin{equation}
 (1/4)J_3 = [0.7, 0.775]  \ni \theta.
\label{defj3b}
\end{equation}

Remember that confidence arcs on a circle are being considered.
On the circle the arc $[1.2, 1.8]$ is equivalent to the arc $[0.2, 0.8]$, as are 
$[2.2, 2.8], [3.2, 3.8]$ $ \dots$.
Similarly, $[2.6, 3.2]$ is equivalent to $[0.6, 1.2]$.

The symbol $\subset _1$ will be used to signify that a confidence arc on the circle is a subset of another confidence arc on the circle. Similarly, the symbol $\in_1$ will be used to signify that a point is contained within an arc on the circle, e.g. $0.3 \in_1 [1.2, 1.8]$.
The previous example was rather simple in that 
$[0.3, 0.6] \subset_1 [1.2, 1.8]$ and $[0.8, 1.1] \subset_1 [2.6, 3.2]$.

Consider the following example for which, $L_{k+1} \not \subset_1 2 J_{k}$.
(Note that $L_1 = J_1$.)
\\
\\
\noindent {\bf Example 2}
\\
Suppose that after doing some measurements of $U_\theta$, $U_\theta^2$ and $U_\theta^4$ it is found that
\begin{eqnarray}
L_1 &=& [0.1, 0.4 ]  \ni \theta \label{ex2l1} \\
L_2 &=& [0.7, 1.0 ] \ni (2\theta)_{\mathrm{mod \, 1}} \label{ex2l2} \\
L_3 &=& [0.9, 1.2 ] \ni (4\theta)_{\mathrm{mod \, 1}}. \label{ex2l3}
\end{eqnarray}
It follows from (\ref{ex2l1}) that 
\begin{equation}
 2J_1= [0.2, 0.8] \ni 2\theta.
\label{2ex2l1}
\end{equation}
Now $L_2 \not \subset_1  2 J_1$.
From (\ref{ex2l2}) and  (\ref{2ex2l1}) it follows that
\begin{equation}
 [ 0.7, 0.8] \ni 2\theta.
\label{2defj21}
\end{equation}
However, for simplicity, the confidence arcs $J_k$ will be kept of equal length (in this example $0.3$, in the more general algorithm $1/3$).
There is no unique way to do this.
A convenient way is to keep $J_k \subset 2J_{k-1}$ and $J_k$ of length $0.3$. Thus for this example 
 the upper bound for $2\theta$ remains as $0.8$ and the lower bound is chosen to be $0.8-0.3 = 0.5$.
This gives
\begin{equation}
J_2 =  [ 0.5, 0.8]  \ni 2\theta .
\label{2defj21b}
\end{equation}
From (\ref{2defj21b}) it follows that
\begin{equation}
2J_2 =  [1.0, 1.6]  \ni 4\theta.
\label{2j21b}
\end{equation}
Now, again $L_3 \not \subset_1  2 J_2$. To keep $J_k \subset_1 2J_{k-1}$ and $J_k$ of length $0.3$, the lower bound remains as $1.0$ and the upper bound becomes $1.0 + 0.3 = 1.3$,
\begin{equation}
J_3 =  [1.0, 1.3] \ni 4\theta.
\label{defj3b2}
\end{equation}
A confidence arc for $2^{3-1} \theta$ has been found of length $0.3$ and coverage probability $1$. This gives  a confidence arc for $\theta$ of length $0.3/2^{3-1}=0.075$ and coverage probability $1$, namely
\begin{equation}
 (1/4)J_3 = [0.25, 0.325] \ni \theta .
\label{defj3b22}
\end{equation}
{ \bf General Algorithm}
\\
The general algorithm will now be presented. Confidence arcs are now of length $1/3$ rather than $0.3$, and coverage probability $1$,
\begin{eqnarray}
L_k &=& [x(k), x(k) + 1/3], \qquad x(k) \in [0,1)\\
J_k &=& [ z(k), z(k) + 1/3].
\end{eqnarray}
As in the examples, $2J_k$ and $L_{k+1}$ are used to find a confidence arc $J_{k+1}$, with $J_{k+1} \subset 2J_k$. For $J_{k+1} \subset 2J_k$ it is required that
$z(k+1) \in [ 2z(k), 2z(k)+1/3]$. Assuming that $J_k \ni 2^{k-1} \theta$ and $L_{k+1} \ni (2^k \theta)_{\mathrm{mod \, 1}} $, there are three possibilities.
For each possibility a figure is given (showing, for simplicity, a line instead of an arc), with a small vertical line representing the choice of the lower boundary $z(k+1)$ of $J_{k+1}$. Note that $J_1 = L_1$.
 \begin{figure}[htp]\centering\includegraphics[totalheight=0.12\textheight]{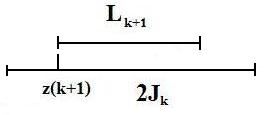}
 \caption[a]{Situation (i)}
 \label{fig:conf1}\end{figure}

 \begin{figure}[htp]\centering\includegraphics[totalheight=0.12\textheight]{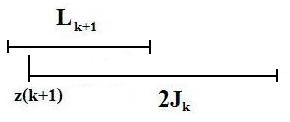}
 \caption[a]{Situation (ii)}
 \label{fig:conf2}\end{figure}
 
  \begin{figure}[htp]\centering\includegraphics[totalheight=0.12\textheight]{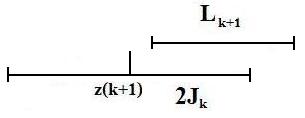}
 \caption[a]{Situation (iii)}
 \label{fig:conf3}\end{figure}
 \noindent (i) The simplest possibility is that $L_{k+1} \subset_1 2J_k$. This occurs when 
\begin{equation*}
 ( x(k+1) - 2z(k) )_{\mathrm{mod \, 1}} \in [0,1/3).
 \end{equation*}
In this case the lower boundary of $J_{k+1}$ is taken to be 
\begin{equation*}
z(k+1)=2z(k)+  (x(k+1) - 2z(k))_{\mathrm{mod \, 1}}.
\end{equation*}
(ii) Another possibility is that $x(k+1) \not \in_1 2J_k$ but $x(k+1) +1/3 \in_1 2J_k$. 
This occurs when 
\begin{equation*}
 ( x(k+1) - 2z(k) )_{\mathrm{mod \, 1}} \in [2/3,1).
 \end{equation*}
In this case the lower boundary of $J_{k+1}$ is taken to be 
\begin{equation*}
z(k+1)=2z(k).
\end{equation*}
(iii) The final possibility is that $x(k+1) \in_1 2J_k$ but $x(k+1) +1/3 \not \in_1 2J_k$. 
This occurs when 
\begin{equation*}
 ( x(k+1) - 2z(k) )_{\mathrm{mod \, 1}} \in [1/3,2/3).
 \end{equation*}
In this case the lower boundary of $J_{k+1}$ is taken to be 
\begin{equation*}
z(k+1)=2z(k) + \frac{1}{3}.
\end{equation*}

This iterative scheme gives the confidence arc $J_l = [z(l), z(l)+1/3]$
for $2^{l-1}\theta$ of length $1/3$ and coverage probability $1$.
This gives a confidence arc for $\theta$ of length $1/(2^{l-1} \times 3)$,
and coverage probability $1$, namely $(1/2^{l-1}) J_l = [z(l)/2^{l-1}, (z(l)+1/3)/2^{l-1}]$.
The centre of this interval, modulo $1$, is taken as the final estimate $\hat \theta$ of $\theta$, i.e.
\begin{equation*}
\hat \theta = \left(\frac{z(l)+1/6}{2^{l-1}}\right)_{\mathrm{mod \, 1}}.
\end{equation*}
The final confidence arc for $\theta$ of length $1/(2^{l-1} \times 3)$ contains $\theta$ if 
$L_k \ni (2^{k-1}\theta)_{\mathrm{mod \, 1}}$, for every $k=1, \dots, l$.
If, for every $k=1, \dots, l$, $L_k$ has coverage probability at least $1-\epsilon/l$, the coverage probability of the final confidence arc is at least $1-\epsilon$. 
 
 \subsection{Finding $L_k$}
 The following function will be used:
 \begin{equation*}
\mathrm{atan2}(x,y) =  \left\{ \begin{array}{lcc}
\ \mathrm{arctan}(y/x) & x >0, & \\
\ \mathrm{arctan}(y/x) + \pi & x < 0, & y \geq 0, \\
\ \mathrm{arctan}(y/x) - \pi & x < 0, & y < 0, \\
\ \pi/2 & x=0, & y > 0, \\
\ -\pi/2 & x=0, & y < 0, \\
\ \mathrm{undefined} & x=0, & y=0.  \end{array} \right.
\end{equation*}
 Here, details are given for calculating confidence arcs $L_k$ for $(2^{k-1} \theta)_{\mathrm{mod \, 1}}$ of length $1/3$ and coverage probability at least $1-\epsilon/l$.
  First it will be shown how to compute a confidence arc of length $1/3$,  then, how to make the coverage probability at least $1-\epsilon/l$.
 
The problem of finding a confidence arc for $\theta$ will be considered. The analysis is exactly the same as for $(2^{k-1} \theta)_{\mathrm{mod \, 1}}$, except that in the latter case the experimenter lets $U_\theta$ act $2^{k-1}$ times on the same $| \psi_x \rangle$.
 
The experimenter lets $U_\theta$ act on $| \psi_x \rangle$ and then measures in $x$. Outcome $0$ is observed with probability $p_x(0;\theta) = (1+\cos(2\pi\theta))/2$.
The state $U_\theta | \psi_x \rangle$ is measured in $x$ a total of $N$ times and outcome $0$ is observed $N_{x=0}$ times. This gives an estimate $2N_{x=0}/N -1$ of $\cos(2\pi\theta)$.

The experimenter lets $U_\theta$ act on $| \psi_x \rangle$ and measures in $y$. Outcome $0$ is observed with probability
 $p_y(0;\theta) = (1+\sin(2\pi\theta))/2$.
The state $U_\theta | \psi_x \rangle$ is measured in $y$ a total of $N$ times and outcome $0$ is observed $N_{y=0}$ times. This gives an estimate $2N_{y=0}/N -1$ of $\sin(2\pi\theta)$.
Estimates of $\sin(2\pi\theta)$ and $\cos(2\pi\theta)$ give the estimate
 \begin{equation}
 \hat \theta_1 = \frac{1}{2\pi} \left(\mathrm{atan2}\left(\frac{2N_{y=0}}{N} -1, \frac{2N_{x=0}}{N} -1\right)\right)_{\mathrm{mod \, 2\pi}}
  \end{equation}
  of $\theta$.
The confidence arc is
 \begin{equation}
 L_1 = ( (\hat \theta_1 -1/6)_{\mathrm{mod \, 1}}, (\hat \theta_1 -1/6)_{\mathrm{mod \, 1}} +1/3).
 \end{equation}
 More generally, an estimate $(2^{k-1} \hat \theta_k)_{\mathrm{mod \, 1}}$ of $(2^{k-1} \theta)_{\mathrm{mod \, 1}}$ gives the confidence arc 
  \begin{equation}
 L_k = \bigg( \left((2^{k-1} \hat \theta_k)_{\mathrm{mod \, 1}} -1/6\right)_{\mathrm{mod \, 1}}, \left((2^{k-1} \hat \theta_k)_{\mathrm{mod \, 1}} -1/6\right)_{\mathrm{mod \, 1}} +1/3\bigg). \qquad
 \end{equation}
It is necessary to find the accuracy needed for the estimates of $p_x(0;\theta)$ and  $p_y(0;\theta)$ to ensure that $| \hat \theta - \theta |_1 \leq 1/6$, and hence $L_1 \ni \theta$. 
  
Put $x = \cos(2\pi \theta)$,  $y = \sin(2\pi \theta)$, $x_0 = 2N_{x=0}/N-1$, $y_0 = 2N_{y=0}/N-1$ and $\phi(x,y) = \mathrm{atan}(y,x)$. Define 
\begin{equation}
| \hat \phi - \phi|_{2\pi} = \mathrm{min}  \left( (\hat \phi - \phi )_{\mathrm{mod \, 2\pi}}, (\phi - \hat \phi  )_{\mathrm{mod \, 2\pi}} \right).  
\end{equation}
Given that 
\begin{eqnarray} 
| x- x_0| &\leq&  \alpha,  \label{xalpha1} \\
| y- y_0| &\leq&  \alpha, \label{yalpha1}
\end{eqnarray}
 an upper bound is sought on $| \phi(x,y) -  \phi(x_0,y_0) |_{2 \pi}$.
 This will be done in steps.
\\
(i) 
\begin{eqnarray}
|\phi(x,y) -  \phi(x_0,y_0)|_{2 \pi}  &=& | [  \phi(x,y) -  \phi(x,y_0)]   \nonumber \\
  &+& [  \phi(x,y_0) -  \phi(x_0,y_0)] |_{2 \pi} \nonumber \\
  &\leq& | \phi(x,y) -  \phi(x,y_0)|_{2\pi} \nonumber  \\
  &+& | \phi(x,y_0) -  \phi(x_0,y_0)|_{2\pi}. \qquad 
  \label{somenumber}
 \end{eqnarray}
Put 
\begin{eqnarray*}
\psi_1 &=& | \phi(x,y) -  \phi(x,y_0)|_{2\pi} \\
\psi_2 &=& | \phi(x,y_0) -  \phi(x_0,y_0)|_{2\pi}.
\end{eqnarray*}
(ii) Consider the triangle $T_1$ given by the points $(0,0), (x,y)$ and $(x,y_0)$, with $y_0$ satisfying (\ref{yalpha1}). The angle at the point $(0,0)$ is $\psi_1$, and is opposite a side of length $| y-y_0|$. The angle, say $\psi_A$, at the point $(x,y_0)$ will be opposite a side of length $1$. 
Using the sine rule for $T_1$ gives
\begin{equation}
\frac{\sin\psi_1}{|y - y_0|} = \frac{\sin\psi_A}{1} .
\label{sine1}
\end{equation}
For any triangle the angles add up to $\pi$. The largest angle will be opposite the longest side. For any angle, $\psi^*$ say, not opposite the longest side, $ \psi^* \in [0,\pi/2]$. If $\alpha \leq 1/2$ then from (\ref{yalpha1}) $| y - y_0| \leq 1$ and so $ \psi_1 \in [0,\pi/2]$. Thus $\psi_1 \leftrightarrow \sin \psi_1$, and hence  $\psi_1 = \arcsin  \beta$, with $\beta = |y - y_0| \times \sin\psi_A$.
As $\psi_A \in [0, \pi]$, consequently $\sin \psi_A \in [0,1]$, and using (\ref{yalpha1}) it follows that
$\beta \in [0, \alpha]$. Since $\arcsin$ is a monotone function on $[0,\alpha]$, it follows that
\begin{equation}
\psi_1 \leq  \arcsin(\alpha).
\label{sine1b}
\end{equation}
(iii) Consider the triangle $T_2$ given by the points $(0,0), (x,y_0)$ and $(x_0,y_0)$, with $x_0$ and $y_0$ satisfying (\ref{xalpha1}) and (\ref{yalpha1}) respectively. The angle at the point $(0,0)$ is $\psi_2$ and is opposite a side of length $| x-x_0|$. The angle, say $\psi_B$, at the point $(x_0,y_0)$ is opposite a side of length $r$, where 
 \begin{eqnarray}
r &=& \sqrt{x^2 + y_0^2 } \nonumber \\
&\geq& \min_\Delta\sqrt{ x^2 + (y+\Delta)^2 }, \quad  \Delta \in [-\alpha, \alpha]\nonumber \\
&=& \min_\Delta \sqrt{ x^2 + y^2 + 2y\Delta + \Delta^2 }\nonumber \\
&=& \min_\Delta \sqrt{ 1 + 2y\Delta + \Delta^2 }\nonumber \\
&\geq&  \min_\Delta \sqrt{ 1 -  2|\Delta| + |\Delta|^2 }\nonumber \\
&=&  \min_\Delta  \quad 1 -  |\Delta| \nonumber \\
&=& 1 -\alpha. \label{alphawac}
\end{eqnarray}
Using the sine rule for $T_2$ gives
\begin{equation}
\frac{\sin\psi_2}{|x - x_0|} = \frac{\sin\psi_B}{r} .
\label{sine1c}
\end{equation}
If $\alpha \leq 1/2$ then $\alpha \leq 1 - \alpha$ and so from (\ref{xalpha1}) and (\ref{alphawac}), $| x - x_0| \leq r$. It follows that $\psi_2 \in [0, \pi/2]$ and so $\psi_2  \leftrightarrow \sin\psi_2$.
Using (\ref{xalpha1}), (\ref{alphawac}), (\ref{sine1c}) and monoticity of $\arcsin$ on $[0,1]$ gives 
\begin{equation}
\psi_2 \leq  \arcsin\left(\frac{\alpha}{1-\alpha}\right).
\label{sine1d}
\end{equation}
\begin{theorem}
Given (\ref{xalpha1}) and (\ref{yalpha1}) for $\alpha \leq 1/2$,
\begin{equation}
| \phi(x,y) -  \phi(x_0,y_0)|_{2 \pi} \leq \arcsin(\alpha) + \arcsin\left(\frac{\alpha}{1- \alpha}\right).\qquad 
\end{equation}
\end{theorem}
\proof
This follows from (\ref{somenumber}),(\ref{sine1b}) and (\ref{sine1d}).
\\
\\
For the iterative algorithm it is required that $|\hat \theta - \theta|_1 \leq 1/6$, which is equivalent to
\begin{equation}
| \phi(x,y) -  \phi(x_0,y_0) |_{2 \pi} \leq \frac{\pi}{3}.
 \label{pi/3}
\end{equation}
If $\alpha = 0.3794$ then (\ref{pi/3}) holds, and (\ref{xalpha1}) and  (\ref{yalpha1}) are equivalent to  \begin{eqnarray}
 | N_{x=0}/N - p_x(0;\theta) | &\leq& 0.1897, \label{hatpx}\\
  | N_{y=0}/N - p_y(0;\theta) | &\leq& 0.1897. 
  \label{hatpy}
 \end{eqnarray}
  It follows that if
  \begin{equation}
\mathrm{Pr} \bigg(  | N_{x=0}/N - p_x(0;\theta) | \leq 0.1897 \bigg) \geq \sqrt{1-\epsilon/l} \label{hatpx1}
\end{equation}
and
\begin{equation}
 \mathrm{Pr} \bigg( | N_{y=0}/N - p_y(0;\theta) | \leq 0.1897  \bigg) \geq \sqrt{1-\epsilon/l}, 
  \label{hatpy1}
 \end{equation}
 then
 \begin{equation}
 \mathrm{Pr} \left( L_1 \ni \theta \right) \geq 1 - \epsilon/l.
 \label{lkooh}
 \end{equation} 
An analogous result holds for $L_k, k=2, \dots,l$.
 In Section \ref{subsec:berny} it is shown that if $N=24.437 \log (4l/\epsilon)$ then (\ref{hatpx1}) and (\ref{hatpy1}) hold.

\subsection{Number of measurements needed} \label{subsec:berny}
The following Bernstein inequality \citep{encylo} will be used:
\begin{theorem}
If the equations
\begin{equation*}
E[Y_j] =0, \quad E[Y_j^2] =b_j,\quad j=1,\dots,n,
\end{equation*}
hold for the independent random variables $Y_1,\dots, Y_n$ with
\begin{equation}
E[|Y_j|^l]  \leq \frac{b_j}{2} H^{l-2} l!
\label{hhh}
\end{equation}
(where $l>2$ and $H$ is a constant independent of $j$), then the following inequality holds for the sum $S_n = \sum_{j=1}^n Y_j$:
\begin{equation}
\mathrm{Pr}( | S_n| >r) \leq 2 \exp\left(-\frac{r^2}{2(B_n+Hr)}\right),
\label{eq.Bern}
\end{equation}
where $B_n = \sum_{j=1}^n b_j$.
\end{theorem}

The observed measurement outcomes from a single measurement in $x$ have distribution and moments
\begin{equation*}
X_j \sim \mathrm{Bin}(1,p), \quad E[X_j] =p, \quad E[X_j^2]= p(1-p), 
\end{equation*}
where $p =(1+\cos(2\pi\theta))/2$. Put $b_j = p(1-p)$ for $j=1, \dots, N$ and consider the random variable $R_j = X_j -p$, which has moments
\begin{equation*}
 E[R_j] =0, \quad E[R_j^2]= p(1-p)=b_j. 
\end{equation*}
Now, for $l > 2$,  
\begin{eqnarray}
E[|R_j|^l]  &=& p|1-p|^l+ (1-p)|0-p|^l \nonumber \\
&\leq& p(1-p)^2+ (1-p)p^2 \nonumber  \\
&=& p(1-p)  \nonumber \\
&=& b_j.
 \label{eq.bj}
\end{eqnarray} 
Thus, comparing (\ref{eq.bj}) with (\ref{hhh}), $H=1$ is a suitable choice. Substituting  $B_N = \sum_{j=1}^N b_j = N p(1-p)$
 and  $S_N=\sum_{j=1}^NR_j= N_{x=0}-Np$ into (\ref{eq.Bern}) gives
 \begin{equation*}
\mathrm{Pr}( |N_{x=1}-Np | >r) \leq 2 \exp\left(-\frac{r^2}{2(Np(1-p)+r)}\right).
\end{equation*}
 Putting $r=N\delta$, gives
  \begin{eqnarray}
\mathrm{Pr}( |N_{x=1}/N-p | > \delta) &\leq& 2 \exp\left(-\frac{N \delta^2}{2(p(1-p)+\delta)}\right) \nonumber \\
& \leq& 2\exp\left(-\frac{N \delta^2}{2(1/4+\delta)}\right). \label{eq.3939}
\end{eqnarray}
The inequality $\mathrm{Pr}( |N_{x=0}/N-p | < \delta) \geq \sqrt{1-\epsilon/l}$, is equivalent to the inequality
$\mathrm{Pr}( |N_{x=0}/N-p | > \delta) \leq 1- \sqrt{1-\epsilon/l}$, which holds if 
 $\mathrm{Pr}( |N_{x=0}/N-p | > \delta) \leq \epsilon/(2l)$.
 Substituting $\delta=0.1897$ into (\ref{eq.3939}), it can be found that (\ref{hatpx1}) holds if  \begin{equation}
 N = 24.437 \ln (4l/\epsilon)
 \label{eq.Nx}
 \end{equation}
 measurements in $x$ are performed at each stage. The analysis is exactly the same for measurements in $y$,
 and so a total number of 
 \begin{equation}
 N_{tot} = 48.874 \ln (4l/\epsilon)
 \label{eq.ntot}
 \end{equation}
 measurements are required at each stage. This ensures that (\ref{hatpx1}) and (\ref{hatpy1}) hold, and consequently (\ref{lkooh}) holds.

\subsection{The behaviour of the fidelity} \label{534}
The behaviour of $1- \langle F(U_{\hat \theta}, U_\theta) \rangle$ will be analysed as a function of the number $n$ of times $U_\theta$ is used. 
As in \cite{rudolph03}, the worst--case value of $1 - \langle F(U_{\hat \theta}, U_\theta) \rangle$ will be sought. 
That is, if the final confidence arc does not contain $\theta$ then $\hat \theta = (\theta +1/2)_{\mathrm{mod} \, 1}$, and if it does then $\theta$ lies on the boundary of the confidence arc, i.e.\ 
 $|\hat \theta - \theta |_1 = 1/(2^{l} \times 3)$.
This gives
\begin{eqnarray*}
 1- \langle F(U_{\hat \theta}, U_\theta) \rangle &\leq& 1- \left((1 - \epsilon) \frac{1 + \cos(2\pi/(2^{l} \times 3))}{2} +  \epsilon \times 0\right) \\
&\approx& \epsilon +  \frac{\pi^2}{2^{2l}\times 9} - \frac{\epsilon \pi^2}{2^{2l}\times 9}.
\end{eqnarray*}
If $\epsilon = 1/2^{2l}$, then $ 1- \langle F(U_{\hat \theta}, U_\theta) \rangle = O(1/2^{2l})$.
This requires a total of
\begin{equation}
N_{tot} = 48.874  \log (4l\times 2^{2l} )
\label{eq.totn}
\end{equation}
measurements  at each stage.
The number of times $U_\theta$ is used is $n = N_{tot} (2^l-1)$,
and so $1/2^{l} \approx N_{tot}/n$. The number of measurements, (\ref{eq.totn}), made at each stage is $O(l)$;  noticing that $\log n$ is also $O(l)$, it follows that
\begin{equation}
 1- \langle F(U_{\hat \theta}, U_\theta) \rangle = O\left(\left(\frac{\log n}{n}\right)^2\right).
\label{eq.logn}
\end{equation}

\section{Simulations}
The analysis in Section \ref{534} concentrated on optimizing the worst-case asymptotic scaling of $1- \langle F \rangle$ with respect to $n$.
Here a more pragmatic line will be taken. Of interest is the minimum 
number of measurements needed such that the final confidence arc contains $\theta$ a satisfactory proportion of the time.

The iterative algorithm will now be investigated using simulations with the computer package MAPLE. 
A value for the parameter $\theta \in[0,1)$ is given by a random variable with a uniform distribution.
Measurement results can be simulated, since the number of times outcome $0$ is observed has a Binomial distribution. For example, at the $k$th iterative stage, measuring in $x$,  $N_{x=0} \sim \mathrm{Bin}(N, (1+\cos(2^{k} \pi \theta))/2)$.
From the simulated results of measurements in $x$ and $y$ for stages $1, \dots, l$, an estimate of $\theta$ is obtained using the iterative algorithm given in section \ref{32}. It can then be checked whether the final confidence arc contains $\theta$. This is done for $100,000$ randomly chosen values of $\theta$, and the number of times the final interval contains $\theta$ is recorded.

For most recent iterative schemes the total number of iterations is reasonably small:
$6$ in \cite{higgins07} and $7$ in \cite{liu07}. Simulations were performed with the number of iterations varying between $6$ and $9$. Table \ref{nni2b.tab} gives the number of times the final confidence arc contains the true value of $\theta$. 

\begin{table}[th]
\centering
\begin{tabular}{|c|c|c|c|c|}
\hline
&\multicolumn{4}{c|}{Number of iterative stages $(l)$}\\
\cline{2-5}
$N_{tot}$    &  6 & 7 & 8 & 9  \\
\hline
20 &  99,792 & 99,729 & 99,747 & 99,712  \\
30  &  99,993 & 99,987 & 99,982 & 99,978  \\
40 &  99,999 & 100,000 & 99,998 & 99,999  \\
50  & 100,000 & 100,000 & 99,999  & 100,000 \\
\hline 

\end{tabular} 
\caption{Numbers of trials out of 100,000 with $| \hat \theta - \theta| \leq 1/(2^l \times 3)$.} \label{nni2b.tab}
\end{table} 
It seems a waste to use $N_{tot} = 48.874 \log (2l \times 2^{2l})$ measurements at each stage, since the simulations suggest that for practical purposes it is sufficient to use fewer measurements -- even as few as $20$ or $30$.

 \subsection{Estimating the coverage probability}
 Using the above simulations the coverage probability can be estimated, i.e.\ the probability that, using the iterative algorithm, the known true value $\theta$ is contained in the final confidence interval.

Suppose the true (unknown) coverage probability is $p$.
For the $i$th trial put 
\begin{eqnarray*}
W_i  &=& 1 \quad  \mathrm{if \, interval \, covers}  \, \theta \\
  &=& 0 \quad \mathrm{if \, not.}
\end{eqnarray*}
Then $W_1, \dots, W_M$ are independent identically distributed Bernoulli random variables, i.e.\ $W_i \sim \mathrm{Bin}(1,p)$.
Thus
\begin{equation*}
W_1 +  \cdots + W_M \sim  \mathrm{Bin}(M,p).
\end{equation*}
If $m$ out of $M$ intervals cover $\theta$ then $p$ is estimated by $m/M$.
An approximate $95 \%$ confidence interval for $p$ is
\begin{equation*}
\frac{m}{M} \pm 1.96 \sqrt{\frac{\frac{m}{M} \left(1 - \frac{m}{M} \right)}{M}}. 
\end{equation*}
The longest confidence interval ($0.00066$) is that for using $9$ iterative stages and a total of $20$ measurements at each stage.
Using the half-length of this confidence interval, the confidence interval 
\begin{equation*}
\frac{m}{100,000} \pm 0.00033
\end{equation*}
can be computed from the results given in Table \ref{nni2b.tab}. It has coverage probability at least $95\%$.

\section{The noisy case}
It is known that  when even a small amount of noise is present
the performance of phase estimation schemes is greatly reduced \citep{huelga97,shaji07}.  

This section investigates the performance of the iterative estimation algorithm when depolarizing noise is present. The channel
\begin{equation}
\rho_0 \mapsto (1-r) U_\theta \rho_0 U_\theta^\dagger + \frac{r}{2}\mathbb{I}_2, \qquad 0 < r < 1,
\label{eq.noisy}
\end{equation}
is considered, where $U_\theta$ is the same as before, (\ref{U}), and $\rho_0 = | \psi_x \rangle \langle \psi_x|$. (The channel (\ref{eq.noisy}) is identical to  $U_\theta \rho_0 U_\theta^\dagger$ undergoing phase damping with $\lambda = r(2-r)$  \cite[p.\ 383]{chuang00}.)
\cite{zheng06} gave the very interesting result that if $r >0$,  then the optimal asymptotic rate at which $1- \langle F(U_{\hat \theta}, U_\theta) \rangle$ approaches zero is given by the standard quantum limit.

Defining $n'$ as the maximum number of times the experimenter lets $U_\theta$ act on the same input state, \cite{zheng06} argued that if $(1-r)^{n'}$ is close to $1$, and thus $n' r << 1$, then it is still possible to estimate $\theta$ as before with the rate $O((\log n/n)^2)$. 

The whole point of using an iterative scheme is that the distinguishability of
$\theta$  from $\cos(n 2 \pi \theta)$, with $n >>1$, is considerably greater than from $\cos(2 \pi\theta)$.

To measure distinguishability, the quantity $F_\theta^M/m$ will be used, where $m$ is the number of times $U_\theta$ acts on the same input state.
This is because of interest is to maximize the distinguishability of $\theta$ per use of the channel.

If there is no noise, and the experimenter lets $U_\theta$ act $m$ times on the input state and measures in $x$, then outcome $0$ is observed with probability $p(0;\theta) = (1 + \cos (m 2 \pi \theta))/2$ and $1$ with probability  $p(1;\theta) = 1 - p(0;\theta)$. The Fisher information from this measurement is $F_\theta^{M_x} = 4 \pi^2 m^2$, which is equal to the SLD quantum information. Measuring in $y$ gives the same Fisher information. Thus $F_\theta^{M_x}/m = F_\theta^{M_y}/m = 4 \pi^2 m$.
At the $k$th stage of the iterative procedure, the experimenter lets $U_\theta$ act $m=2^{k-1}$ times on the input state, and so $F_\theta^{M_x}/m = F_\theta^{M_y}/m = \pi^2 2^{k+1}$. Thus $F_\theta^{M}/m$ (where $M$ is an arbitrary measurement in $x$ or $y$) increases exponentially with $k$.

In the noisy case, when the experimenter lets $U_\theta$ act $m$ times on the output state and then measures in $x$, outcome $0$ is observed  with probability $p(0;\theta) = (1 + (1-r)^m \cos (m2 \pi \theta))/2$ and $1$ with probability  $p(1;\theta) = 1 - p(0;\theta)$. Measuring in $y$, outcome $0$ is observed with probability $p(0;\theta) = (1 + (1-r)^m \sin (m2 \pi \theta))/2$ and $1$ with probability  $p(1;\theta) = 1 - p(0;\theta)$.
This gives
\begin{eqnarray*}
F^{M_x}_\theta &=& \frac{4\pi^2m^2 (1-r)^{2m} \sin^2(2m \pi\theta)}{1-(1-r)^{2m} \cos^2(2m \pi\theta)}\\
 F^{M_y}_\theta &=& \frac{4\pi^2m^2 (1-r)^{2m} \cos^2(2m \pi\theta)}{1-(1-r)^{2m} \sin^2(2m \pi\theta)} \\
 H_\theta &=& 4\pi^2m^2 (1-r)^{2m}. 
\end{eqnarray*}
Notice that
\begin{equation*}
F^{M_x}_\theta + F^{M_y}_\theta \approx  H_\theta. 
\end{equation*}
Thus measuring both in $x$ and $y$, the average Fisher information from a single measurement $M$ is approximately $H_\theta/2$. 

 The maximal value of $F_\theta^M/m$, taken over $m$, will occur close to the maximal value of $H_\theta/m$.
When $r >0$, $H_\theta/m$, and hence $F_\theta^M/m$, does not increase indefinitely with $m$.  Instead it reaches its maximum at
\begin{equation}
m = - \frac{1}{2\log (1-r)},
\end{equation}
after which it decreases.
When $r$ is small, this maximum is obtained at 
\begin{equation}
m  \approx \frac{1}{2r}.
\label{bestrm}
\end{equation}
Thus in the noisy case the number of iterative stages that should be performed is limited by the amount of noise.
The number of stages that can be performed, for small $r$,  such that $H_\theta/m$, and hence $F_\theta^M/m$, increases at each stage is approximately $l \approx - \log_2 r$.

\begin{figure}[htp]
 \centering\includegraphics[totalheight=0.42\textheight]{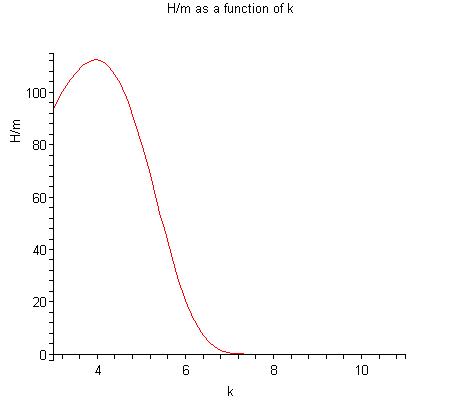}
 \caption[a]{$H_\theta/m$ at the $k$th iterative stage, with $r=2^{-4}$.}
 \label{fig:P4}
 \end{figure}

\begin{figure}[htp]
 \centering\includegraphics[totalheight=0.42\textheight]{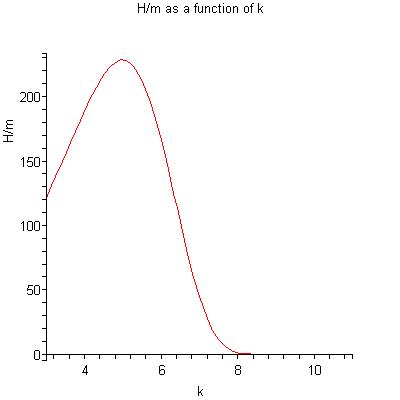}
 \caption[a]{$H_\theta/m$ at the $k$th iterative stage, with $r=2^{-5}$.}
 \label{fig:P5}
 \end{figure}

\begin{figure}[htp]
 \centering\includegraphics[totalheight=0.42\textheight]{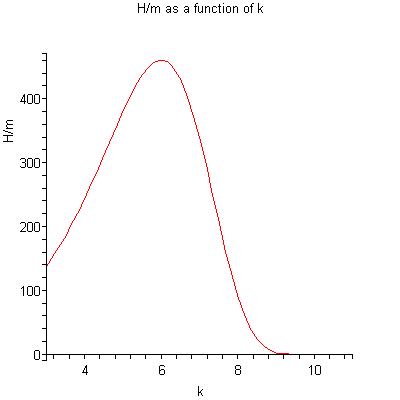}
 \caption[a]{$H_\theta/m$ at the $k$th iterative stage, with $r=2^{-6}$.}
 \label{fig:P6}
 \end{figure}

\begin{figure}[htp]
 \centering\includegraphics[totalheight=0.42\textheight]{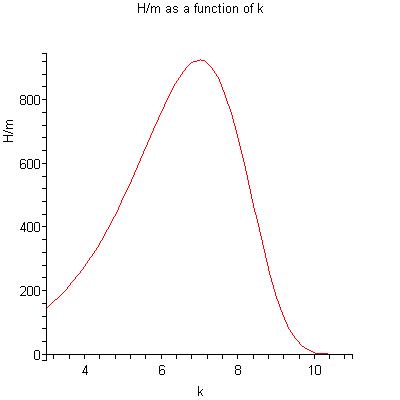}
 \caption[a]{$H_\theta/m$ at the $k$th iterative stage, with $r=2^{-7}$.}
 \label{fig:P7}
 \end{figure}

\begin{figure}[htp]
 \centering\includegraphics[totalheight=0.42\textheight]{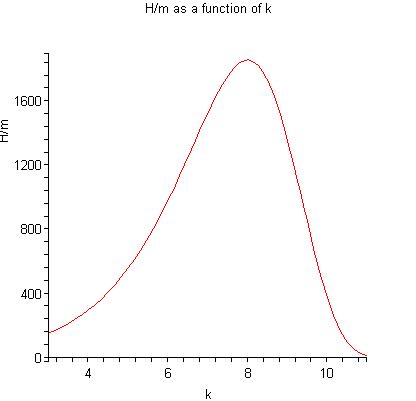}
 \caption[a]{$H_\theta/m$ at the $k$th iterative stage, with $r=2^{-8}$.}
 \label{fig:P8}
 \end{figure}

Figures \ref{fig:P4} -- \ref{fig:P8}  give $H_\theta/m$ at the $k$th iterative stage.  It can be seen that $H_\theta/m$ increases up to $k = -\log_2 r$, decreases slightly near $k = -\log_2 r +1$ and falls rapidly for $k > -\log_2 r +1$.

Tables \ref{nni20.tab} --  \ref{nni200.tab} contain the results of simulations, for  magnitudes of noise $r=2^{-4},2^{-5}, \dots, 2^{-8}$ and total number of iterative stages $l = 4, \dots,9$ --  the number of measurements at each stage is fixed.
Consider the diagonals of Tables \ref{nni20.tab} --  \ref{nni200.tab}, from $r = 2^{-4}$, $l =4$ to $r = 2^{-8}$, $l =8$. This corresponds 
to the experimenter performing $l = -\log_2 r$ iterative stages, which involves going up to the iterative stage at which $F_\theta^M/m$ is maximized. 
Similarly, the diagonal from $r = 2^{-4}$, $l =5$ to $r = 2^{-8}$, $l =9$  corresponds 
to the experimenter performing $l = -\log_2 r + 1$ iterative stages etc.
It is interesting to note that when $l >  -\log_2 r$, there is a significant decrease in the number of  confidence intervals containing $\theta$, and when $l >  -\log_2 r +1$, an even greater  decrease in the number of  confidence intervals containing $\theta$. For example, using $30$ measurements at each stage, if the experimenter performs $l = -\log_2 r$ iterative stages then the final confidence interval contains $\theta$ approximately $98\%$ of the time; if the experimenter increases to $l = -\log_2 r +1$ iterative stages, then the final confidence interval contains the true value of $\theta$ approximately  $89\%$ of the time. If the experimenter increases to $l = -\log_2 r +2$ iterative stages, then approximately $61\%$ of the time the final confidence interval contains $\theta$ -- a considerable drop in performance. It can be seen from Table \ref{nni200.tab}, for which $200$ measurements are performed at each stage,  that this drop in performance does not just occur when performing relatively small numbers of measurements at each stage. 

It is interesting to see that the drop off in performance, in terms of the coverage probability -- which can be calculated from Tables \ref{nni20.tab} --  \ref{nni200.tab},  occurs at the same point as the drop in performance as measured by $H_\theta/m$, and consequently $F_\theta/m$ -- seen in Figures \ref{fig:P4} -- \ref{fig:P8}.

Since $F_\theta^M/m$  starts to decrease after $l = -\log_2 r$ iterative stages, it makes no sense to choose $l > -\log_2 r$.
The simulations also suggest that it is safer to do no more than $l = -\log_2 r$ iterative stages. This is equivalent to letting $U_\theta$ act no more than $n' = 1/(2r)$ times on the same input state.
Thus for a given level of noise the experimenter can let $U_\theta$ act on an input state more times than $n'$ satisfying  $n' r << 1$ (though the $O((\log n/n)^2)$ rate may not be kept).
A sensible suggestion is, more generally, that for the channel (\ref{eq.noisy}) the optimum number of iterative stages, where at the $k$th stage $U_\theta$ is used $2^{k-1}$ times,  is $l = \lfloor-\log_2 r  \rfloor$. 

A related question was considered in \cite{rubin07}, where the `stopping point', was $N$ the number of entangled photons to be included in the NOON input states. Rubin and Kaushik found that the optimal precision in measurement occurred for $N=1.279/L$, where $L$ is the magnitude of loss (analogous to the point, $n'=1/(2r)$, at which $F_\theta^M/m$ is maximized).  

If $l = -\log_2 r$ iterative stages are performed and the final confidence interval contains $\theta$, this corresponds to a precision $| \hat \theta - \theta|_1 \leq r/3$.
If the experimenter desires greater precision in his final estimate than $| \hat \theta - \theta|_1 \leq  r/3$, then it seems sensible for him to perform more measurements at the final iterative stage.


\begin{table}[th]
\centering
\begin{tabular}{|c|c|c|c|c|c|c|}
\hline
&\multicolumn{6}{c|}{Number of iterative stages $(l)$}\\
\cline{2-7}
$r$ &   4 & 5 & 6 & 7 & 8 & 9  \\
\hline
$2^{-4}$ &  94,601 & 78,896 & 48,831 & 24,429 & 11,514 & 5,430  \\
$2^{-5}$ &  98,804 & 94,854 &  79,207 & 49,625 & 24,891 & 11,738  \\
$2^{-6}$ &   99,608 & 98,728 &  94,840 & 79,428 & 50,121 & 24,887  \\
$2^{-7}$ &   99,779 & 99,571 &  98,768 & 94,917 & 79,544 & 50,130  \\
$2^{-8}$  &  99,823 & 99,719 & 99,571  & 98,764  & 94,715  & 79,745  \\
\hline 

\end{tabular} 
\caption{Numbers of trials out of 100,000 with $| \hat \theta - \theta|_1 \leq 1/(2^l \times 3)$,
with $N_{tot} =20$.}
 \label{nni20.tab}
\end{table} 

\begin{table}[th]
\centering
\begin{tabular}{|c|c|c|c|c|c|c|}
\hline
&\multicolumn{6}{c|}{Number of iterative stages $(l)$}\\
\cline{2-7}
$r$ &   4 & 5 & 6 & 7 & 8 & 9  \\
\hline
$2^{-4}$ &  98,290 & 88,340 & 60,423 & 32,445 & 16,059 & 8,042  \\
$2^{-5}$ &  99,804 & 98,408 &  88,537 & 61,293 & 32,756 & 16,460  \\
$2^{-6}$ &   99,967 & 99,807 &  98,430 & 88,708 & 61,148 & 32,595  \\
$2^{-7}$ &   99,985 & 99,955 &  99,802 & 98,476 & 88,895 & 61,699  \\
$2^{-8}$  &  99,988 & 99,977 &  99,962 & 99,812 & 98,467 & 88,864  \\
\hline 

\end{tabular} 
\caption{Numbers of trials out of 100,000 with $| \hat \theta - \theta|_1 \leq 1/(2^l \times 3)$,
with $N_{tot} =30$.}
 \label{nni30.tab}
\end{table} 

\begin{table}[th]
\centering
\begin{tabular}{|c|c|c|c|c|c|c|}
\hline
&\multicolumn{6}{c|}{Number of iterative stages $(l)$}\\
\cline{2-7}
$r$ &   4 & 5 & 6 & 7 & 8 & 9  \\
\hline
$2^{-4}$ &  99,336 & 91,501 & 63,433 & 33,429 & 16,130 & 7,940  \\
$2^{-5}$ &  99,972 & 99,349 &  91,753 & 64,469 & 34,079 & 16,262  \\
$2^{-6}$ &   99,999 & 99,962 &  99,391 &  92,139 & 64,768 & 33,861  \\
$2^{-7}$ &   99,999 & 99,993 &  99,960 & 99,388 &  92,190 & 64,744  \\
$2^{-8}$  &  99,998 & 99,999  & 99,997 & 99,957 & 99,371   &  92,287 \\
\hline 

\end{tabular} 
\caption{Numbers of trials out of 100,000 with $| \hat \theta - \theta|_1 \leq 1/(2^l \times 3)$,
with $N_{tot} =40$.}
 \label{nni40.tab}
\end{table} 

\begin{table}[th]
\centering
\begin{tabular}{|c|c|c|c|c|c|c|}
\hline
&\multicolumn{6}{c|}{Number of iterative stages $(l)$}\\
\cline{2-7}
$r$ &   4 & 5 & 6 & 7 & 8 & 9  \\

\hline
$2^{-4}$ &  99,741 & 94,475 & 68,789 & 37,529 & 18,793 & 9,308  \\
$2^{-5}$ &  99,991 & 99,738 &  94,644 & 69,626 & 37,976 & 19,139  \\
$2^{-6}$ &   99,999 & 99,993 &  99,759 &  94,909 & 70,021 & 38,232  \\
$2^{-7}$ &   100,000 & 99,998 &  99,995 & 99,770 &  95,030 & 70,402  \\
$2^{-8}$  &  100,000 & 100,000  & 99,999 & 99,995 & 99,790  &  94,983 \\
\hline 

\end{tabular} 
\caption{Numbers of trials out of 100,000 with $| \hat \theta - \theta|_1 \leq 1/(2^l \times 3)$,
with $N_{tot} =50$.}
 \label{nni50.tab}
\end{table}

\begin{table}[th]
\centering
\begin{tabular}{|c|c|c|c|c|c|c|}
\hline
&\multicolumn{6}{c|}{Number of iterative stages $(l)$}\\
\cline{2-7}
$r$ &   4 & 5 & 6 & 7 & 8 & 9  \\
\hline
$2^{-4}$ &  99,993 & 98,780 & 78,515 & 43,641 & 21,599 & 10,983  \\
$2^{-5}$ &  100,000 & 99,994 &  98,924 & 79,739 & 44,374 & 22,150  \\
$2^{-6}$ &   100,000 & 100,000 &  99,999 & 98,904 & 79,899 & 44,762  \\
$2^{-7}$ &   100,000 & 100,000 &  100,000 & 99,997 & 98,966 & 80,004  \\
$2^{-8}$  &  100,000 & 100,000 &  100,000 & 100,000 & 99,998 & 98,989  \\
\hline 

\end{tabular} 
\caption{Numbers of trials out of 100,000 with $| \hat \theta - \theta|_1 \leq 1/(2^l \times 3)$,
with $N_{tot} =100$.}
 \label{nni100.tab}
\end{table} 

\begin{table}[th]
\centering
\begin{tabular}{|c|c|c|c|c|c|c|}
\hline
&\multicolumn{6}{c|}{Number of iterative stages $(l)$}\\
\cline{2-7}
$r$ &   4 & 5 & 6 & 7 & 8 & 9  \\
\hline
$2^{-4}$ &  100,000 & 99,907 & 87,516 & 50,576 & 25,592 & 12,868  \\
$2^{-5}$ &  100,000 & 100,000 &  99,937 & 88,393 & 51,414 & 26,064  \\
$2^{-6}$ &   100,000 & 100,000 &  100,000 & 99,946 & 88,754 & 52,022  \\
$2^{-7}$ &   100,000 & 100,000 &  100,000 & 100,000 & 99,953 & 88,932  \\
$2^{-8}$  &  100,000 & 100,000 &  100,000 & 100,000 & 100,000 & 99,938  \\
\hline 

\end{tabular} 
\caption{Numbers of trials out of 100,000 with $| \hat \theta - \theta|_1 \leq 1/(2^l \times 3)$,
with $N_{tot} =200$.}
 \label{nni200.tab}
\end{table}

\appendix

 \renewcommand{\baselinestretch}{1.00}

\chapter{Notation} \label{notation}
\begin{table}[th]
\centering
\begin{tabular}{|l|l|}
\hline
notation & definition \\
\hline
$| \psi \rangle$ &  finite dimensional complex column vector of unit length (see (\ref{eq.vecpsi})).
\\
$\langle \psi |$ & dual of $| \psi \rangle$ (see (\ref{eq.vecpsid})).
\\
$\langle \psi | \phi \rangle$ & inner product of $| \psi \rangle$ and $| \phi \rangle$ (see (\ref{innerp1})).
\\
$| \phi \rangle \langle \psi |$ & outer product of $| \psi \rangle$ and $| \phi \rangle$ (see (\ref{eq,op1})).
\\
$| 0 \rangle$ & $(1, 0)^T$  ($T$ denotes transpose).
\\
$| 1 \rangle$ & $(0, 1)^T$. 
\\
$z^*$ & complex conjugate of $z$.
\\
$\rho$ & density matrix (see Section \ref{mb:states}).
\\
$\mathbb{I}$ & identity matrix. 
\\
$\mathbb{I}_d$ & $d \times d$ identity matrix. 
\\
$\sigma_x, \sigma_y, \sigma_z$ & Pauli matrices (\ref{eg:Paulim}).
\\
$\sigma_1, \sigma_2, \sigma_3$ & Pauli matrices.
\\
$M$ & POVM  (see Section \ref{measurements}).
\\
\hline

\end{tabular}
\end{table}

\begin{table}[th]
\centering
\begin{tabular}{|l|l|}
\hline
notation & definition \\
\hline
$M_m$ & element of POVM $M$ corresponding to outcome $m$.
\\
$\mathcal{H}$ & Hilbert space.
\\
$\mathcal{H}_{A,B}$ & $\mathcal{H}_A \otimes \mathcal{H}_B$  an extended Hilbert space.
\\
$A^\dagger$ & Hermitian transpose of $A$ (see Section \ref{sysa}).
\\
$| \phi^A \phi^B \rangle$ &  $| \phi^A \rangle \otimes |\phi^B \rangle$.
\\
$S(\mathcal{H})$ & set of states on the Hilbert space $\mathcal{H}$.
\\
$\rho_A$ & reduced state on $S(\mathcal{H}_A)$ (see Section \ref{sec:pt}).
\\
$E_k$ & Kraus operator (see Section \ref{sec:qch}).
\\
$U$ & unitary matrix (any matrix satisfying $U U^\dagger = U^\dagger U = \mathbb{I}$).
\\
$\mathcal{E}$ & quantum channel (see Section \ref{sec:qch}).
\\
$F_\theta$ & Fisher information
(See Section \ref{eq.FI00a}).
\\
$F_\theta^M$ & Fisher information from single measurement using $M$
(See section \ref{eq.FI00a}).
\\
$\leadsto$ & converges in distribution to.
\\
$\langle  F \rangle$ & expectation of $F$.
\\
$\rho_\theta$ & parameterized family of states.
\\
$\lambda_{SLD}$, $\lambda$, $\lambda_\theta$ & SLD quantum score (see (\ref{eq:sldscore})).
\\
$C_E(\theta)$ & Sarovar and Milburn's bound based on arbitrary
\\
&  set of Kraus operators $E = \{ E_k \}$ (see \ref{ce}).
\\
$\Upsilon_k$ & canonical Kraus operator (see before (\ref{eq:condkco})).
 \\
 $C_\Upsilon(\theta)$ & Sarovar and Milburn's bound based on \\
 & canonical Kraus operators $\{ \Upsilon_k \}$.
 \\
$C_L(\theta)$ & metric derived from $C_\Upsilon$ (see (\ref{calebmetric})).
\\
$H^{S}_\theta$, $H_\theta$ & SLD quantum information (see Section \ref{subsec:SLDQFI} ).
\\
$F(\hat U, U)$ & fidelity between $\hat U$ and $U$ (see (\ref{eq.fhuu3})). 
\\
\hline

\end{tabular}
\end{table}

\printindex


\bibliography{Biblio}
\addcontentsline{toc}{chapter}{\bibname}

\end{document}